\documentclass[twocolumn,twocolappendix,trackchanges]{aastex701}

\usepackage{amsmath,amssymb}
\usepackage{physics}
\usepackage{amsfonts}
\usepackage{mathtools}
\usepackage{footmisc}

\usepackage{etoolbox}
\usepackage[refpage]{nomencl}
\usepackage{xcolor}

\providetoggle{nomsort}
\settoggle{nomsort}{true} 

\makeatletter
\iftoggle{nomsort}{%
    \let\old@@@nomenclature=\@@@nomenclature        
        \newcounter{@nomcount} \setcounter{@nomcount}{0}%
        \newcommand{\threedigits}[1]{\ifnum#1<100 0\two@digits{#1} \else \number#1\fi}
        \renewcommand\the@nomcount{\threedigits{\value{@nomcount}}}
        \def\@@@nomenclature[#1]#2#3{
          \addtocounter{@nomcount}{1}%
        \def\@tempa{#2}\def\@tempb{#3}%
          \protected@write\@nomenclaturefile{}%
          {\string\nomenclatureentry{\the@nomcount\nom@verb\@tempa @[{\nom@verb\@tempa}]%
          \begingroup\nom@verb\@tempb\protect\nomeqref{\theequation}%
          |nompageref}{\thepage}}%
          \endgroup
          \@esphack}%
      }{}
\makeatother

\newcommand{\mynomone}[3][section]{%
  \begingroup\edef\x{\endgroup
  \unexpanded{\nomenclature{#2}}%
    {\unexpanded{#3} \hspace*{\fill}  (\csname the#1\endcsname)}}\x}

\newcommand{\mynomtwo}[4][section]{%
  \begingroup\edef\x{\endgroup
  \unexpanded{\nomenclature[#2]{#3}}%
    {\unexpanded{#4} \hspace*{\fill}  (\csname the#1\endcsname)}}\x}

\renewcommand\nomgroup[1]{%
  \item[\bfseries
  \ifstrequal{#1}{A}{Acronyms}{%
  \ifstrequal{#1}{S}{Symbols}{%
  \ifstrequal{#1}{C}{Other Symbols}{}}}%
]}

\newcounter{logglabel}

\makenomenclature

\newcommand{\eq}[1]{Eq.~(\ref{#1})}

\shorttitle{Dynamical Tides in Mean Motion Resonances}
\shortauthors{Farhat \& Chiang}
\graphicspath{{./}{figures/}}

\begin{document}

\title{Rattling Resonant Chains with Dynamical Tides and Parametric Resonance Passage}

\author[0000-0001-7864-6627]{Mohammad Farhat}
\altaffiliation{Miller Fellow}
\email{farhat@berkeley.edu}

\affiliation{Department of Astronomy, University of California, Berkeley, Berkeley, CA 94720-3411, USA  }
\affiliation{Department of Earth and Planetary Science, University of California, Berkeley, Berkeley, CA 94720-4767, USA}

\author[0000-0002-6246-2310]{Eugene Chiang}
\email{echiang@astro.berkeley.edu}
\affiliation{Department of Astronomy, University of California, Berkeley, Berkeley, CA 94720-3411, USA  }
\affiliation{Department of Earth and Planetary Science, University of California, Berkeley, Berkeley, CA 94720-4767, USA}



\begin{abstract}
Close-in sub-Neptunes are born in mean-motion resonant chains that break apart on a timescale of $\sim$100 Myr. If the chains are first ``rattled'' --- have their free eccentricities excited to a few percent --- then breaking occurs on the required timescale. We propose that chains are rattled by internal oscillations tidally driven by the star in the innermost planet. Unlike other proposals, there is no need to invoke any reservoir of planetesimals or external perturbers --- the chain unravels from within. We develop a Hamiltonian theory (and an accompanying toy model) which predicts a parametric instability whereby chain librations exponentially amplify when the tidal forcing frequency equals the sum of the internal mode frequency and the chain libration frequency. We also introduce a new eigenmode solver for rotating planets, \texttt{MAQAM}, that in combination with the evolutionary code \texttt{ORCHARD} enables us to track the frequency evolution of pressure, gravity, and inertial modes in cooling sub-Neptunes. Inertial modes in the gas envelope and fluid mantle can evolve to become momentarily parametrically resonant. Passage through a single parametric resonance can raise the free eccentricities in the chain to a few percent within a few tens of Myrs. 

\end{abstract}


\section{Introduction}
The youngest close-in sub-Neptunes are predominantly in 
mean-motion resonant or near-resonant configurations, unlike their mature counterparts  (\citealt{dai2024prevalence}; see also \citealt{izidoro2017breaking,izidoro2021formation} for prescient theoretical models). 
The fraction of systems containing at least one pair near a first- or second-order mean-motion resonance (MMR) falls from $86\%\pm13\%$ at ages from 10 to 100 Myr, to $38\%\pm12\%$ between 100 Myr and 1 Gyr, to $23\%\pm3\%$ at ages $> 1$ Gyr, based on data from the Transiting Exoplanet Survey Satellite (TESS) and Kepler. Transit-timing variations, presumably driven by proximity to MMRs, are detected 3--4 times more often among young TESS-detected sub-Neptunes than in the predominantly older Kepler sample \citep{lopez2026searching}. 
These findings indicate that resonances are established early, on timescales $\lesssim 10$ Myr when sub-Neptunes are forming \citep[e.g.,][]{lee2016breeding}, and subsequently broken over a timescale of order $100\,{\rm Myr}$. 

The MMRs almost certainly originate from planets convergently migrating in their natal circumstellar gas disks  \citep[e.g.,][]{goldreich1980disk,lee2002dynamics,terquem2007migration,choksi2020sub}. Disk gas damps planet orbital eccentricities and drives the planets deep into resonance, close to equilibrium points --- so close that the chains, if left undisturbed, would survive for Gyrs \citep[e.g.,][]{matsumoto2012orbital,li2025resonant}. 
How strongly do chains need to be ``rattled''  \citep[][]{hadden2026rattle} to eventually break? \citet{choksi2026two} have computed that chains disintegrate (planets cross orbits and collide) over the observed timescale of 100 Myr if their free eccentricities (the components of eccentricities not forced by the resonances) are somehow excited beforehand to values of a few percent.

Many theories lay responsibility for rattling/breaking the chains on bodies that are not members of the chain, present when the planetary system was young and more crowded. These other bodies can be planetesimals or small rocky planets \citep[e.g.,][]{ogihara2026formation,li2026intruder,hadden2026rattle,choksi2026two}, or other planets in other chains \citep[e.g.,][]{goldberg2026close,lorusso2026planetesimal}. These scenarios, depending as they do on objects that are no longer there or are otherwise unobservable, are hard to test, and fine-tuning problems are present to varying degrees. One observable is the period ratio distribution of planet pairs in young, as yet unbroken resonant chains. Some such pairs are wide of commensurability (e.g.~having period ratios of 1.51--1.52 $>$ 3/2), but others are narrow (e.g.~having period ratios 1.48--1.49 $<$ 3/2). The latter is difficult to explain from convergent migration, but could be ascribed to collisions with non-resonant bodies, provided they have the right numbers and masses \citep[][]{choksi2026two}.

In this paper, we consider the possibility that the chain contains within it the seeds of its own destruction. The chain may be rattled by eccentricity tides, raised by the host star on the innermost planet of the chain. Energy and angular momentum may shuttle between the planet's tidally excited internal oscillations and the chain's orbits. Unlike previous studies that treated only dissipative equilibrium tides and their effects on MMRs \citep[e.g.,][]{papaloizou2011tidal,lithwick2012resonant,batygin2013dissipative,delisle2014tidal,delisle2014resonance}, we consider here dynamical tides --- how active ``sloshing'' within the planet can feed into MMR libration. Oscillations can be driven inside both the inflated gaseous envelopes of sub-Neptunes \citep[e.g.,][]{lopez2014understanding,owen2020constraining} and the magma oceans that possibly lie beneath
\citep[e.g.,][]{nicholls2024magma,tang2025reassessing,calder2026most}. Chain librations modulate the orbit of the innermost planet and thus the stellar tide; conversely, deformations inside the planet change how the orbit is tidally forced, and those orbital changes propagate back into the chain via the MMR. We show below that this reciprocal interaction creates a parametric resonance between tidal forcing, chain libration, and planetary internal modes (of which the most relevant will be inertial modes restored by Coriolis forces). As the innermost planet cools over tens of Myrs, its internal mode frequency evolves such the system passes through the parametric resonance. Resonance passage can impart free eccentricities of a few percent to all the planets in the chain --- just the amount required for chain destruction over hundreds of Myrs \citep[][]{choksi2026two}.

Before developing the formal theory, we first sketch a reduced model that illustrates the basic physics. The model comprises two simple harmonic oscillators, one which is forced and which represents the planetary internal mode forced by the stellar tide, and a second representing chain libration. When the two oscillators are coupled, there can be parametric resonance and instability.

\subsection{A Two-Oscillator Toy Model}
\label{sec:toy_model}

Consider a planet with an internal oscillation mode driven by one harmonic of the stellar tide. Characterizing the internal mode by a displacement $x(t)$ and ignoring damping, we have 
\begin{equation}
\ddot x+\omega_\alpha^2x=F_0\cos\sigma t,
    \label{eq:toy_single_planet}
\end{equation}
where $\omega_\alpha$ is the mode's natural frequency, $\sigma$ is the tidal forcing frequency, and $F_0$ is the forcing amplitude. Now add a second planet in mean-motion resonance (MMR) with the first. The MMR has an equilibrium about which there can be a small libration. Represent the libration by a second undamped oscillator,
\begin{equation}
\ddot q+\omega_a^2q=0
    \label{eq:toy_free_libration}
\end{equation}
with natural frequency $\omega_a \ll \omega_\alpha$ and displacement $q(t)$ (a proxy for any librating quantity such as the MMR argument).

The MMR libration changes the inner planet's eccentricity and semimajor axis, which in turn changes the stellar tide acting on that planet. Consider a change in tidal forcing amplitude proportional to $q$, and neglect changes in the tidal forcing phase.  
The interaction potential 
is then
\begin{equation}
U_{\rm int}=-x(F_0+\kappa_{\rm t}q)\cos\sigma t,
    \label{eq:toy_interaction_energy}
\end{equation}
where the coupling coefficient $\kappa_{\rm t}>0$ measures the change in forcing amplitude per unit $q$. The forces $-\partial U_{\rm int}/\partial x$ and $-\partial U_{\rm int}/\partial q$ couple our two oscillators: 
\begin{align} \label{eq:toy_coupled1}
\ddot x+\omega_\alpha^2x&=(F_0+\kappa_{\rm t}q)\cos\sigma t,\\
\ddot q+\omega_a^2q&=\kappa_{\rm t}x\cos\sigma t.
    \label{eq:toy_coupled2}
\end{align}    
Thus MMR libration changes the tide  driving the planet's internal oscillation; conversely, changes in the planet's internal structure change how the orbit is tidally forced and by extension how the planets librate in the MMR. We refer to this two-oscillator system as parametrically driven because the driver depends on the system's own parameters, through $q$ and $x$ on the right-hand sides of Eqs.~(\ref{eq:toy_coupled1}) and (\ref{eq:toy_coupled2}) (as distinct from the wholly external force $F_0 \cos \sigma t$). Our model generalizes the familiar parametrically driven pendulum, where periodically varying its length at frequency $\sigma\simeq2\omega$ amplifies oscillations of natural frequency $\omega$ \citep{landau1982mechanics}.

We ask whether the reciprocal interaction between the internal mode and the MMR chain can amplify MMR libration (as needed to eventually break the resonant chain).
We consider for simplicity the case when 
both oscillators start at rest, $x(0)=\dot x(0)=q(0)=\dot q(0)=0$. The external driver $F_0 \cos \sigma t$ sets the system into motion (analogous to how there is an eccentricity tide for the non-zero forced eccentricity of the MMR). Over successive oscillations, energy can be fed coherently and efficiently from the planet's internal mode into MMR libration at either of two resonances
whose existence can be seen 
as follows. 
Momentarily neglecting the coupling term $\kappa_t q$ in \eq{eq:toy_coupled1} gives
\begin{equation}
    x(t)=
X_0
    \left(\cos\sigma t-\cos\omega_\alpha t\right),
\qquad
X_0
=\frac{F_0}{\omega_\alpha^2-\sigma^2}.
    \label{eq:toy_planetary_response}
\end{equation}
Substituting this response into the 
coupling term driving MMR libration (RHS of Eq.~\ref{eq:toy_coupled2}) 
gives
\begin{align}
\kappa_{\rm t}x\cos\sigma t
&=\frac{\kappa_{\rm t}
X_0
}{2} \left(1+\cos2\sigma t\right)
-\frac{\kappa_{\rm t}
X_0
}{2}\big\{\cos[(\omega_\alpha-\sigma)t]\nonumber\\
&\qquad+\cos[(\omega_\alpha+\sigma)t]\big\}.
    \label{eq:toy_returning_force}
\end{align}
The term that varies as $\cos (\omega_\alpha-\sigma)t$ 
can drive MMR libration resonantly when  
either of two ``sideband" conditions $\sigma=\omega_\alpha\pm\omega_a$ is satisfied. The remaining forcing terms are constant or oscillate much faster than the libration.

To study the resonant sideband dynamics, we keep only the resonant driver in Eq.~(\ref{eq:toy_returning_force}) and restore $\kappa_t q$ to Eq.~(\ref{eq:toy_coupled1}). We allow the natural-frequency responses of both oscillators to have slowly varying amplitudes $X(t)$ and $Q(t)$. 
To leading order,\footnote{Using cosine to describe the $x$-oscillation follows from Eq.~\eqref{eq:toy_planetary_response}, which in turn follows from our assumed at-rest initial conditions. Allowing for arbitrary initial conditions (i.e.~writing 
$q=Q_c(t)\cos\omega_a t+Q_s(t)\sin\omega_a t$ and $x=
X_0 
\cos\sigma t+X_c(t)\cos\omega_\alpha t+X_s(t)\sin\omega_\alpha t$) changes none of our conclusions.
}
\begin{align}
x(t)&\simeq 
X_0 
\cos\sigma t+X(t)\cos\omega_\alpha t,\nonumber\\
q(t)&\simeq Q(t)\sin\omega_a t,
    \label{eq:toy_amplitudes}
\end{align}
with $X(0)=
-X_0$ 
and $Q(0)=0$. Substituting into Eq.~\eqref{eq:toy_coupled1} 
leaves
\begin{equation}
\ddot X\cos\omega_\alpha t-2\omega_\alpha\dot X\sin\omega_\alpha t\simeq\kappa_{\rm t}Q\sin\omega_a t\cos\sigma t.
    \label{eq:toy_ddot_X}
\end{equation}
Let $\tau$ denote a characteristic timescale over which $X$ and $Q$ change, such that $\omega_\alpha\gg\omega_a\gg\tau^{-1}$. The term with $\ddot X$ is then smaller than $2\omega_\alpha\dot X$ by a factor of order $(\omega_\alpha\tau)^{-1}$ and can be neglected. Eq.~\eqref{eq:toy_ddot_X} becomes 
\begin{equation}
-2\omega_\alpha\dot X\sin\omega_\alpha t\simeq\frac{\kappa_{\rm t}Q}{2} \left\{
\sin[(\omega_a+\sigma)t]+\sin[(\omega_a-\sigma)t] \right\}.
    \label{eq:toy_d_X}
\end{equation}
We multiply Eq.~\eqref{eq:toy_d_X} by $\sin\omega_\alpha t$ and time-average over an interval $2\pi/\omega_a\ll T\ll\tau$ within which $Q$ and $X$ are constant to leading order. At $\sigma=\omega_\alpha-\omega_a$, the two  terms on the RHS average to 1/2 and 0, respectively. At $\sigma=\omega_\alpha+\omega_a$, these averages are 0 and $-1/2$. The LHS averages to $-\omega_\alpha\dot X$, giving
\begin{equation}
\dot X=
\begin{cases}
-\dfrac{\kappa_{\rm t}}{4\omega_\alpha}Q,
& \sigma=\omega_\alpha-\omega_a,\\[6pt]
+\dfrac{\kappa_{\rm t}}{4\omega_\alpha}Q,
& \sigma=\omega_\alpha+\omega_a.
\end{cases}
    \label{eq:toy_planetary_amplitude}
\end{equation}
Now we work on  Eq.~\eqref{eq:toy_coupled2}. Substituting Eq.~\eqref{eq:toy_amplitudes} into Eq.~\eqref{eq:toy_coupled2} and neglecting $\ddot Q$ under the same slow-evolution assumption gives
\begin{equation}
2\omega_a\dot Q\cos\omega_a t\simeq\kappa_{\rm t}X_0\cos^2\sigma t
+\kappa_{\rm t}X\cos\omega_\alpha t\cos\sigma t.
    \label{eq:toy_orbital_projection}
\end{equation}
Multiplying by $\cos\omega_a t$ and averaging over $T$, the first term contributes zero and the second contributes $\kappa_{\rm t}X/4$ at either sideband resonance. Thus
\begin{equation}
\dot Q=\frac{\kappa_{\rm t}}{4\omega_a}X.
    \label{eq:toy_orbital_amplitude}
\end{equation}
Differentiating Eq.~\eqref{eq:toy_orbital_amplitude} and substituting Eq.~\eqref{eq:toy_planetary_amplitude} yields
\begin{equation}
\ddot Q=
\begin{cases}
-\dfrac{\kappa_{\rm t}^{\,2}}{16\omega_\alpha\omega_a}Q,
& \sigma=\omega_\alpha-\omega_a,\\[6pt]
+\dfrac{\kappa_{\rm t}^{\,2}}{16\omega_\alpha\omega_a}Q,
& \sigma=\omega_\alpha+\omega_a.
\end{cases}
    \label{eq:toy_amplitude_second_order}
\end{equation}
Define
\begin{equation}
r=\frac{\kappa_{\rm t}}{4\sqrt{\omega_\alpha\omega_a}}.
    \label{eq:toy_growth_rate}
\end{equation}
With $Q(0)=0$ and $\dot Q(0)=-\kappa_{\rm t}X_0/(4\omega_a)$, the solutions at $\sigma=\omega_\alpha-\omega_a$ are
\begin{align}
x(t)&\simeq 
X_0
[\cos\sigma t-\cos(rt)\cos\omega_\alpha t],\nonumber\\
q(t)&\simeq-
X_0
\sqrt{\frac{\omega_\alpha}{\omega_a}}\sin(rt)\sin\omega_a t,
    \label{eq:toy_full_difference}
\end{align}\begin{figure}[t]
\centering
\includegraphics[width=\linewidth]{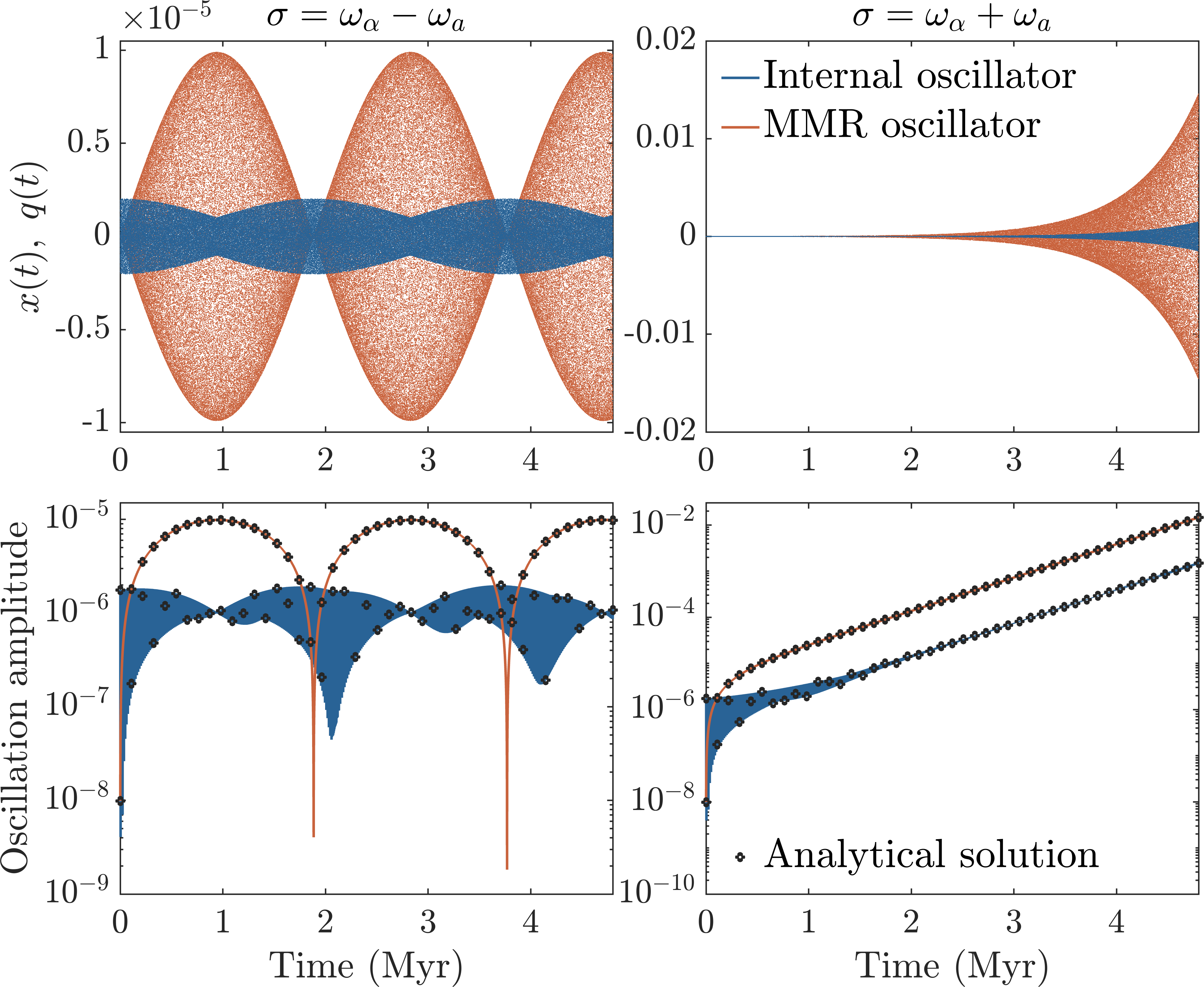}
\caption{The two-oscillator toy model driven at its sideband resonances. Colored curves show numerical solutions of the equations of motion \eqref{eq:toy_coupled1}--\eqref{eq:toy_coupled2} for the two oscillators starting at rest, for the ``difference'' (left) and ``sum'' (right) sidebands. Upper panels show oscillator displacements in scaled units; lower panels show the oscillation amplitudes. Black markers give the analytical solutions from Eqs.~\eqref{eq:toy_full_difference} and \eqref{eq:toy_full_sum}. Parameters used for this calculation are $\omega_\alpha=300\,{\rm yr}^{-1}$, $\omega_a=3.5\,{\rm yr}^{-1}$, $\kappa_{\rm t}=2\times10^{-4}\,{\rm yr}^{-2}$, and $F_0=2\times10^{-3}$ scaled units ${\rm yr}^{-2}$.}
\label{fig:toy_model_integration}
\end{figure}
i.e., $q$ and $x$ undergo bounded oscillations. We shall show in the full theory of Section \ref{sec:theory} that this sideband indeed mediates a bounded 
oscillatory exchange of action between 
planetary internal oscillations and the MMR. By comparison, at the sideband resonance $\sigma = \omega_\alpha + \omega_a$, the solutions are
\begin{align}
x(t)&\simeq 
X_0
[\cos\sigma t-\cosh(rt)\cos\omega_\alpha t],\nonumber\\
q(t)&\simeq-
X_0
\sqrt{\frac{\omega_\alpha}{\omega_a}}\sinh(rt)\sin\omega_a t.
    \label{eq:toy_full_sum}
\end{align}
Here both oscillation amplitudes grow asymptotically as $e^{rt}$. The reciprocal interaction between the planet's internal mode and MMR libration, mediated by stellar tidal forcing, drives a parametric instability 
(whose energy source is ultimately the orbital energy of the innermost planet). Figure~\ref{fig:toy_model_integration} compares numerical integrations of the original equations of motion \eqref{eq:toy_coupled1}-\eqref{eq:toy_coupled2} with the approximate analytical 
solutions \eqref{eq:toy_full_difference}-\eqref{eq:toy_full_sum}, confirming bounded oscillations for $\sigma = \omega_\alpha - \omega_a$, and instability for $\sigma = \omega_\alpha + \omega_a$.

The remainder of this paper develops the theory of sideband resonances and parametric instability for real (or at least more real) resonant chains and  internal oscillation modes. In Section~\ref{sec:theory}, we construct the Hamiltonian coupling collective librations of a generic MMR chain to planetary internal normal modes, and derive the conditions for parametric growth in the presence of damping. Section~\ref{sec:Equilibrium_chain} considers a fiducial chain of four sub-Neptunes, constructs its reduced MMR Hamiltonian, and determines its stable points 
and libration modes. In Section~\ref{sec:fiducial_planet_modal_problem}, we model the thermal and structural evolution of the innermost sub-Neptune using the evolutionary code \texttt{ORCHARD}. We then introduce \texttt{MAQAM}, a normal mode solver for the fluid layers of rotating planets, and compute its internal oscillations. In Section~\ref{sec:chain_instability} we study the dynamics of passage through the two parametric sideband resonances, assessing the degree to which passage can amplify the chain's librations subject to constraints from internal mode evolution, including damping. Readers are pointed in particular to Section \ref{subsec:parametric_excitation_example} where a proof-of-concept calculation of resonance passage and eccentricity excitation is made. We summarize and chart future directions in Section \ref{sec:Summary}.

\section{Theory}
\label{sec:theory}
\subsection{Orbital, Spin, and Internal Degrees of Freedom}
We consider $N$ coplanar planets of masses $m_i$ orbiting a star of mass $M_\star$, with $i=1,\ldots,N$ increasing outwards. Conservative dynamics comprise orbital motion, rotation, and internal planetary oscillations. 
For planet $i$, with semimajor axis $a_i$ and eccentricity $e_i$, define the canonical heliocentric Poincar\'{e} pairs $(\lambda_i,\Lambda_i)$ and $(\gamma_i,\Gamma_i)$, where $\lambda_i$ is the mean longitude and $\gamma_i=-\varpi_i$, with $\varpi_i$ the longitude of periapse. The actions are defined as \citep[e.g.,][]{laskar2000spacing}:
\begin{equation}
 \Lambda_i=\beta_i\sqrt{\mu_i a_i},
 \qquad
 \Gamma_i=\Lambda_i\left(1-\sqrt{1-e_i^2}\right),
 \label{eq:poincare_actions}
\end{equation}
where $\beta_i=M_\star m_i/(M_\star+m_i)$, $\mu_i=\mathcal G(M_\star+m_i)$, and $\mathcal G$ is the gravitational constant. Here $\Lambda_i$ is the circular angular momentum and $\Gamma_i$ the angular momentum deficit (AMD), with $G_i=\Lambda_i-\Gamma_i$ the orbital angular momentum. In these variables, the unaveraged orbital Hamiltonian is written as
\begin{align}
    \mathcal{H}_{\rm orb}
    &=
    -\sum_{i=1}^{N}
    {\mu_i^2\beta_i^3\over 2\Lambda_i^2}
    \nonumber\\
    &\quad
    +
    \sum_{i<j}
    \mathcal{H}_{ij}
    (\Lambda_i,\Lambda_j,\Gamma_i,\Gamma_j,
     \lambda_i,\lambda_j,\gamma_i,\gamma_j),
    \label{eq:Horb_unaveraged}
\end{align}
where the first term is the Keplerian part and $\mathcal{H}_{ij}$ contains the direct and indirect perturbative interactions between planets $i$ and $j$.

The spin angle $\theta_i$ and its conjugate momentum $S_i$ describe rotation. For a principal axis rotator, a one-degree-of-freedom reduction of the rigid body Hamiltonian written in the principal axis frame is \citep[e.g.,][]{touma1994lie}
\begin{equation}
    \mathcal{H}_{\rm spin}
    =
    \sum_{i=1}^{N}
    {S_i^2\over 2C_i},
    \label{eq:Hspin}
\end{equation}
where $C_i$ is the moment of inertia about the spin axis.

Internal to each planet $i$, we consider oscillatory modes with real, nonzero frequencies. Mode $\alpha$ of planet $i$ has displacement eigenfunction $\boldsymbol\xi_{i\alpha}(\boldsymbol r)$ and  frequency $\omega_{i\alpha}$ in the frame rotating with $\boldsymbol\Omega_i=\Omega_i\hat{\boldsymbol z}$, where $\hat{\boldsymbol z}$ is normal to the orbital plane and $\boldsymbol r$ is position relative to the planet's center in that frame. Its azimuthal dependence is $\exp(im_{i\alpha}\phi)$, with harmonic integer order $m_{i\alpha}$ and rotating azimuth $\phi$. The mode's inertial frequency is $\omega_{i\alpha,\rm in}=\omega_{i\alpha}+m_{i\alpha}\Omega_i$. Rotating modes are normalized in phase space \citep[e.g.,][]{Friedman1978langrangian,schenk2001nonlinear} using the real norm
\begin{equation}
 b_{i\alpha}
 =\int_{V_i}\rho_i\boldsymbol\xi_{i\alpha}^{*}\cdot
 \left(2\omega_{i\alpha}\boldsymbol\xi_{i\alpha}
       +2i\boldsymbol\Omega_i\times\boldsymbol\xi_{i\alpha}\right)
 \,d^3r,
 \label{eq:mode_gyroscopic_norm}
\end{equation}
where $\rho_i$ is the equilibrium density and $V_i$ the planetary volume. Let $A_{i\alpha}(t)$ carry the amplitudes and phases of the eigenfunctions $\boldsymbol\xi_{i\alpha}(\boldsymbol r)$. The real displacement is then
\begin{equation}
    \boldsymbol\xi_i^{\mathfrak{Re}}(\boldsymbol r,t)
    =\mathfrak{Re}\sum_\alpha
    A_{i\alpha}(t)\boldsymbol\xi_{i\alpha}(\boldsymbol r).
    \label{eq:physical_mode_amplitude_expansion}
\end{equation}
The quadratic energy of a mode in the rotating frame is $E_{i\alpha}^{\rm rot}=\omega_{i\alpha}b_{i\alpha}|A_{i\alpha}|^2/4$. We express the mode dynamics in canonical action-angle form using the action $J_{i\alpha}=E_{i\alpha}^{\rm rot}/\omega_{i\alpha}$ and phase $\psi_{i\alpha}$ such that
\begin{equation}
    J_{i\alpha}=\frac{b_{i\alpha}}{4}|A_{i\alpha}|^2,
    \qquad
    A_{i\alpha}=2\sqrt{\frac{J_{i\alpha}}{b_{i\alpha}}}\,e^{-i\psi_{i\alpha}}.
    \label{eq:mode_amplitude_action_phase}
\end{equation}
The action measures the strength of the oscillation independently of the arbitrary eigenfunction normalization.

In these variables, the quadratic Hamiltonian of the internal modes is
\begin{equation}
    \mathcal{H}_{\rm NM}
    =\sum_{i=1}^{N}\sum_\alpha \omega_{i\alpha}J_{i\alpha}.
    \label{eq:H_NM_action}
\end{equation}
For free oscillations, Hamilton's equations give $\dot J_{i\alpha}=0$ and $\dot\psi_{i\alpha}=\omega_{i\alpha}$. In the nonrotating limit, $b_{i\alpha}=2\omega_{i\alpha}\mathcal M_{i\alpha}$, where the effective mode mass $\mathcal M_{i\alpha}$ is defined as
\begin{equation}
 \mathcal M_{i\alpha}
 =\int_{V_i}\rho_i|\boldsymbol\xi_{i\alpha}|^2\,d^3r.
 \label{eq:mode_mass}
\end{equation}
For a real eigenfunction, Eqs.~\eqref{eq:physical_mode_amplitude_expansion} and \eqref{eq:mode_amplitude_action_phase} recover the usual oscillator displacement proportional to $\sqrt{2J_{i\alpha}/(\mathcal M_{i\alpha}\omega_{i\alpha})}\cos\psi_{i\alpha}$ \citep[see e.g.,][]{wisdom2016dynamic}.

Adding the orbital (Eq.~\ref{eq:Horb_unaveraged}), spin (Eq.~\ref{eq:Hspin}), and internal oscillatory contributions (Eq.~\ref{eq:H_NM_action}) gives, before any coupling, the Hamiltonian
\begin{equation}
    \mathcal{H}_0
    =\mathcal{H}_{\rm orb}
    +\mathcal{H}_{\rm spin}
    +\mathcal{H}_{\rm NM}.
    \label{eq:H0_uncoupled}
\end{equation}
The stellar tide couples the internal modes to the orbital and spin variables through the interaction derived next.

\subsection{Tidal Coupling}
\label{subsec:hamiltonian_tidal_coupling}

Let $U_i(\boldsymbol r,t)$ be the stellar tidal potential per unit mass in the rotating frame of planet $i$, with its constant and dipole terms removed. Its projection onto mode $\alpha$ defines the generalized force \citep{press1977formation,ivanov2004tidal}
\begin{equation}
F_{i\alpha}=-\int_{V_i}\rho_i\boldsymbol\xi_{i\alpha}^{*}\cdot\nabla U_i\,d^3r.
    \label{eq:generalized_force}
\end{equation}
Although $U_i$ is real, $F_{i\alpha}$ is generally complex. Using Eq.~\eqref{eq:physical_mode_amplitude_expansion}, the interaction energy to first order in the displacement is\footnote{Each mode $(\boldsymbol\xi_{i\alpha},\omega_{i\alpha})$ has a conjugate partner $(\boldsymbol\xi_{i\alpha}^{*},-\omega_{i\alpha})$. The two represent the same real oscillation and have opposite signs of $b_{i\alpha}$. To avoid double counting, we retain only the member with $b_{i\alpha}>0$, for which the action $J_{i\alpha}$ is nonnegative.}
\begin{align}
    \mathcal H_{\rm tide}
    &=\sum_i\int_{V_i}\rho_i\boldsymbol\xi_i^{\mathfrak{Re}}\cdot\nabla U_i\,d^3r
    \nonumber\\
    &=-\sum_{i\alpha}\mathfrak{Re}\!\left(A_{i\alpha}^{*}F_{i\alpha}\right).
    \label{eq:H_int_complex}
\end{align}
The tidal force depends on the instantaneous orbital and spin variables and admits the Fourier expansion
\begin{equation}
    F_{i\alpha}
    =\sum_b F_{i\alpha b}(\Lambda_i,\Gamma_i,S_i)e^{-i\Theta_{ib}},
    \label{eq:F_fourier}
\end{equation}
where $b$ labels the tidal harmonics and $F_{i\alpha b}$ are their Fourier coefficients. The harmonic phases are
\begin{equation}
    \Theta_{ib}
    =c^{(1)}_{ib}\lambda_i+c^{(2)}_{ib}\gamma_i+c^{(3)}_{ib}\theta_i,
    \label{eq:Theta_ib}
\end{equation}
where the integers $c^{(1)}_{ib}$, $c^{(2)}_{ib}$, and $c^{(3)}_{ib}$ specify the harmonic.

For the quadrupolar tide at zero obliquity, all tidal harmonics acting on a given mode share one spatial overlap, which is the projection of their force pattern onto $\boldsymbol\xi_{i\alpha}$ as in Eq.~\eqref{eq:generalized_force}. With the orbital phases in $\Theta_{ib}$, we choose the arbitrary phase of the eigenfunction to make this overlap, and hence all $F_{i\alpha b}$, real. Substituting Eqs.~\eqref{eq:mode_amplitude_action_phase} and \eqref{eq:F_fourier} into Eq.~\eqref{eq:H_int_complex} gives
\begin{equation}
    \mathcal H_{\rm tide}
    =\sum_{i\alpha b}\epsilon_{i\alpha b}\sqrt{J_{i\alpha}}
    \cos(\Theta_{ib}-\psi_{i\alpha}),
    \label{eq:tide_mode_basic}
\end{equation}
with
\begin{equation}
    \epsilon_{i\alpha b}
    =-\frac{2F_{i\alpha b}}{\sqrt{b_{i\alpha}}}.
    \label{eq:epsilon_def}
\end{equation}
The coefficient $\epsilon_{i\alpha b}$ measures the coupling of harmonic $b$ to mode $\alpha$. A larger $|\epsilon_{i\alpha b}|$ reflects stronger tidal forcing, greater spatial overlap, or a larger mode displacement at fixed action. The sign of $\epsilon_{i\alpha b}$ has no significance by itself, since it can be absorbed into a phase shift of the harmonic. $\epsilon_{i\alpha b}\sqrt{J_{i\alpha}}$ has dimensions of energy.

Adding $\mathcal{H}_{\rm tide}$ to the parent Hamiltonian of \eq{eq:H0_uncoupled}, the total Hamiltonian becomes
\begin{align}
    \mathcal{H}&=\mathcal{H}_{\rm orb}+\sum_i {S_i^2\over 2C_i}+\sum_{i\alpha}\omega_{i\alpha}J_{i\alpha}\nonumber\\
    &+\sum_{i\alpha b}\epsilon_{i\alpha b}\sqrt{J_{i\alpha}}\cos(\Theta_{ib}-\psi_{i\alpha}).
    \label{eq:H_full_explicit}
\end{align}
At fixed planetary structure, this Hamiltonian is autonomous when the orbital and spin angles and their conjugate actions are retained. The same interaction governs mode excitation and the back reaction on the orbit and spin.

\subsection{Resonant Chain Oscillators}
\label{subsec:resonant_chain_oscillators}

We now reduce the orbital dynamics near a chain of MMRs to calculate how collective chain librations modulate the tidal forcing. We adopt the $N$-planet resonant chain formalism of \citet{delisle2017analytical}, which extends the Hamiltonian expansions of \citet{laskar1995stability} and the  reducing transformations for resonant dynamics \citep[e.g.,][]{henrard1986reducing,wisdom1986canonical} to multi planet chains. 

We transform the Poincar\'{e} variables to resonant variables and average $\mathcal H_{\rm orb}$ over the nonresonant fast angles. After reduction by cyclic coordinates, the retained variables are $\mathbf Z=(\boldsymbol\chi,\mathbf I)$, where $\boldsymbol\chi$ is an independent set of slow angles from which the two-body resonant arguments and Laplace combinations can be reconstructed, and $\mathbf I$ is the vector of their conjugate actions. The reduced Hamiltonian is $\mathcal H_{\rm ch}(\mathbf Z;\boldsymbol\eta)$, with $\boldsymbol\eta$ denoting the actions conserved in the averaged orbital problem.

At fixed $\boldsymbol\eta$, a resonant chain equilibrium $\mathbf Z_\star$ satisfies
\begin{equation}
\left.\nabla_{\mathbf Z}\mathcal H_{\rm ch}\right|_{\mathbf Z_\star}=0.
    \label{eq:chain_fixed_point}
\end{equation}
The local Hamiltonian structure near such an equilibrium follows by writing $\delta{\bf Z}={\bf Z}-{\bf Z}_\star$ and expanding
\begin{equation}
\mathcal H_{\rm ch}=\mathcal H_\star+\frac{1}{2}\delta\mathbf Z^{\mathsf T}\mathbf K_{\rm ch}\delta\mathbf Z+\mathcal O(\delta\mathbf Z^3),
    \label{eq:Hch_quadratic}
\end{equation}
where the linear term vanishes by construction, $\mathcal H_\star=\mathcal H_{\rm ch}(\mathbf Z_\star;\boldsymbol\eta)$, and $\mathbf K_{\rm ch}=\left.\nabla_{\mathbf Z}\nabla_{\mathbf Z}\mathcal H_{\rm ch}\right|_{\mathbf Z_\star}$ is the Hessian evaluated at the fixed point. The linearized Hamiltonian equations are therefore
\begin{equation}
\delta\dot{\mathbf Z}=\mathbf J_{\rm ch}\mathbf K_{\rm ch}\delta\mathbf Z\,
    \label{eq:linearized_chain}
\end{equation}
where ${\bf J}_{\rm ch}$ is the canonical symplectic matrix,
\begin{equation}
    {\bf J}_{\rm ch}= \begin{pmatrix}
        \bf 0 & \mathbb I \\
        -\mathbb I & \bf 0
    \end{pmatrix},
\end{equation}
with $\mathbb I$ and $\mathbf 0$ denoting identity and zero matrices, respectively.

We consider an elliptic equilibrium with nonzero libration frequencies. A real canonical transformation resolves its linearized motion into independent oscillations \citep{arnold1989mathematical},
\begin{equation}
\delta\mathbf Z=\sum_a\sqrt{2J_a}\left[\mathbf u_a\cos\psi_a+\mathbf v_a\sin\psi_a\right].
    \label{eq:chain_mode_expansion}
\end{equation}
Here $J_a\geq0$ and $\psi_a$ are the action and phase of collective oscillation mode $a$. The real vectors $\mathbf u_a$ and $\mathbf v_a$ describe its shape and are normalized so that $(\psi_a,J_a)$ is a canonical pair. We provide their construction in Appendix~\ref{app:chain_mode_normalization}.  For the elliptic equilibrium, the eigenvalues of ${\bf J}_{\rm ch}{\bf K}_{\rm ch}$ occur in pairs $\lambda_a=\pm i|\omega_a|$. The quadratic Hamiltonian (second term in Eq.~\ref{eq:Hch_quadratic}) is then written in the normal form:
\begin{equation}
\mathcal H_{\rm ch}^{(2)}=\sum_a\omega_aJ_a.
    \label{eq:Hch_normal}
\end{equation}
Here $\omega_a$ is the signed frequency governing $\dot\psi_a=\omega_a$. Its magnitude is the libration frequency and its sign is that of the quadratic mode energy. It is noteworthy that an elliptic equilibrium need not be an energy minimum. Together with the inverse resonant transformation, the mode vectors $(\mathbf u_a,\mathbf v_a)$ translate $J_a$ and $\psi_a$ into variations of the planets' orbital elements. These relations determine how each chain libration will modulate the tidal forcing as we detail next, and how much orbital excitation a given action represents.

\subsection{Chain Modulation of the Tidal Interaction}
\label{subsec:chain_modulation}

Chain librations change the tidal forcing through variations of the planets' semimajor axes, eccentricities, and orbital phases.  To first order in the chain oscillation amplitudes, we write
\begin{align}
\Theta_{ib}&=\Theta_{ib}^{\star}(t)+\delta\Theta_{ib},
    \label{eq:Theta_expand_chain}\\
\epsilon_{i\alpha b}&=\epsilon_{i\alpha b}^{\star}+\delta\epsilon_{i\alpha b}.
    \label{eq:epsilon_expand_chain}
\end{align}
Here the superscript $\star$ denotes the reference motion at the chain equilibrium, and the perturbations are evaluated from the oscillatory corrections of the orbital variables; namely:
\begin{align}
\delta\Theta_{ib}
&=\left.\nabla_{\mathbf Z}\Theta_{ib}\right|_\star
  \cdot\delta\mathbf Z
+\left.\nabla_{\boldsymbol\beta}\Theta_{ib}\right|_\star
  \cdot\delta\boldsymbol\beta,\\
\delta\epsilon_{i\alpha b}
&=\left.\nabla_{\mathbf Z}\epsilon_{i\alpha b}\right|_\star
  \cdot\delta\mathbf Z.
\end{align}
Here ${\boldsymbol\beta}$ denotes the set of fast and cyclic angles removed in the orbital reduction (see Appendix~\ref{app:chain_modulation_coefficients}). We define the unmodulated phase of tidal harmonic $b$ relative to mode $\alpha$ as $\Phi_{i\alpha b}=\Theta_{ib}^{\star}-\psi_{i\alpha}$. The tidal Hamiltonian (Eq.~\ref{eq:tide_mode_basic}) then separates into the forcing at MMR equilibrium and its leading modulation,
\begin{equation}
\mathcal H_{\rm tide}=\sum_{i\alpha b}\epsilon_{i\alpha b}^{\star}\sqrt{J_{i\alpha}}\cos\Phi_{i\alpha b}+\mathcal H_{\rm mod},
    \label{eq:H_tide_plus_mod}
\end{equation}
where
\begin{equation}
\mathcal H_{\rm mod}=\sum_{i\alpha b}\sqrt{J_{i\alpha}}\left[\delta\epsilon_{i\alpha b}\cos\Phi_{i\alpha b}-\epsilon_{i\alpha b}^{\star}\delta\Theta_{ib}\sin\Phi_{i\alpha b}\right].
    \label{eq:H_mod_raw}
\end{equation}
Terms quadratic in the chain displacement are omitted.

Following Eq.~\eqref{eq:chain_mode_expansion}, each perturbation is a sum of terms proportional to $\sqrt{J_a}\cos\psi_a$ and $\sqrt{J_a}\sin\psi_a$. Periodic modulation therefore generates tidal sidebands. Analogous libration-induced modulation of tidal forcing is established for spin libration \citep{frouard2017tides}. The resulting interaction is
\begin{equation}
\mathcal H_{\rm mod}=\sum_{i\alpha ab}\sum_{s=\pm1}g^{(s)}_{i\alpha ab}\sqrt{J_aJ_{i\alpha}}\cos\!\left(\Phi_{i\alpha b}+s\psi_a+\chi^{(s)}_{i\alpha ab}\right).
    \label{eq:H_mod_harmonics}
\end{equation}
The index $s=\pm1$ labels the two sidebands. The coefficient $g^{(s)}_{i\alpha ab}\geq0$, with dimensions of frequency, measures the coupling of chain oscillation $a$ to internal mode $\alpha$ of planet $i$ through tidal harmonic $b$. The  offset $\chi^{(s)}_{i\alpha ab}$ measures the phase of the coupling. Both coefficients are constructed in Appendix~\ref{app:chain_modulation_coefficients}.

Let $\sigma_{ib}=\dot\Theta_{ib}^{\star}$ define the tidal frequency on the reference MMR orbit. Each chain oscillation generates forcing components at $\sigma_{ib}\pm\omega_a$.  The drift of the sideband phase relative to the internal oscillation is
\begin{equation}
\Delta_s=\sigma_{ib}-\omega_{i\alpha}+s\omega_a.
    \label{eq:detuning_s}
\end{equation}
Thus a sideband is resonant when
\begin{equation}
\omega_{i\alpha}\simeq\sigma_{ib}+s\omega_a.
\label{eq:sideband_resonance_condition}
\end{equation}
Near this condition, the relative phase evolves slowly, allowing the interaction to accumulate over many oscillations.  These two sideband resonances are the focus of the rest of the paper.

Collecting the contributions derived insofar, the Hamiltonian governing the local dynamics around the chain equilibrium becomes 
\begin{align}
    \mathcal{H}_{\rm loc}&=\mathcal{H}_\star+\sum_i {S_i^2\over 2C_i}+\sum_a\omega_aJ_a+\sum_{i\alpha}\omega_{i\alpha}J_{i\alpha}
    \nonumber\\
    &+\sum_{i\alpha b}\epsilon_{i\alpha b}^{\star}\sqrt{J_{i\alpha}}\cos\Phi_{i\alpha b}
    \nonumber\\
    &+\sum_{i\alpha ba}\sum_{s=\pm1}
    g^{(s)}_{i\alpha ab}\sqrt{J_aJ_{i\alpha}}\cos\left(\Phi_{i\alpha b}+s\psi_a
    +\chi^{(s)}_{i\alpha ab}\right)\nonumber\\
    &+\mathcal{O}(\delta{\bf Z}^2\sqrt{J_{i\alpha}})+\mathcal{O}(\delta{\bf Z}^3).
    \label{eq:H_local_full}
\end{align}
The first line describes the uncoupled dynamics of the reference MMR motion, the spin, and the chain and internal oscillations. The second line is the tidal forcing evaluated at the chain equilibrium. The third line is the leading correction to the tides produced by chain librations. As it contains both $J_a$ and $J_{i\alpha}$, it allows for action exchange between the chain and internal oscillations. Its $s=+1$ component transfers action between the chain and internal mode, whereas $s=-1$ permits their actions to grow or decay together. The final line collects higher-order tidal and orbital terms. We next examine the dynamics of an isolated sideband resonance.

\subsection{Resonant Two-Oscillator Normal Form}
\label{subsec:two_oscillator_normal_form}

\begin{figure*}
\centering
\includegraphics[width=\textwidth]{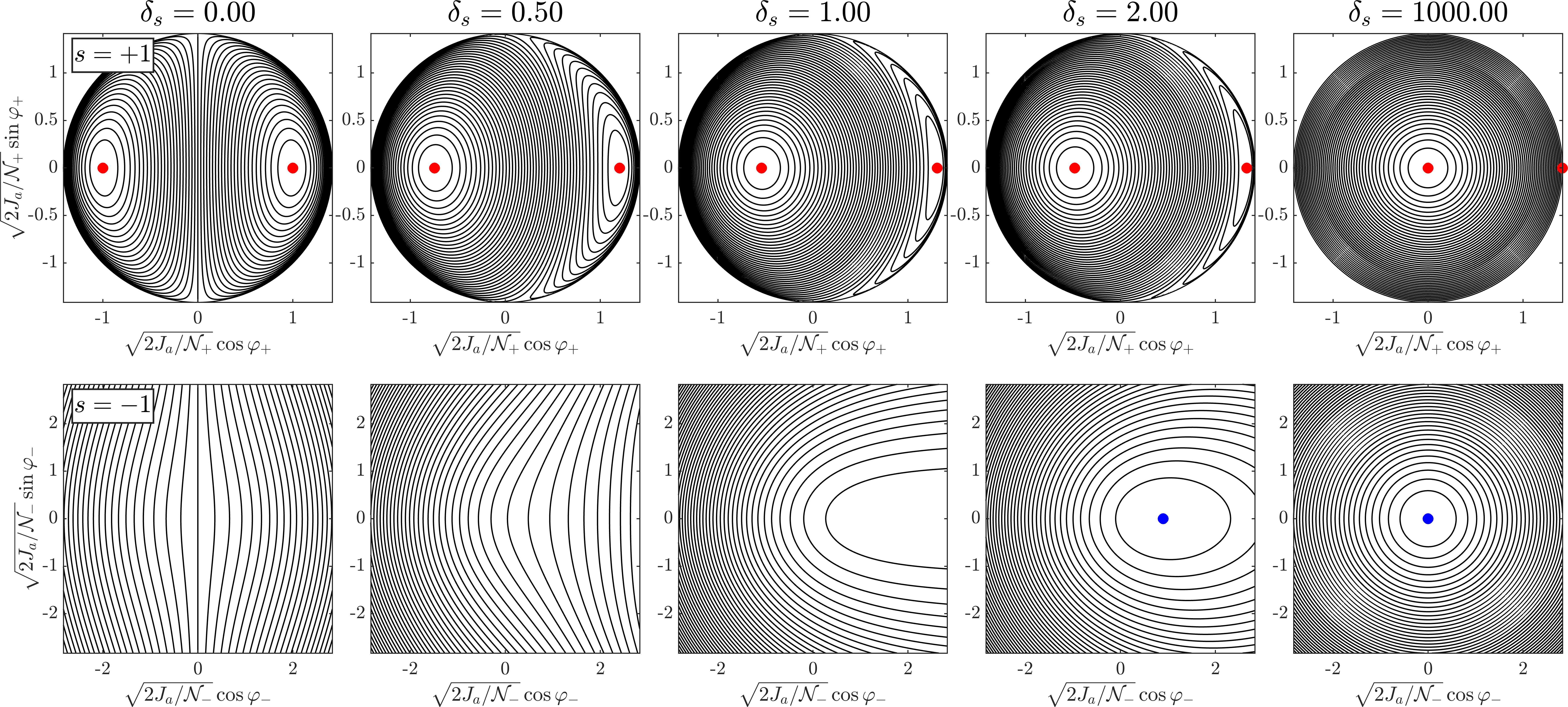}
\caption{Level curves of the two-oscillator Hamiltonian in Eq.~\eqref{eq:two_oscillator_hamiltonian_s} at normalized detuning $\delta_s=\Delta_s/g_s$. Upper row: the $s=+1$ action bounded exchange interaction at fixed $\mathcal N_+>0$. Lower row: the $s=-1$ pair coupling and parametric instability at fixed $\mathcal N_->0$. Colored points mark solutions to the equations of motion~(\ref{eq:chain_action_eom_s})-(\ref{eq:resonant_phase_eom_s}). }
\label{fig:two_oscillator_phase_portrait}
\end{figure*}

We isolate one internal mode $i\alpha$, one chain mode $a$, and one tidal harmonic $b$, retaining a single sideband $s$. We retain the slowly varying term of \eq{eq:H_mod_harmonics} and average over the remaining harmonics. For compactness, we define $g_s=g^{(s)}_{i\alpha ab}$ and $\chi_s=\chi^{(s)}_{i\alpha ab}$.  We introduce the resonant canonical variables, replacing $(\psi_a,J_a)$ and $(\psi_{i\alpha},J_{i\alpha})$ with $(\varphi_s,J_a)$ and $(\psi_{i\alpha},\mathcal N_s)$. The resonant phase
\begin{equation}
\varphi_s=\psi_a-s\psi_{i\alpha}+s\Theta_{ib}^{\star}+s\chi_s
    \label{eq:resonant_phase_s}
\end{equation}
is conjugate to $J_a$. The remaining angle $\psi_{i\alpha}$ is cyclic, with conserved conjugate action
\begin{equation}
\mathcal N_s=J_{i\alpha}+sJ_a.
    \label{eq:channel_invariants_pm}
\end{equation}
Eliminating $J_{i\alpha}=\mathcal N_s-sJ_a$ gives, up to an additive constant, the autonomous, resonant normal form Hamiltonian\footnote{The conserved $\mathcal K_s$ is a rotating frame invariant. The  tide represents energy exchange with the reference orbital and spin motion, whose reaction is omitted in this reduction.}
\begin{equation}
\mathcal K_s=s\Delta_sJ_a+g_s\sqrt{J_a(\mathcal N_s-sJ_a)}\cos\varphi_s.
    \label{eq:two_oscillator_hamiltonian_s}
\end{equation}
Hamilton's equations give
\begin{equation}
 \dot J_a
 =
 g_s
 \sqrt{
 J_a\left(\mathcal N_s-sJ_a\right)
 }
 \sin\varphi_s,
 \label{eq:chain_action_eom_s}
\end{equation}
and
\begin{equation}
 \dot\varphi_s
 =
 s\Delta_s
 +
 \frac{g_s}{2}
 \frac{\mathcal N_s-2sJ_a}
 {\sqrt{J_a\left(\mathcal N_s-sJ_a\right)}}
 \cos\varphi_s.
 \label{eq:resonant_phase_eom_s}
\end{equation}
The upper row of Figure~\ref{fig:two_oscillator_phase_portrait} captures the dynamics of the $s=+1$ resonance through the phase space of $\mathcal{K}_+$. Since $\mathcal N_+$ is an action sum, $0\leq J_a\leq\mathcal N_+ $, and its reduced phase space is compact. Radially outward motion corresponds to action transfer between the modes. At exact resonance, $\Delta_+=0$, the fixed points occur at equal partition, $J_a=J_{i\alpha}=\mathcal N_+/2$, and surrounding trajectories describe periodic exchange of a fixed action budget between the modes. Increasing $|\Delta_+|/g_+$ moves the equilibria toward unequal partitions and reduces the action excursion accessible from an initially cold chain. The maximum transferable action to the chain is
\begin{equation}
J_a^{\max}=\frac{g_+^2}{g_+^2+\Delta_+^2}J_{i\alpha}(0).
\label{eq:splus_maximum_transfer}
\end{equation}
Thus exchange through the $s=+1$ sideband requires the internal mode to store the action ultimately delivered to the chain. Damping, ignored in this section, reduces the available total action.  For $|\Delta_+|\gg g_+$, the contours become nearly circular, corresponding to nearly constant $J_a$: the coupling term is present, but the phase mismatch is too rapid for appreciable net transfer.

For $s=-1$, conservation of $\mathcal N_-$ allows both actions to grow together. On the displayed $\mathcal N_->0$ surface (bottom row of Figure~\ref{fig:two_oscillator_phase_portrait}), the level curves are open for $|\Delta_-|\leq g_-$, admitting exponential growth of the actions for $|\Delta_-|<g_-$ and quadratic growth at $|\Delta_-|=g_-$. For $|\Delta_-|>g_-$, the contours close around an elliptic equilibrium. Within this frozen, linear model, at very large detuning, the coupling again averages away and the actions remain nearly constant. Next, we discuss this interaction in more detail.

\subsection{Parametric Instability of the Resonant Chain}
\label{subsec:parametric_instability}

In the $s=-1$ channel, the stellar tidal forcing acts as a periodic pump that couples the planetary mode and chain libration, and permits their actions, near the resonance $\sigma_{ib}=\omega_{i\alpha}+\omega_a$, to grow together. The corresponding term in the coupling Hamiltonian (Eq.~\ref{eq:H_local_full}) is
\begin{equation}
\mathcal H_-=g_-\sqrt{J_aJ_{i\alpha}}
\cos\left(\Theta_{ib}^\star-\psi_{i\alpha}-\psi_a+\chi_-\right),
    \label{eq:sminus_resonant_term}
\end{equation}
where $g_-=g^{(-)}_{i\alpha ab}$ and $\chi_-=\chi^{(-)}_{i\alpha ab}$. 
 
This is a two-mode, or nondegenerate, parametric instability, found in simplified form in Section \ref{sec:toy_model}. We call it parametric because the stellar tide exerts periodically varying forces on the two oscillators that depend on the oscillations themselves.  To see this, define oscillator coordinates $Q_\nu=\sqrt{2J_\nu}\cos\psi_\nu$ and $P_\nu=-\sqrt{2J_\nu}\sin\psi_\nu$, for $\nu=a,i\alpha$. For $\Phi=\Theta_{ib}^\star+\chi_-$, the interaction in Eq.~\eqref{eq:sminus_resonant_term} contributes $\dot P_a=-(g_-/2)Q_{i\alpha}\cos\Phi  + (g_-/2)P_{i\alpha}\sin\Phi$, and similarly with $a$ and $i\alpha$ interchanged. Thus we see how one oscillator's forcing depends on the other oscillator's coordinates, with coupling coefficients given by $g_-\cos(\Theta_{ib}^\star + \chi_-)$ and $g_-\sin(\Theta_{ib}^\star + \chi_-)$ that vary over the tidal cycle.  
Analogous two-mode resonances occur in rotating fluids, where periodic elliptical strain couples pairs of inertial waves \citep[e.g.,][]{kerswell2002elliptical,barker2016non}, and in stellar oscillations, where a parent mode serves as the pump for a pair of daughter modes \citep[e.g.,][]{kumar1995nonlinear,wu2001gravity,weinberg2012nonlinear,van2024non}.

The open contours in the lower row of Figure~\ref{fig:two_oscillator_phase_portrait} describe unbounded growth in the absence of damping. We now include damping to determine when the tidal pump can sustain this growth. Let $c_a$ and $c_{i\alpha}$ denote the slowly varying complex canonical amplitudes in the resonant frame, normalized by
\begin{equation}
|c_a|^2=J_a,\qquad |c_{i\alpha}|^2=J_{i\alpha}.
\end{equation}
Absorbing the tidal forcing and coupling phases into these amplitudes, with $\arg(c_ac_{i\alpha})=-\varphi_-$, gives
\begin{equation}
\mathcal H_-=\frac{g_-}{2}\left(c_ac_{i\alpha}+c_a^\ast c_{i\alpha}^\ast\right).
    \label{eq:sminus_complex_hamiltonian}
\end{equation}
Hamilton's equations $i\dot c_\nu=\partial\mathcal H_-/\partial c_\nu^\ast$, with $\nu\in\{a,i\alpha\}$, give at exact resonance $\dot c_a=-ig_-c_{i\alpha}^\ast/2$ and $\dot c_{i\alpha}^\ast=ig_-c_a/2$. Without coupling, the phase of $c_ac_{i\alpha}$ drifts at the detuning frequency $\Delta_-$. Let $\gamma_a$ and $\gamma_{i\alpha}$ be the positive amplitude damping rates of the chain and internal modes, respectively. Including damping, the two amplitudes evolution follows
\begin{equation}
\frac{d}{dt}\begin{pmatrix}c_a\\c_{i\alpha}^{\ast}\end{pmatrix}=\begin{pmatrix}-\gamma_a+i\Delta_-/2 & -ig_-/2\\ig_-/2 & -\gamma_{i\alpha}-i\Delta_-/2\end{pmatrix}\begin{pmatrix}c_a\\c_{i\alpha}^{\ast}\end{pmatrix}.
    \label{eq:sminus_linear_system}
\end{equation}
For fixed coefficients, the eigenvalues are
\begin{equation}
\lambda_{\pm}=-\frac{\gamma_a+\gamma_{i\alpha}}{2}\pm\left[\frac{g_-^2}{4}+\frac{\left(\gamma_{i\alpha}-\gamma_a+i\Delta_-\right)^2}{4}\right]^{1/2},
    \label{eq:sminus_eigenvalues}
\end{equation}
For positive damping rates, parametric instability requires $\mathfrak{Re}\{\lambda_+\}>0$, or
\begin{equation}
\frac{g_-^2}{4}>\gamma_a\gamma_{i\alpha}\left[1+\left(\frac{\Delta_-}{\gamma_a+\gamma_{i\alpha}}\right)^2\right].
    \label{eq:sminus_instability_threshold}
\end{equation}
At exact resonance, this reduces to $g_-^2>4\gamma_a\gamma_{i\alpha}$. In the absence of damping, growth occurs for $|\Delta_-|<g_-$, reproducing the interval of open trajectories in Figure~\ref{fig:two_oscillator_phase_portrait}. Unequal damping broadens the instability window beyond $|\Delta_-|=g_-$, permitting growth at detunings that are stable in the undamped system shown in Figure~\ref{fig:two_oscillator_phase_portrait}.

Once the instability sets in, both oscillations grow together in the growing eigenmode, with amplitudes proportional to $\exp(\mathfrak{Re}\{\lambda_+\}t)$ and actions proportional to $\exp(2\mathfrak{Re}\{\lambda_+\}t)$. Let $J_{a,{\rm crit}}$ denote the critical chain action required for seeding its delayed instability, and let $e_{\rm free,rms}^{\rm crit}$ be the corresponding rms free eccentricity of the planets. The time required for the instability to amplify the chain oscillation from $J_a(t=0)$ to $J_{a,{\rm crit}}$ is 
\begin{equation}
\begin{aligned}
t_{\rm crit}
&=\frac{1}{2\mathfrak{Re}\{\lambda_+\}}
\ln\left[\frac{J_{a,{\rm crit}}}{J_a(t=0)}\right]\\
&=\frac{1}{\mathfrak{Re}\{\lambda_+\}}
\ln\left[
\frac{e_{\rm free,rms}^{\rm crit}}
     {e_{\rm free,rms}^{\max}(t=0)}
\right].
\end{aligned}
\label{eq:sminus_critical_time}
\end{equation}

\section{Equilibrium Chain}

\label{sec:Equilibrium_chain}
We apply the theory to a fiducial chain of four $5M_\oplus$ planets orbiting $M_\star=1\,M_\odot$. We here recover the chain's equilibria and collective librations.
\subsection{Reduced Hamiltonian for a Four-Planet Chain}
\label{subsec:fiducial_mixed_chain_hamiltonian}
The reference configuration has nominal mean motions
\begin{equation}
\bar n_1:\bar n_2:\bar n_3:\bar n_4=1:\frac{3}{4}:\frac{1}{2}:\frac{3}{8}.
\label{eq:fiducial_mean_motion_ratios}
\end{equation}
We set the nominal inner period to $\bar P_1=2\pi/\bar n_1=7$ days. Bars denote quantities in this exactly commensurable configuration. 

Following \citet{delisle2017analytical}, we average over the fast synodic angle $\lambda_3-\lambda_4$ and reduce by rotational symmetry. The eight orbital degrees of freedom then reduce to six canonical pairs: $(\phi_1,L_1)$, $(\phi_2,L_2)$, and $(\sigma_i,\Gamma_i)$ for $i=1,\ldots,4$, where
\begin{align}
\phi_1&=\lambda_1-2\lambda_2+\lambda_3,\nonumber\\
\phi_2&=\lambda_2-3\lambda_3+2\lambda_4,\nonumber\\
\sigma_i&=4\lambda_4-3\lambda_3-\varpi_i,
\label{eq:laplace_angles_mixed}
\end{align}
and
\begin{equation}
L_1=\Lambda_1,\qquad L_2=2\Lambda_1+\Lambda_2.
\label{eq:L123_mixed}
\end{equation}
Here $\{\phi_1,\phi_2\}$ are the slow Laplace angles and $\sigma_i$ are the resonant angles involving the periapses. The reduction leaves two conserved actions,
\begin{align}
\mathcal J_{\rm sc}&=8\Lambda_1+6\Lambda_2+4\Lambda_3+3\Lambda_4,\label{eq:Jfast_mixed}\\
G_{\rm tot}&=\sum_{i=1}^{4}(\Lambda_i-\Gamma_i).
\label{eq:Gtot_mixed}
\end{align}
For fixed masses and resonance ratios, $\mathcal J_{\rm sc}$ sets the nominal orbital scale of the chain, while $G_{\rm tot}$ is the total orbital angular momentum. Since $G_{\rm tot}$ is conserved, the quantity
\begin{equation}
\delta_{\rm res}=\sum_{i=1}^{4}\bar\Lambda_i-G_{\rm tot},
\label{eq:delta_res_fiducial}
\end{equation}
defining the angular momentum deficit relative to the circular reference chain is also conserved. 

We expand the Keplerian Hamiltonian to second order in $\hat\Lambda_i=\Lambda_i-\bar\Lambda_i$ and retain the leading planetary perturbation to first order in eccentricity, with its coefficients evaluated at the nominal semimajor axes. Dropping an additive constant, the reduced Hamiltonian is
\begin{equation}
\mathcal H_{\rm ch}=\mathcal H_{\rm Kep}+\mathcal H_{\rm res},
\label{eq:Hch_fiducial_mixed}
\end{equation}
with
\begin{equation}
\mathcal H_{\rm Kep}=-\frac{3}{2}\sum_{i=1}^{4}\frac{\bar n_i}{\bar\Lambda_i}\hat\Lambda_i^2.
\label{eq:Hkep_fiducial_mixed}
\end{equation}
The linear Keplerian term vanishes at fixed $\mathcal J_{\rm sc}$. The resonant part is
\begin{equation}
\mathcal H_{\rm res}=\sum_{(i,j)\in\mathcal R}\big[C_{i,j}\sqrt{\Gamma_i}\cos\vartheta_{ij,i}
+C_{j,i}\sqrt{\Gamma_j}\cos\vartheta_{ij,j}\big],
\label{eq:Hres_fiducial_compact}
\end{equation}
where $\mathcal R=\{(1,2),(2,3),(3,4),(1,3),(2,4)\}$ is the set of retained resonant pairs. For $k=i,j$, the phases are
\begin{equation}
\vartheta_{ij,k}=\begin{cases}
\sigma_k-3\phi_1-2\phi_2, & (i,j)=(1,2),\\
\sigma_k-2\phi_2, & (i,j)=(2,3),\\
\sigma_k, & (i,j)=(3,4),\\
\sigma_k-\phi_1-2\phi_2, & (i,j)=(1,3),\\
\sigma_k-\phi_2, & (i,j)=(2,4),
\end{cases}
\label{eq:fiducial_pair_resonant_angle}
\end{equation}
Thus the nonadjacent $2:1$ interactions enter at the same eccentricity order as the adjacent resonances. The real coefficients $C_{i,j}$ and $C_{j,i}$ set the strengths of the inner  and outer planet eccentricity terms, including direct and indirect contributions. Secular terms of second order in eccentricity and higher order resonances are omitted. We provide the inverse canonical transformation, expanded resonant Hamiltonian, and coefficient formulas in Appendix~\ref{app:fiducial_chain_hamiltonian}.

\subsection{Equilibria and Normal Modes}
\label{subsec:chain_equilibria_modes}
\label{sec:action_energy_requirements}
\begin{figure}[t]
\centering
\includegraphics[width=\linewidth]{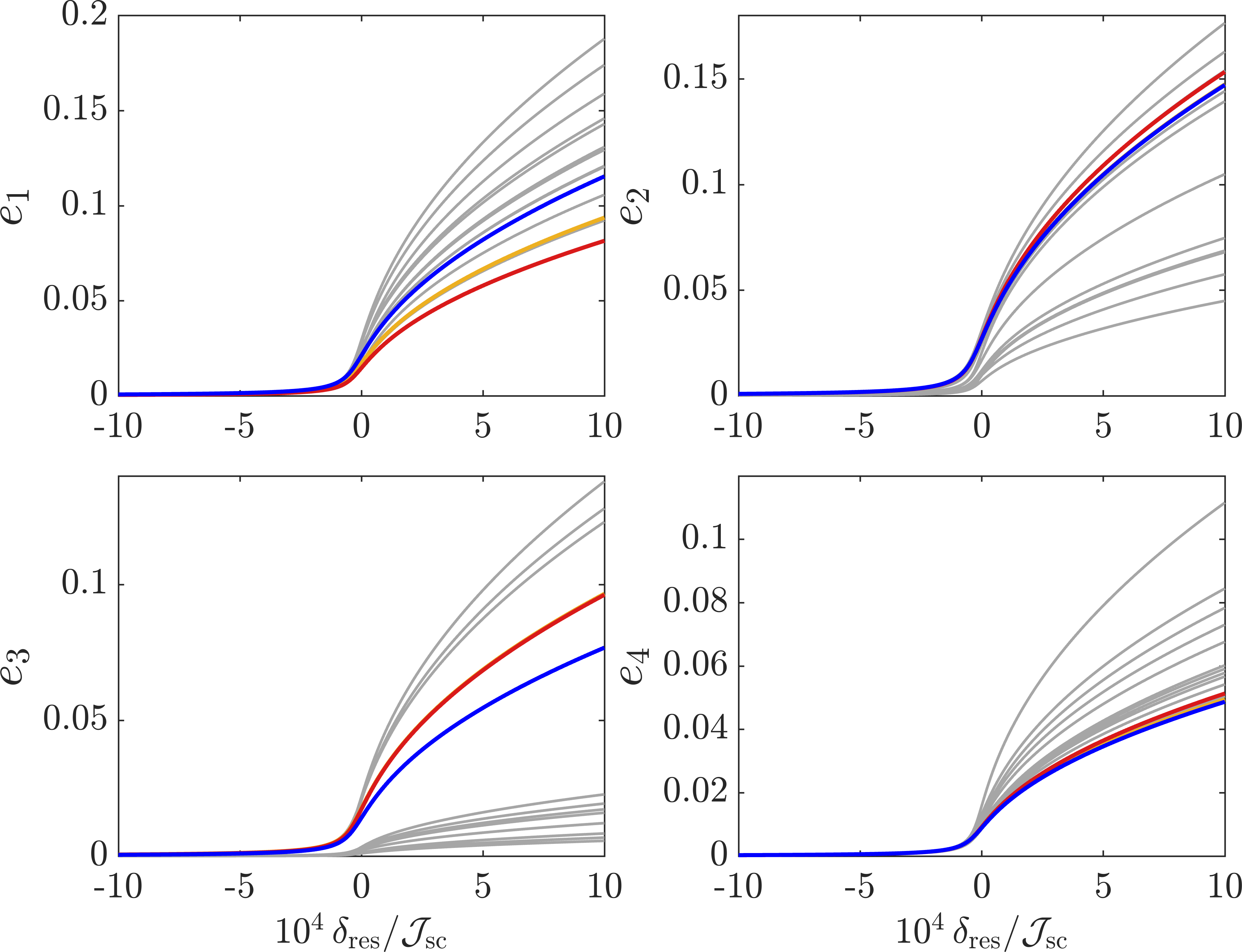}
\caption{Equilibrium eccentricities of the fiducial $3:4:6:8$ chain as functions of the conserved AMD offset $\delta_{\rm res}/\mathcal J_{\rm sc}$. Each panel shows one planet. Colored curves identify the six elliptic families; gray curves show the remaining unstable families. The equilibria satisfy Eq.~\eqref{eq:chain_fixed_point} for the Hamiltonian in Eq.~\eqref{eq:Hch_fiducial_mixed}.}
\label{fig:chani_ecc}
\end{figure}

\begin{figure}[t]
\centering
\includegraphics[width=\linewidth]{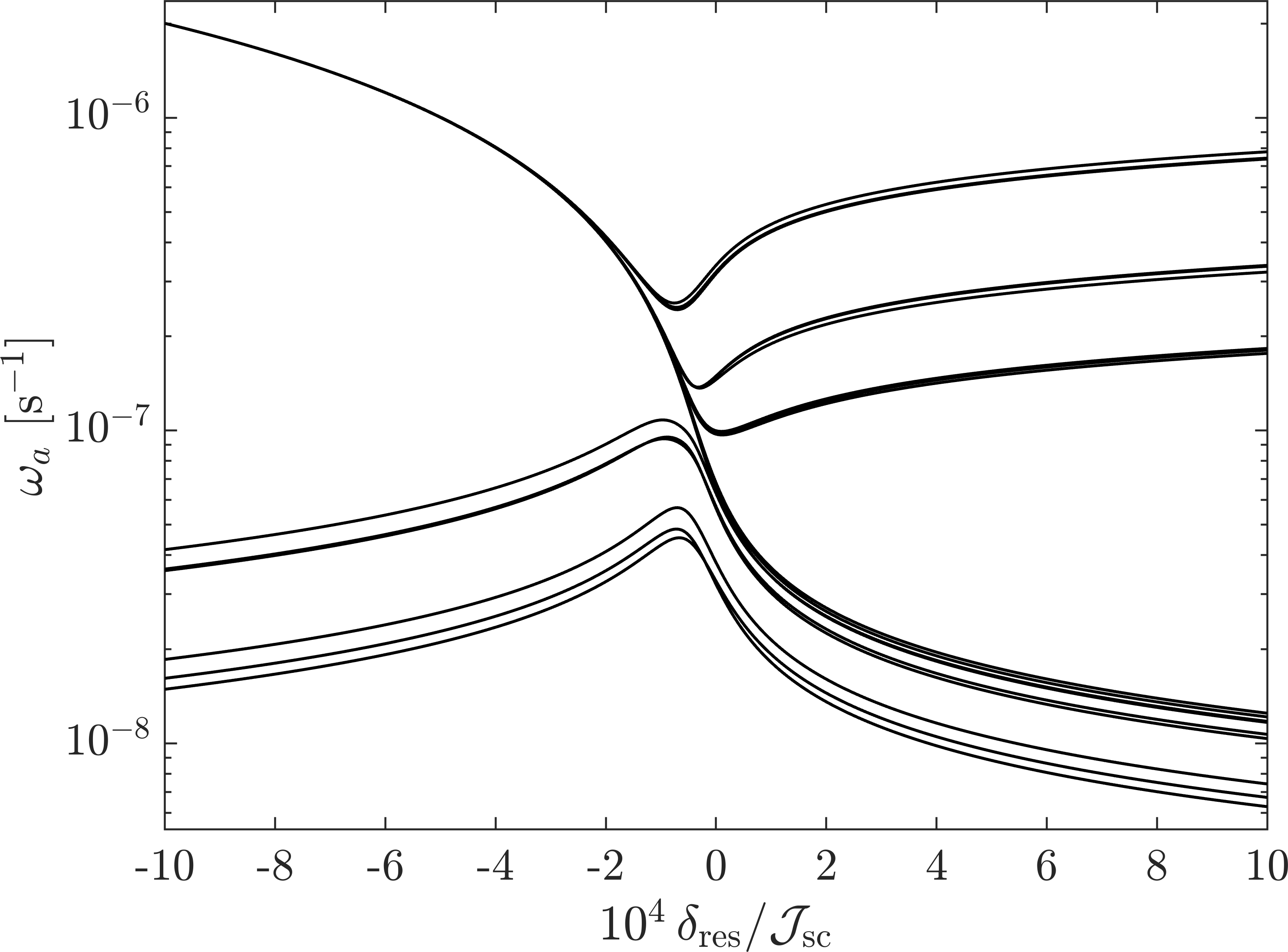}
\caption{Libration frequencies $\omega_a$ along the elliptic equilibrium families in Figure~\ref{fig:chani_ecc}. Each equilibrium has six oscillatory modes, obtained from the eigenvalues of the linearized Hamiltonian flow in Eq.~\eqref{eq:linearized_chain}.}
\label{fig:chain_modes}
\end{figure}

At fixed $\mathcal J_{\rm sc}$ and $\delta_{\rm res}$, we obtain the resonant equilibria by solving
\begin{equation}
    {\partial \mathcal{H}_{\rm ch}\over\partial \Gamma_i}=0,
    \qquad
    {\partial \mathcal{H}_{\rm ch}\over\partial \sigma_i}=0,
    \qquad i=1,\ldots,4,
    \label{eq:fixed_sigma_gamma_mixed}
\end{equation}
\begin{equation}
    {\partial \mathcal{H}_{\rm ch}\over\partial L_k}=0,
    \qquad
    {\partial \mathcal{H}_{\rm ch}\over\partial \phi_k}=0,
    \qquad k=1,2.
    \label{eq:fixed_phi_L_mixed}
\end{equation}
We find 24 solution families, of which six are elliptic over broad intervals of $\delta_{\rm res}/\mathcal J_{\rm sc}$. Each elliptic equilibrium supports six normal modes, obtained from Eq.~\eqref{eq:linearized_chain}. Figures~\ref{fig:chani_ecc} and \ref{fig:chain_modes} show the equilibrium eccentricities and the corresponding frequency spectra, with six libration frequencies for each elliptic family.

For the subsequent calculations, we select an elliptic equilibrium at $\delta_{\rm res}/\mathcal J_{\rm sc}=3.61\times10^{-4}$, with  $(e_1,e_2,e_3,e_4)_\star=(0.0775,0.0989,0.0516,0.0327).$
Linearizing the Hamiltonian flow about this equilibrium gives the libration frequencies 
$\omega_a =(5.76,10.17,10.84,32.53,57.83,96.24)\times10^{-8}\ {\rm s}^{-1},$
corresponding to periods between $75.6$ and $1263$ days.


\begin{figure}[t]
\centering
\includegraphics[width=\linewidth]{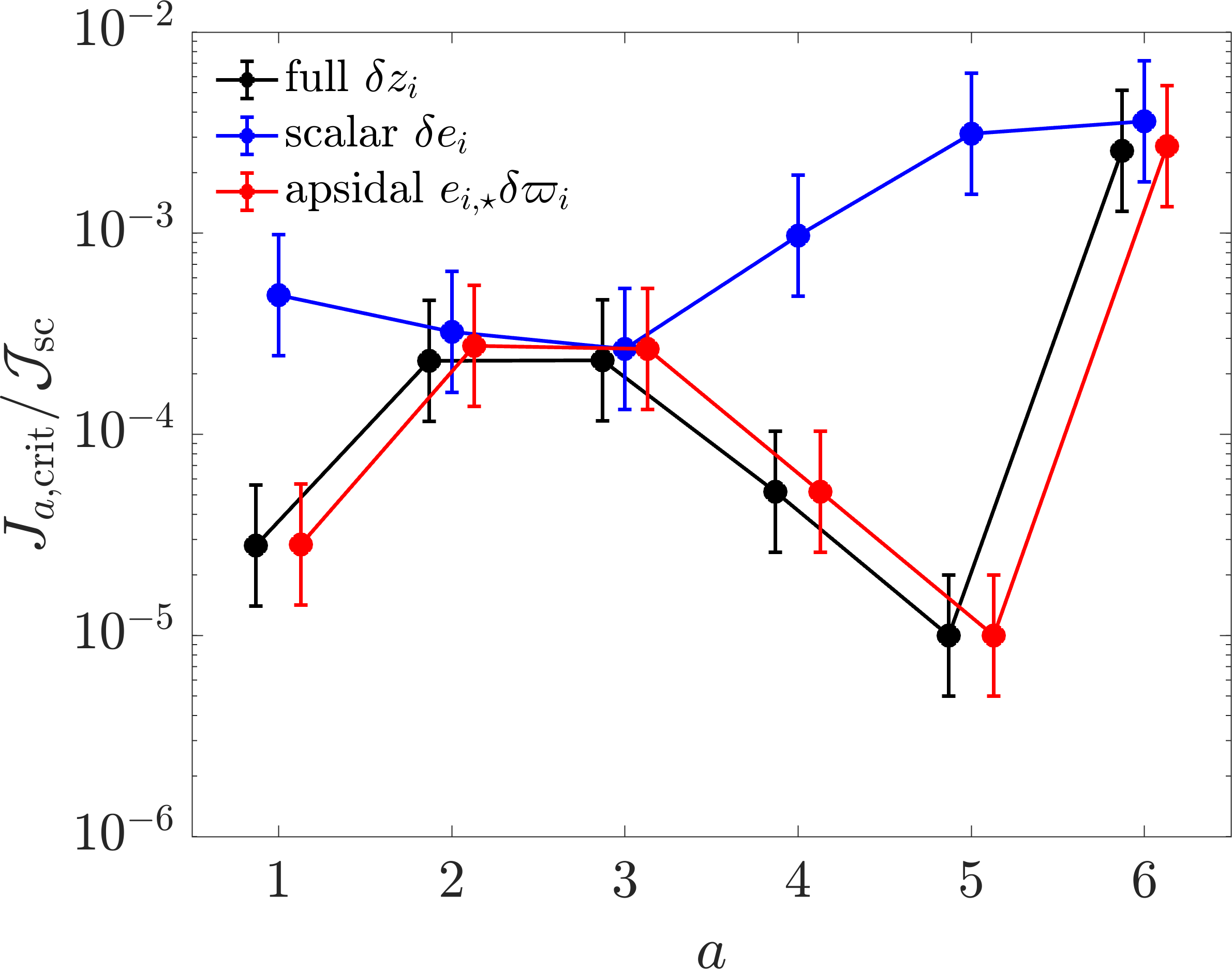}
\caption{Mode action required to reach the target rms free eccentricity if each chain normal mode is excited separately. Blue uses the scalar eccentricity criterion in Eq.~\eqref{eq:Ja_crit_rms}; black and red use the complex displacement $|\delta z_i|$ and its apsidal contribution $|e_{i,\star}\Delta\varpi_i|$, respectively. Each measure uses the rms over planets, maximized over the common libration phase. Bars span target amplitudes of $e_{\rm free, rms}^{\rm crit}=0.025-0.05$.}
\label{fig:action_requirement}
\end{figure}

The mode vectors in Eq.~\eqref{eq:chain_mode_expansion}, together with the inverse resonant transformation, determine the orbital response to a given action. For a single excited mode $a$, let $\Delta\Lambda_i^{(a)}$ and $\Delta\Gamma_i^{(a)}$ denote the departures from the selected equilibrium. To linear order, the scalar eccentricity displacement is
\begin{equation}
\Delta e_i^{(a)}=\frac{\sqrt{1-e_{i,\star}^2}}{\Lambda_{i,\star}e_{i,\star}}
\times\left(\Delta\Gamma_i^{(a)}-\frac{\Gamma_{i,\star}}{\Lambda_{i,\star}}\Delta\Lambda_i^{(a)}\right).
\label{eq:Deltae_from_DeltaGamma}
\end{equation}
We define the eccentricity amplitude per unit $\sqrt{J_a}$ as
\begin{equation}
\mathcal E_{ia}=\frac{\max_{\psi_a}|\Delta e_i^{(a)}|}{\sqrt{J_a}},
\label{eq:Eia_mode_shape}
\end{equation}
so that planet $i$ has the linear eccentricity envelope $e_{i,\star}\pm\mathcal E_{ia}\sqrt{J_a}$. To measure changes in eccentricity for the whole chain, we use the rms free eccentricity
\begin{equation}
\begin{aligned}
e_{\rm free,rms}&=\left[\frac{1}{4}\sum_{i=1}^{4}\left(\Delta e_i^{(a)}\right)^2\right]^{1/2}\\
&=\mathcal C_{a,\rm rms}(\psi_a)\sqrt{J_a}.
\end{aligned}
\label{eq:e_free_rms_def}
\end{equation}
Motivated by \citet{choksi2026two}, we adopt $e_{\rm free,rms}^{\rm crit}=0.025-0.05$ as a fiducial trigger for subsequent orbital instability. The chain action needed to reach this trigger is
\begin{equation}
J_{a,\rm crit}=\left[\frac{e_{\rm free,rms}^{\rm crit}}{\max_{\psi_a}\mathcal C_{a,\rm rms}(\psi_a)}\right]^2.
\label{eq:Ja_crit_rms}
\end{equation}
This scalar criterion excludes changes due solely to apsidal motion. To distinguish these changes, define the complex eccentricity $z_i=e_i e^{i\varpi_i}$. Relative to the precessing equilibrium $z_i^\star(t)$, its displacement $\delta z_i=z_i-z_i^\star$ is, to linear order,
\begin{equation}
\delta z_i=e^{i\varpi_{i,\star}}\left(\Delta e_i+i e_{i,\star}\Delta\varpi_i\right),
\label{eq:dz_scalar_apsidal_split}
\end{equation}
where $\Delta e_i=e_i-e_{i,\star}$ and $\Delta\varpi_i=\varpi_i-\varpi_{i,\star}(t)$. Figure~\ref{fig:action_requirement} compares the action needed to reach the same target rms oscillation under the scalar, complex, and apsidal measures, if each chain mode is excited separately. The complex measure requires the least action through mode $a=5$, but that is set chiefly through apsidal motion, whereas the scalar measure favors $a=3$. We therefore adopt $a=3$ and the scalar criterion in what follows.

\section{Planetary Modes}
\label{sec:fiducial_planet_modal_problem}

We now turn to compute internal planetary oscillations.   The frequencies, eigenfunctions, and tidal couplings of planetary modes depend on the evolving interior. Focusing on the innermost planet, we first construct a model of its evolution in Sec.~\ref{subsec:orch_ps0_evolution} then calculate its free oscillations in Sec.~\ref{subsec:gms_main_text}.

\subsection{Inner Sub-Neptune Evolution}
\label{subsec:orch_ps0_evolution}

We evolve the innermost sub-Neptune with the numerical code \texttt{ORCHARD}, which solves the one-dimensional hydrostatic, thermal, and compositional evolution of differentiated planets \citep{arevalo2026orchard}. The planet has total mass $M_{\rm p}=5.0\,M_\oplus$, comprising a $1.50\,M_\oplus$ iron core, a $3.00\,M_\oplus$ Mg$_2$SiO$_4$ mantle, and a homogeneous envelope of mass $M_{\rm env}=0.50\,M_\oplus$.   The envelope's hydrogen, helium, and water mass fractions are $(X,Y,Z_{\rm H_2O})=(0.45,0.15,0.40)$.

We impose constant irradiation equivalent to a circular seven day orbit about a $1\,M_\odot$, $1\,L_\odot$ star. A Bond albedo of $0.3$ and complete redistribution of absorbed radiation give an equilibrium temperature $T_{\rm eq}=951\,{\rm K}$. The irradiated atmosphere model in \texttt{ORCHARD} sets the intrinsic cooling flux. Energy transport includes convection and radiative diffusion in the envelope, convective and conductive transport in the compact interior, radiogenic heating in the silicate mantle, and latent heat associated with interior phase transitions. We evolve the spherical, nonrotating background to $5\,{\rm Gyr}$, without atmospheric mass loss. The equations of state, transport prescriptions, atmospheric boundary, and numerical settings of this simulation are specified in Appendix~\ref{app:orch_ps0_reproducibility}.

Figure~\ref{fig:orch_ps0_evolution} shows the resulting structural and thermal evolution. The radius $R_{\rm p}$ contracts from $3.64$ to $2.50\,R_\oplus$ during the first $100\,{\rm Myr}$ and reaches $2.40\,R_\oplus$ at $5\,{\rm Gyr}$, while the mantle-envelope interface remains near $1.55\,R_\oplus$. Contraction raises the dynamical frequency $\omega_{\rm dyn}=(\mathcal G M_{\rm p}/R_{\rm p}^{3})^{1/2}$ from $4.0\times10^{-4}$ to $7.0\times10^{-4}\,{\rm s^{-1}}$ over the first $100\,{\rm Myr}$. The intrinsic effective temperature $T_{\rm int}$ falls from $337$ to $70\,{\rm K}$ over the same interval, and the intrinsic luminosity $L_{\rm int}=4\pi R_{\rm p}^{2}\sigma_{\rm SB}T_{\rm int}^{4}$ declines by three orders of magnitude; $\sigma_{\rm SB}$ is the Stefan--Boltzmann constant. 

The temperature $T_{\rm m}$ at the top of the silicate mantle remains near $6000\,{\rm K}$. The full mantle temperature profile stays above the adopted melting curve throughout the simulation, leaving a global fluid mantle at all ages. During the first $100\,{\rm Myr}$, the modeled envelope has no stably stratified layer and hence no buoyancy shell. With rotation included, both the gaseous envelope and the fluid silicate mantle can support acoustic, surface gravity, and inertial oscillations. We move next to compute the spectrum of these modes.
\begin{figure}[t]
    \centering
    \includegraphics[width=\linewidth]{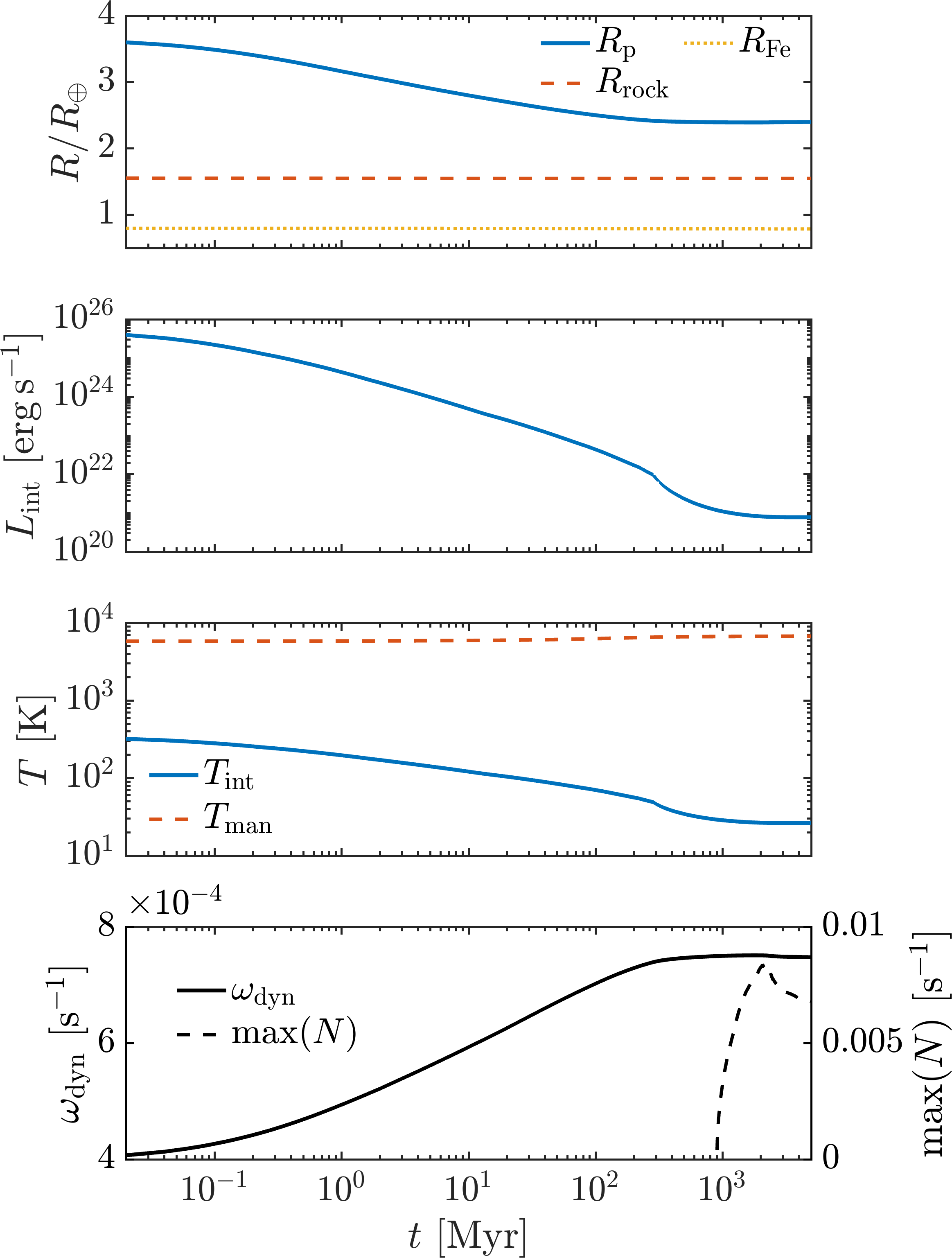}
    \caption{Thermal and structural evolution of the fiducial
    inner sub-Neptune described in Section~\ref{subsec:orch_ps0_evolution} over $5\,{\rm Gyr}$.  From top to bottom: radii of the planet, mantle-envelope interface, and iron core; intrinsic luminosity; mantle-top temperature and intrinsic effective temperature; and the dynamical frequency together with the maximum envelope Brunt--V\"ais\"al\"a frequency. }
    \label{fig:orch_ps0_evolution}
\end{figure}

\subsection{\texttt{MAQAM}: Computing Internal Planetary Modes}
\label{subsec:gms_main_text}

We introduce \texttt{MAQAM}\footnote{Named after the \textit{maqam}, a framework in Arabic music that defines a repertoire of pitches and melodic patterns with a distinctive character.}, a solver for the linear, adiabatic free oscillations of fluid layers in rotating planets, and apply it to snapshots of the planet's evolution. Each snapshot supplies the equilibrium pressure $P(r)$, density $\rho(r)$, gravitational potential $\Phi_0(r)$, and adiabatic exponent $\Gamma_1(r)$ as functions of radius $r$, together with the material boundaries (Appendix~\ref{app:orchard_background}). We assign a uniform rotation $\boldsymbol\Omega=\Omega\hat{\boldsymbol z}$ then solve the free oscillation problem, described below, in both the fluid mantle and gas envelope. 

We describe a perturbation using the Lagrangian displacement $\boldsymbol\xi$ proportional to $\exp(im\phi-i\omega_{\rm rot}t)$, where $m$ is the azimuthal order, $\phi$ is the rotating azimuth, and $\omega_{\rm rot}$ is the signed rotating frame frequency. The linear, adiabatic, inviscid perturbation equations are \citep{lynden1967stability,unno1979nonradial}
\begin{align}
 -\omega_{\rm rot}^2\rho{\boldsymbol\xi}
 -2i\omega_{\rm rot}\rho{\boldsymbol\Omega}
      \mathbin{\times}{\boldsymbol\xi}
 &=
 -{\boldsymbol\nabla}\delta p
 -\delta\rho\,{\boldsymbol\nabla}\Phi_0
 -\rho{\boldsymbol\nabla}\delta\Phi,
 \nonumber\\\nonumber
 \delta\rho
 &=-{\boldsymbol\nabla}\mathbin{\cdot}
       \left(\rho{\boldsymbol\xi}\right),\\
 \Delta p&=c_s^2\Delta\rho,
 \nonumber\\
 \nabla^2\delta\Phi&=4\pi \mathcal{G}\delta\rho.
 \label{eq:gms_main_primitive_system}
\end{align}
Here $\delta p$, $\delta\rho$, and $\delta\Phi$ are Eulerian perturbations of pressure, density, and gravitational potential. For any background scalar $f$, its Lagrangian perturbation is $\Delta f=\delta f+\boldsymbol\xi\cdot\boldsymbol\nabla f$, and $c_s^2=\Gamma_1P/\rho$ is the squared adiabatic sound speed.

\texttt{MAQAM} retains two pieces of physics that are particularly important for our mechanism.  The complete Coriolis acceleration is essential for rotational waves in the convective envelope, with frequencies satisfying $|\omega_{\rm rot}|\lesssim2\Omega$ \citep[e.g.,][]{ogilvie2004tidal}. We retain its coupling between spherical harmonic degrees \citep[see e.g.,][]{papaloizou1978non,lee1987low,savonije1995non,ogilvie2009tidal}. We do not invoke the traditional approximation of rotation, which simplifies the angular problem by neglecting part of the rotation vector and is best suited to low frequency waves in strongly stratified media \citep[e.g.,][]{unno1979nonradial,townsend2003asymptotic}.  Second, the self-gravity perturbation $\delta\Phi$ is retained, so the Cowling approximation is not used \citep{cowling1941non}. This is important for low degree global modes and for the exterior gravitational multipoles from which tidal overlaps are subsequently obtained.

We assume the planet's equilibrium remains spherical and measure rotation by
\begin{equation}
q_{\rm rot}=\frac{\Omega}{\omega_{\rm dyn}}.
\label{eq:gms_main_rotation_parameter}
\end{equation}
The omitted centrifugal deformation is $\mathcal{O}(q_{\rm rot}^2)$. Thus the spherical approximation requires $q_{\rm rot}^2\ll1$ \citep{dewberry2021constraining,dewberry2022dynamical}.

We solve the eigenproblem at azimuthal order  ($m=2$) in both equatorial parity sectors\footnote{Sphericity forces  symmetry under reflection across the equatorial plane, $\theta\rightarrow\pi-\theta$, thus the eigenproblem separates into two independent sectors: scalar perturbations are symmetric in the even spherical harmonics sector, and antisymmetric in the odd sector.}, whose scalar perturbations are symmetric or antisymmetric about the equator. Only the symmetric sector couples to the aligned tidal forcing. The $m=-2$ modes have conjugate eigenfunctions and opposite signed frequencies. 

The envelope and fluid mantle are treated as separate eigenproblems. The envelope oscillates above a rigid, impenetrable mantle and has a free outer boundary. Its gravitational perturbation is continued through the rigid interior and matched to the exterior vacuum. The mantle oscillates above a rigid iron core, with its top treated as a free surface and its gravitational perturbation matched directly to vacuum. The eigensolver therefore does not recover dynamically coupled mantle-envelope and interface modes. The two  eigenproblems and their boundary conditions are described in Appendices~\ref{app:gms_continuum} and \ref{app:gms_magma_ocean}.

Numerically, we first expand the displacement and scalar perturbations in vector and scalar spherical harmonics up to a maximum degree $\ell_{\max}$. The harmonic projection and resulting radial equations are given in Appendix~\ref{app:gms_continuum}. The projected radial system is then discretized using mapped summation by parts operators following \citet{mattsson2017diagonal}, and assembled into a sparse quadratic eigenvalue problem (Eq.~\ref{eq:gms_quadratic_pencil}). After linearization \citep{tisseur2001quadratic}, we recover candidate modes using overlapping shift-invert Arnoldi searches across the rotational, dynamical, and acoustic frequency intervals \citep{sorensen1992implicit}. The complete numerical framework is detailed in Appendix~\ref{app:gms_numerics}.

Numerical acceptance and resolution tests are described in Appendicies~\ref{subsec:gms_acceptance}-\ref{app:benchmark_GMSB0}. Candidate modes must satisfy fourteen operator, mode quality, and resolution checks. In addition, a more stringent angular test is applied to the collective quadrupolar response. Figures~\ref{fig:lmax_convergence} and \ref{fig:Imk22_residuals} motivate our adoption of $\ell_{\max}=52$ for this response. Together, these tests define the final catalogue of accepted mode frequencies, eigenfunctions, gyroscopic normalizations, and exterior gravitational perturbations. Examples of such modal spectra are shown in Figure~\ref{fig:gms_fp_spectrum} for high frequency surface-gravity and acoustic modes, and Figure~\ref{fig:gms_inertial_surface_spectrum} for low frequency inertial modes.

\section{Driving a Resonant Chain toward Instability}
\label{sec:chain_instability}
We use the inner sub-Neptune's evolutionary snapshot from Figure~\ref{fig:orch_ps0_evolution} at $10\,{\rm Myr}$ and recover the oscillation modes with $q_{\rm rot}=10^{-3}$, $10^{-2}$, and $10^{-1}$ corresponding to sub-, pseudo-, and super-synchronous rotation. The envelope supports surface gravity $f$-modes, acoustic $p$-modes, and low frequency rotational modes restored by the Coriolis force. The fluid mantle contributes both rotational and high frequency modes. The envelope has no stably stratified propagation cavity and supports no buoyancy $g$-modes. We test the parametric resonance excitation through its two channels against the critical action, $J_{a,{\rm crit}}$, corresponding to the free eccentricity range $e_{\rm free,rms}^{\rm crit}=0.025-0.05$ (Section~\ref{subsec:chain_equilibria_modes}).

\subsection{Coherent Exchange: Sufficient Forcing, Insufficient Mode Capacity}
\label{subsec:modal_action_requirement}
The $s=+1$ resonant sideband interaction (upper row of Fig.~\ref{fig:two_oscillator_phase_portrait}) supports coherent action exchange between the planetary oscillation and the libration of the chain.  We ask: \textit{i)} whether the stellar tide can supply the planet with the needed action $J_{a,{\rm crit}}$, and \textit{ii)} whether the planetary mode can store it before transferring it to the chain. 

\begin{figure}
\centering
\includegraphics[width=\columnwidth]{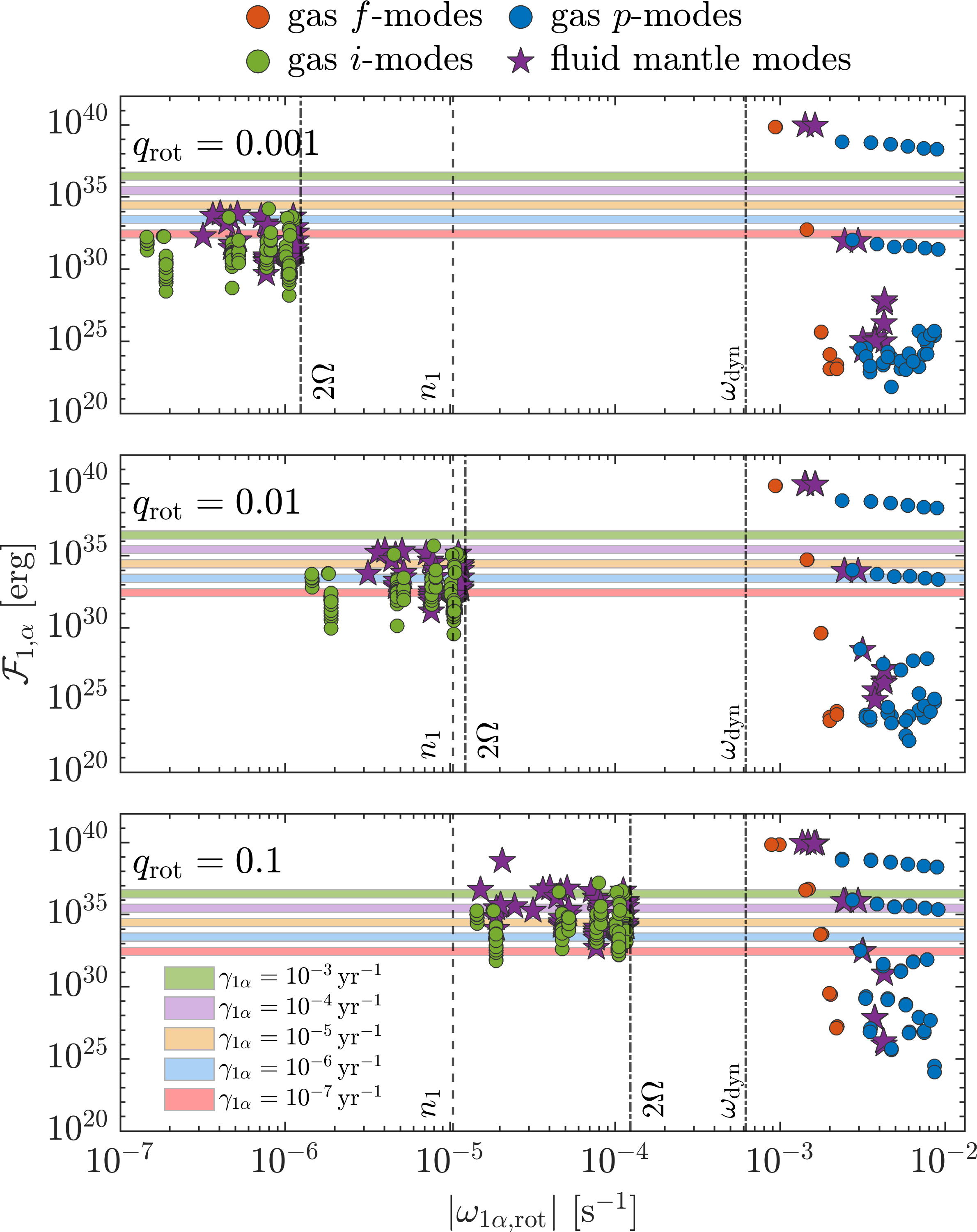}
\caption{Tidal forcing scales (Eq.~\ref{eq:maximum_canonical_forcing_scale}) of gas envelope and fluid mantle modes at three rotation rates. Horizontal bands span the required forcing amplitudes in Eq.~\eqref{eq:canonical_forcing_threshold} over $e_{\rm free,rms}^{\rm crit}=0.025-0.05$ and the six chain modes, for the indicated damping rates and  transfer fraction $\eta=0.5$. A mode above a band's upper edge satisfies all action requirements represented by that band.}
\label{fig:canonical_forcing_qrot}
\end{figure}
At the equilibrium orbit, tidal harmonic $b$ has  forcing coefficient $\epsilon^\star_{1\alpha b}$ (Eq.~\ref{eq:epsilon_def}). We characterize the strongest forcing by
\begin{equation}
\mathcal F_{1,\alpha}=\max_b|\epsilon^\star_{1\alpha b}|\sqrt{\mathcal J_{\rm sc}}.
\label{eq:maximum_canonical_forcing_scale}
\end{equation}
$\mathcal F_{1,\alpha}$ has units of energy but measures forcing strength. Granting exact resonance with this harmonic, the mode action is
\begin{equation}
J_{1\alpha}=\frac{\mathcal F_{1,\alpha}^2}{4\gamma_{1\alpha}^2\mathcal J_{\rm sc}},
\label{eq:maximum_resonant_mode_action}
\end{equation}
where $\gamma_{1\alpha}$ is the amplitude damping rate of mode $\alpha$. If a fraction $0<\eta\leq1$ of this action reaches the chain, the action requirement $\eta J_{1\alpha}\geq J_{a,{\rm crit}}$ becomes
\begin{equation}
\mathcal F_{1,\alpha}\geq\mathcal F_{a,{\rm crit}}=2\gamma_{1\alpha}\left(\frac{\mathcal J_{\rm sc}J_{a,{\rm crit}}}{\eta}\right)^{1/2}.
\label{eq:canonical_forcing_threshold}
\end{equation}
Our normal mode solver, \texttt{MAQAM}, is  inviscid and does not determine $\gamma_{1\alpha}$.  We explore 
$\gamma_{1\alpha}=10^{-7}-10^{-3}\,{\rm yr}^{-1}$, 
a range motivated by calculations for gas giant planets. 
Convective viscosity enhanced by differential rotation can damp $f$- and low-order $p$-modes over $10^5-10^6\,{\rm yr}$, while radiative diffusion gives $p$-mode damping times of $10^3$--$10^7\,{\rm yr}$ \citep{fuller2026excitation}. 
Low-order $f$- and $p$-modes in young giant planets can persist for several Myr \citep{zanazzi2025seismic}. Calculations of Jovian inertial modes also yield amplitude damping rates within our interval, with substantially stronger damping at higher spatial orders \citep{wu2005origin}. These estimates motivate the adopted range, but the true damping rates of our sub-Neptune's envelope and mantle modes remain to be calculated.

Figure~\ref{fig:canonical_forcing_qrot} shows that the strongest $f$- and low-order $p$-modes, together with several mantle modes, satisfy the tidal forcing requirement even at the largest damping rate. Their predominantly spheroidal displacements produce large quadrupolar overlaps, i.e., large $\epsilon^\star_{1\alpha b}$. Inertial modes are confined to the Coriolis interval $|\omega_{1\alpha {\rm rot}}|\leq2\Omega$. Their displacements contain mixed spheroidal and toroidal motion, but only the even parity $\ell=2$ spheroidal component couples directly to the $Y_2^2$ forcing.  Their low frequency motion produces relatively weaker quadrupolar density perturbations than $f$- or $p$-modes, but this suppression weakens as rotation increases. We also recover a Rossby mode, but its odd parity does not couple to the even parity $Y_2^2$ forcing, and it is thus absent from the figure.

Stellar tidal forcing is therefore sufficient, in the ideal resonance limit, to pump gravity, acoustic, inertial, and fluid mantle modes to the needed action, $J_{a,{\rm crit}}/\eta$. However, a coherent exchange episode conserves $J_{1\alpha}+J_a$ (Eq.~\ref{eq:channel_invariants_pm}). Exciting an initially cold chain therefore requires the internal mode to have the capacity to store at least $J_{a,{\rm crit}}/\eta$ before transferring it to the chain. The corresponding energy requirement is
\begin{equation}
 E_{1\alpha,{\rm crit}}^{(+)}
 =\frac{|\omega_{1\alpha,{\rm rot}}|J_{a,{\rm crit}}}{\eta}.
 \label{eq:critical_mode_energy}
\end{equation}
We compare this requirement with an estimate of the mode's physical energy capacity, $E_{1\alpha,{\rm cap}}$. Different treatments of mode saturation have been adopted in the literature. \citet{wu2001gravity} calculate saturation of overstable stellar modes through resonant energy transfer to damped daughter modes. \citet{vick2019chaotic} instead prescribe rapid nonlinear dissipation when the mode energy reaches $0.1GM_{\rm p}^2/R_{\rm p}$.

Here we estimate the capacity geometrically  by scaling the mode until its mass-weighted rms physical displacement equals the mass-weighted rms radius of the oscillating fluid:
\begin{equation}
E_{1\alpha,{\rm cap}}=\frac{|\omega_{1\alpha,{\rm rot}}|b_{1\alpha}}{2\mathcal M_{1\alpha}}\int_{\mathcal V}\rho r^2\,d^3r.
\label{eq:geometric_mode_energy_capacity}
\end{equation}
Here $\mathcal V$ is the fluid volume, $\rho$ its equilibrium density, $r$ the distance from the planet's center, $\mathcal M_{1\alpha}=\int_{\mathcal V}\rho|\boldsymbol\xi_{1\alpha}|^2\,d^3r$, and $b_{1\alpha}>0$ is the gyroscopic norm in Eq.~\eqref{eq:mode_gyroscopic_norm}.
Thus, a mode with $E_{1\alpha,{\rm crit}}>E_{1\alpha,{\rm cap}}$ requires fluid displacement larger than the characteristic radius of the oscillating fluid.

\begin{figure}
\centering
\includegraphics[width=\columnwidth]{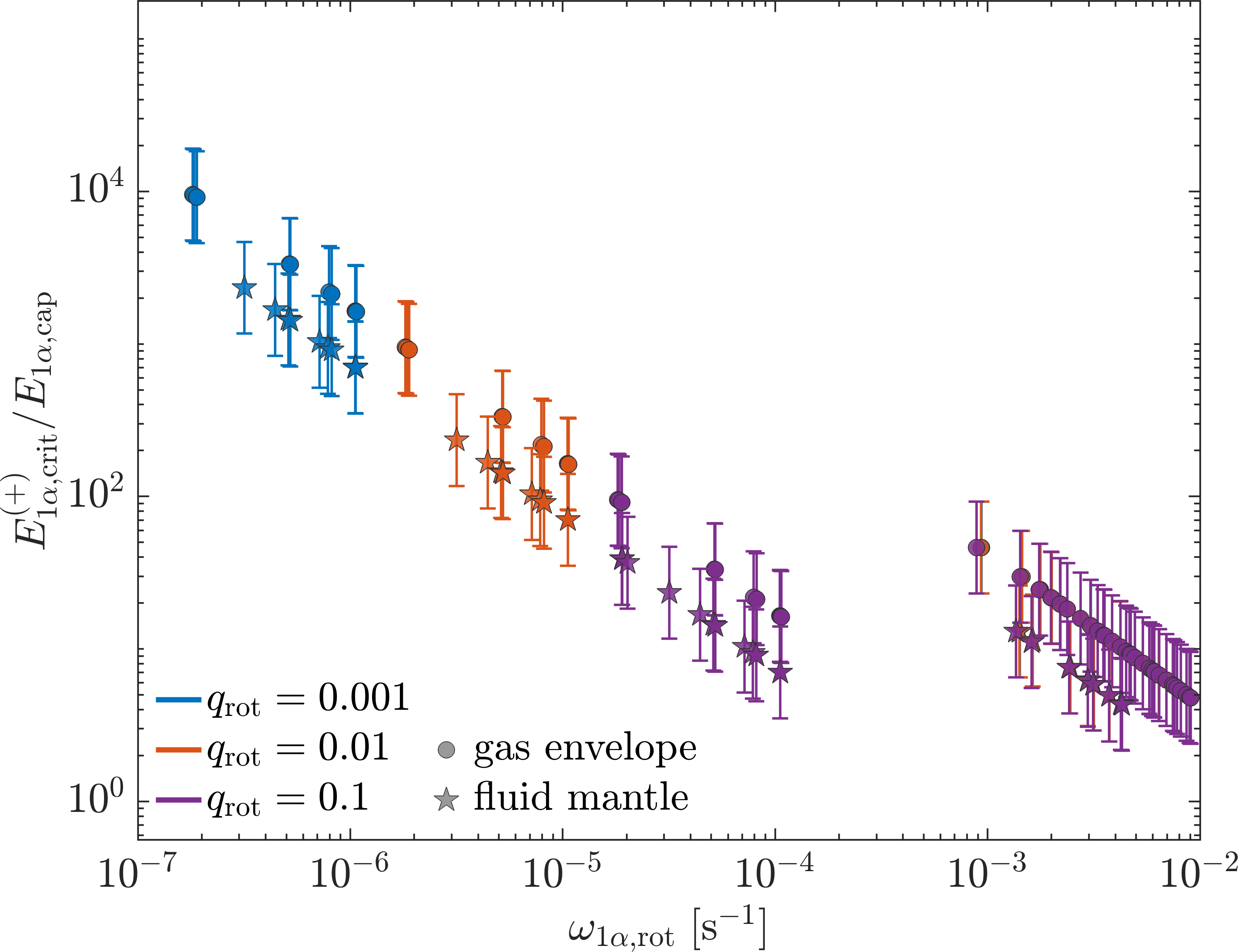}
\caption{Internal mode energy required for coherent transfer of $J_{a,\rm crit}$ into the chain, relative to the mode's geometric capacity (Eq.~\ref{eq:geometric_mode_energy_capacity}). We assume the internal mode transfers all the action it tidally receives ($\eta=1$). The same modes of Figure~\ref{fig:canonical_forcing_qrot} are used here. Vertical bars span $e_{\rm free,rms}^{\rm crit}=0.025$--$0.05$. Colors distinguish rotation rates; circles and pentagrams denote envelope and fluid mantle modes, respectively. }
\label{fig:energy_reqs}
\end{figure}

Figure~\ref{fig:energy_reqs} shows that the required  mode energy exceeds the geometric capacity for every tested mode, even for complete transfer ($\eta=1$). Lower frequency modes need less energy to carry the critical action (Eq.~\ref{eq:critical_mode_energy}), but their capacities fall more steeply with decreasing frequency. Faster rotation reduces the ratio of required energy to capacity for inertial modes, but never below unity. Therefore, while tidal forcing can drive the critical action in many modes, none of these modes can serve as the reservoir needed for disruptive coherent exchange with the chain.

\subsection{Parametric Growth While Staying Within Capacity}
\label{subsec:parametric_feasibility}
In the $s=-1$ resonant sideband interaction (see Section~\ref{subsec:parametric_instability} and bottom panels of Figure~\ref{fig:two_oscillator_phase_portrait}), the tidal pump can amplify the planetary oscillation and chain libration together. Unequal damping of the two oscillations allows the chain to accumulate action while the planetary mode carries a smaller share. We test the conditions of the onset of instability and whether its subsequent growth in the chain can reach $J_{a,{\rm crit}}$ while keeping the planetary mode below its geometric capacity.

We calculate the coupling coefficients $g^{(-)}_{1\alpha ab}$ (Eq.~\ref{eq:gchi_def}) for all modes shown in Figure~\ref{fig:canonical_forcing_qrot} with chain mode $a=3$, for  $\gamma_{1\alpha}=10^{-7}-10^{-3}~{\rm yr}^{-1}$.   Tidal dissipation within the innermost planet also damps the collective chain librations. Using the small eccentricity circularization rate for nearly synchronous rotation, we estimate
\begin{equation}
\begin{aligned}
\gamma_a \simeq{}&
4.1\times10^{-9}\,f_a
\left(\frac{|\mathfrak{Im}\{k_{2,1}
(\sigma,\gamma_{1\alpha})\}|}{10^{-3}}\right)
\left(\frac{R_1}{3R_\oplus}\right)^5
\\
&\times\left(\frac{m_1}{5M_\oplus}\right)^{-1}
\left(\frac{M_\star}{M_\odot}\right)^{-2/3}
\left(\frac{P_1}{7\,{\rm days}}\right)^{-13/3}
\,{\rm yr}^{-1}.
\end{aligned}
\label{eq:chain_damping_estimate}
\end{equation}
Here $k_{2,1}$ is the complex quadrupolar Love number of the inner planet (Eq.~\ref{Eq:Love_Number}), and $f_a$ is an order-unity dimensionless coefficient projecting the inner planet's eccentricity damping onto chain mode $a$; i.e., $\gamma_a=f_a |\dot e_1/e_1|$ \citep[see in analogy the collective damping of secular modes in][]{hansen2015secular}. 
For the strongly dissipative response in Figure~\ref{fig:Imk22_residuals}, $|\mathfrak{Im}\{k_{2,1}\}|\sim10^{-3}$, and Eq.~(\ref{eq:chain_damping_estimate}) gives $\gamma_a\sim10^{-9}\,{\rm yr}^{-1}$. We explore $\gamma_a=10^{-10}$--$10^{-8}\,{\rm yr}^{-1}$ to allow for uncertainty in the tidal theory. 
It will be seen that for such small values of $\gamma_a$,  the system response in the parametrically unstable resonance will be controlled by $\gamma_{1\alpha}$ and not $\gamma_a$.

\begin{figure}
\centering
\includegraphics[width=\columnwidth]{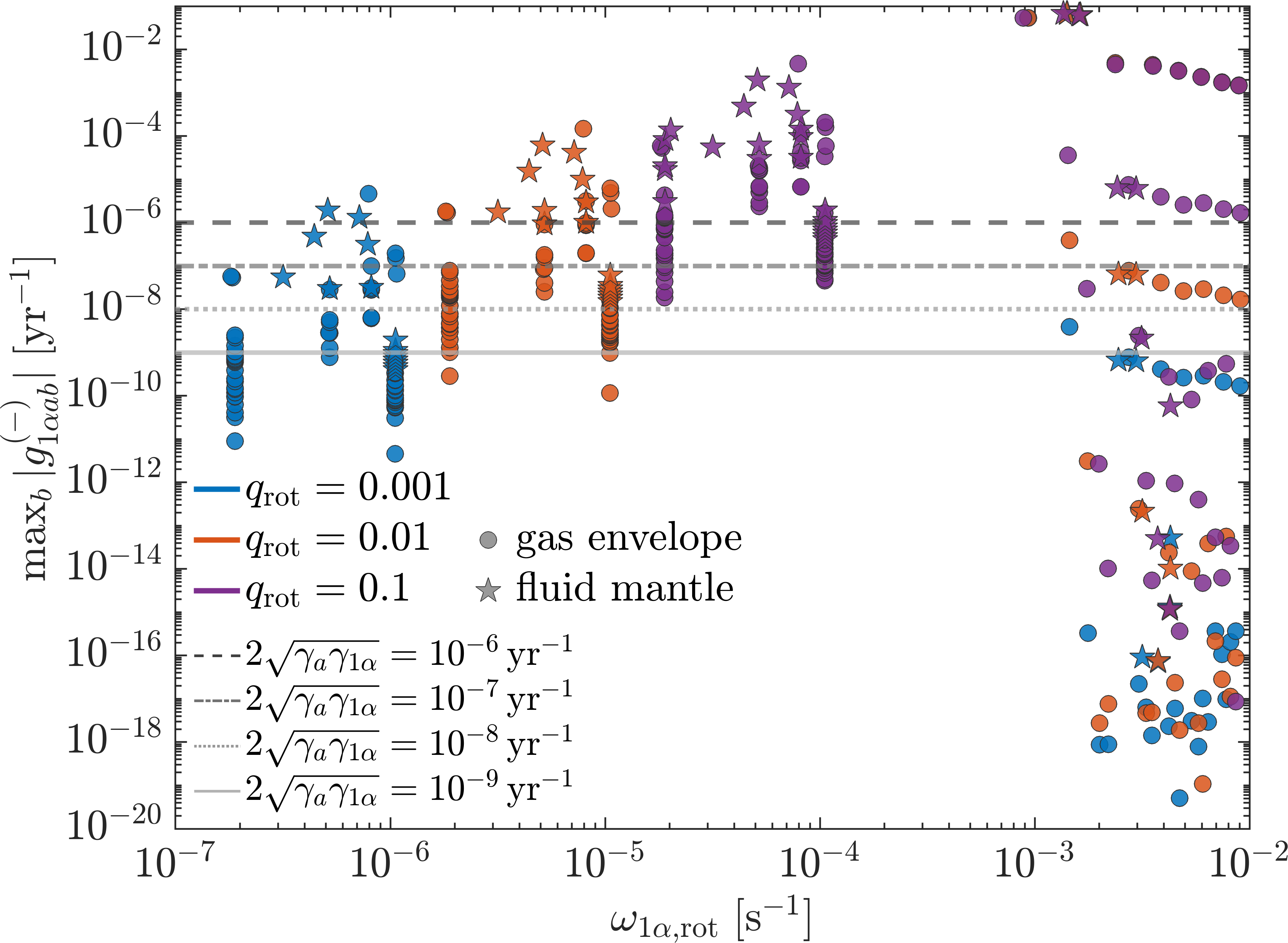}
\caption{Onset of parametric instability. Parametric coupling strength of internal modes to chain mode $a=3$, maximized over the available tidal harmonics. Colors distinguish rotation rates; circles and pentagrams denote envelope and fluid mantle modes, respectively. Horizontal lines give $2\sqrt{\gamma_a\gamma_{1\alpha}}$, the threshold above which a mode satisfies the instability onset condition when the corresponding interaction is resonant. Note that low-order $f$- and $p$-mode points nearly coincide across rotation rates. }
\label{fig:sminus_coupling}
\end{figure}

At the exact $s=-1$ sideband resonance, instability arises when the eigenvalue $\lambda_+$ (Eq.~\ref{eq:sminus_eigenvalues}) is real and positive, which occurs when
\begin{equation}
|g^{(-)}_{1\alpha ab}|>2\sqrt{\gamma_a\gamma_{1\alpha}}.
\label{eq:sminus_resonant_onset}
\end{equation}
Figure~\ref{fig:sminus_coupling} shows that the strongest $f$- and $p$-mode couplings exceed the damping thresholds of Eq.~\eqref{eq:sminus_resonant_onset} for all damping rates considered. Frequency matching, however, remains essential for establishing the resonance. For near-synchronous rotation, the dominant tidal forcing have frequencies of order the orbital frequency, well below the high frequencies of $f$- and $p$-modes. Accessing these high-frequency branches resonantly would require $q_{\rm rot}{\sim}0.7$.\footnote{Our normal mode solver \texttt{MAQAM} includes Coriolis but neglects centrifugal distortion of the equilibrium, an $\mathcal O(q_{\rm rot}^2)$ effect. While $q_{\rm rot}=0.7$ cannot be precluded, recovering the modes of such fast rotators requires using an oblate equilibrium with rotation dependent density and gravitational fields \citep[see e.g.,][]{dewberry2022dynamical}.} In contrast, low frequency inertial and fluid mantle modes admit growth and are accessible to resonant forcing with low-order tidal harmonics. Their coupling strengths increase with rotation, bringing a broader portion of the low frequency spectrum above the damping thresholds.

\begin{figure}
\centering
\includegraphics[width=\columnwidth]{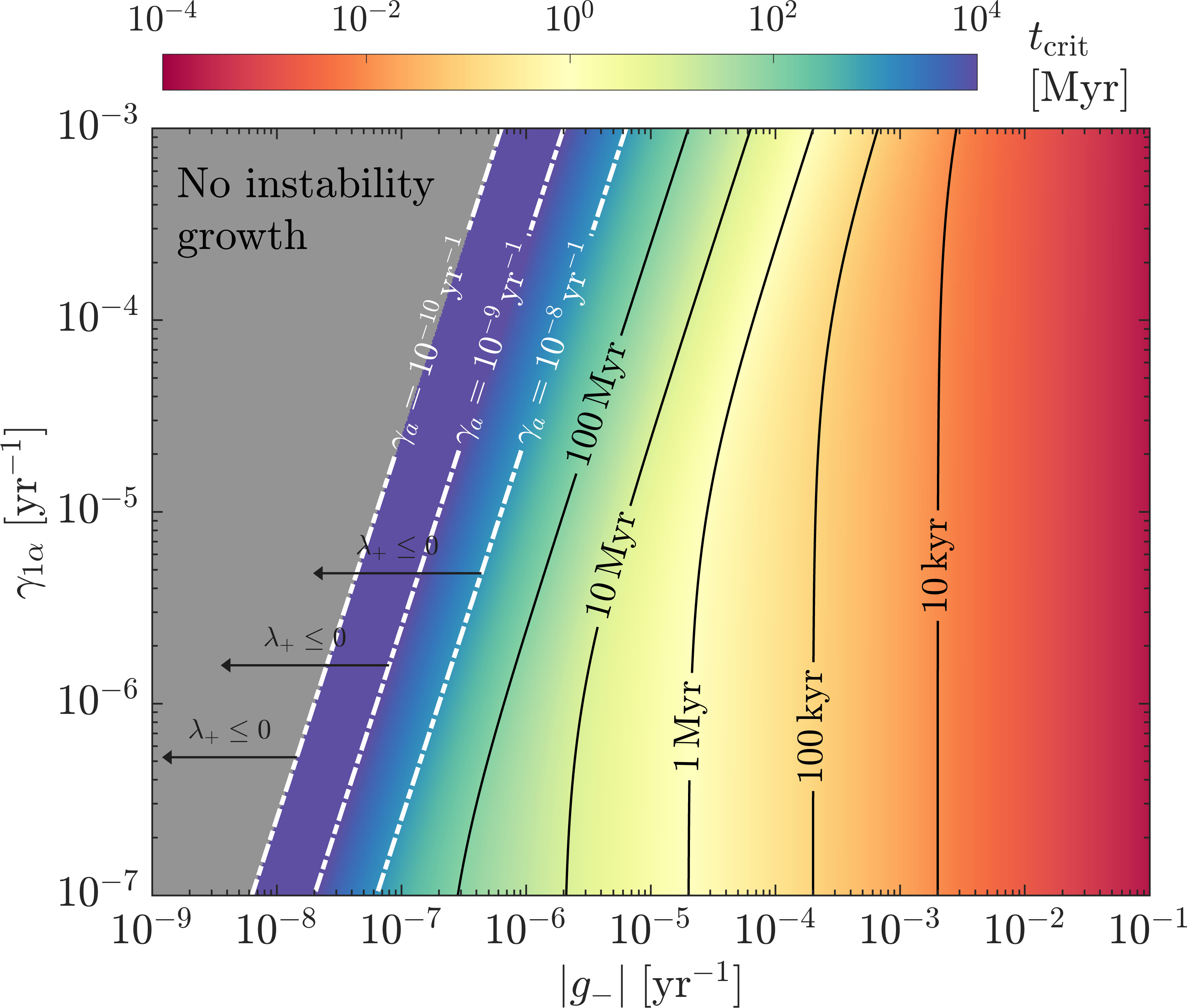}
\caption{Instability growth timescale. Time for the chain to reach the critical action $J_{a, \rm crit}$ after twenty action e-foldings (i.e., reaching $e_{\rm free,rms}^{\rm crit}=0.025-0.05$ from $e_{\rm free,rms}^{\rm crit}(t=0)=10^{-6}$), as a function of the parametric coupling strength (see Figure~\ref{fig:sminus_coupling}) and internal mode damping rate. The surface assumes  $\gamma_a=10^{-10}\,{\rm yr}^{-1}$. Changing $\gamma_a$ mainly changes the instability onset boundary (white curves), but not the values of $t_{\rm crit}$ far from it. Growth is absent to the left of each boundary.}
\label{fig:sminus_growth_time}
\end{figure}

Once the two-mode instability is activated, the time needed for the chain to reach $J_{a, \rm crit}$, denoted by $t_{\rm crit}$, is given by Eq.~\eqref{eq:sminus_critical_time}. Figure~\ref{fig:sminus_growth_time} evaluates $t_{\rm crit}$  for twenty action e-foldings. This corresponds to growth of a free eccentricity seed of order $10^{-6}$ into $e_{\rm free,rms}^{\rm crit}=0.025-0.05$. Many unstable low-frequency modes in Figure~\ref{fig:sminus_coupling} admit growth times below $100\,{\rm Myr}$. At pseudo-synchronous rotation, unstable couplings with $|g^{(-)}_{1\alpha ab}|\sim10^{-5}$--$10^{-4}\,{\rm yr}^{-1}$ give growth times of $10^5$--$10^7\,{\rm yr}$. These times assume exact resonance and fixed coefficients. As we show below, exact resonance is not required: the chain can reach $J_{a, \rm crit}$ through growth accumulated within the instability window.

The remaining question is whether the instability succeeds where coherent exchange fails: how much energy the planetary mode must sustain when the chain reaches its critical action. In the growing solution of Eq.~\eqref{eq:sminus_linear_system}, the planetary mode and chain libration grow at the same exponential rate, maintaining a fixed ratio of their actions. When the chain reaches $J_{a,{\rm crit}}$, the planetary mode has energy 
\begin{equation}
E_{1\alpha,{\rm crit}}^{(-)}=|\omega_{1\alpha,{\rm rot}}|J_{a,{\rm crit}}\frac{\mathfrak{Re}\{\lambda_+\}+\gamma_a}{\mathfrak{Re}\{\lambda_+\}+\gamma_{1\alpha}}.
\label{eq:sminus_critical_mode_energy}
\end{equation}
The last factor here is the ratio $J_{1\alpha}/J_a$, and it is what relaxes the energy demand of the internal mode compared to coherent exchange (recall Eq.~\ref{eq:critical_mode_energy}).  If growth dominates both damping rates, the internal and chain mode actions become comparable and this reduction disappears. However, if $\gamma_{1\alpha}\gg(\mathfrak{Re}\{\lambda_+\},\gamma_a)$, the planetary mode carries much less action than the chain: the stellar tidal forcing continually replenishes the damped internal oscillation while libration accumulates action. 

\begin{figure}
\centering
\includegraphics[width=\columnwidth]{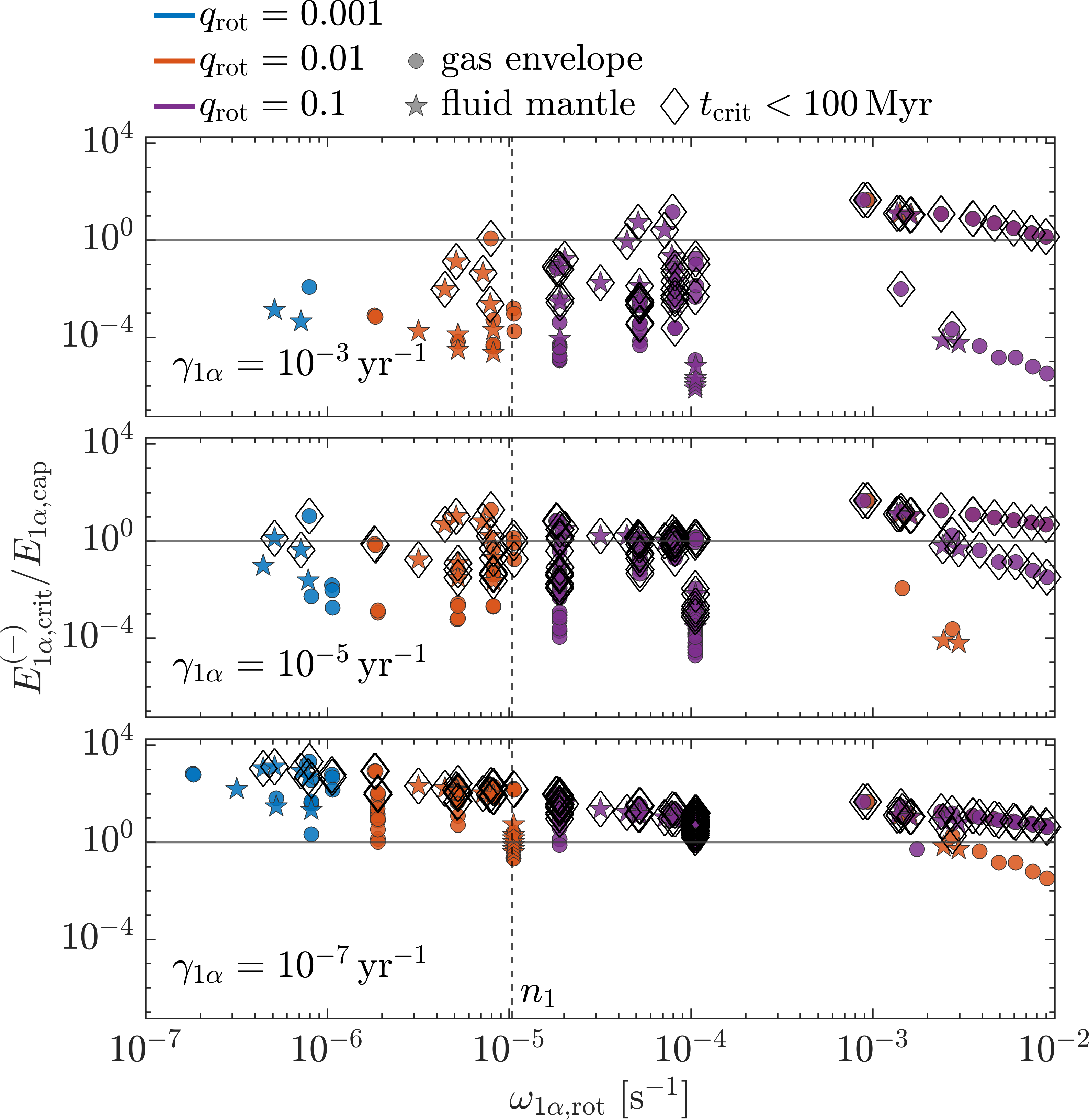}
\caption{Planetary mode energy at the critical chain action, relative to the geometric capacity in Eq.~\eqref{eq:geometric_mode_energy_capacity}. We use the couplings shown in Figure~\ref{fig:sminus_coupling}, 
$\gamma_a=10^{-9}\,{\rm yr}^{-1}$, and the internal damping rates indicated in each panel. Only modes satisfying the parametric instability are plotted. Modes with ordinate values below unity additionally satisfy the geometric energy capacity requirement. Colors distinguish rotation rates; circles and pentagrams denote envelope and fluid mantle modes. Black diamonds identify modes whose growth timescale $t_{\rm crit}<100~{\rm Myr}.$}
\label{fig:sminus_energy_capacity}
\end{figure}

Figure~\ref{fig:sminus_energy_capacity} shows how the additional factor in \eq{eq:sminus_critical_mode_energy} leaves many low-frequency modes below capacity and renders the parametric instability energetically permissible. Only parametrically unstable modes are plotted. For $\gamma_{1\alpha}=10^{-3}\,{\rm yr}^{-1}$, their required energies can be only $10^{-5}$ of the geometric capacity. As damping weakens, more modes become unstable, but the reduction in stored action diminishes. Black diamonds identify modes whose $t_{\rm crit}<100~{\rm Myr}$.  Thus, diamond-enclosed modes below the unity horizontal line satisfy the three desired conditions: \textit{i)} their resonant interaction with the chain is parametrically unstable; \textit{ii)} the chain reaches $J_{a,{\rm crit}}$ within $100\,{\rm Myr}$; and \textit{iii)} their energies at that point remains below its adopted geometric capacity. For  weak damping, $\gamma_{1\alpha}=10^{-7}\,{\rm yr}^{-1}$, no mode satisfies these conditions simultaneously. 

\subsection{Demonstration of Parametric Resonance Passage and Chain Excitation}
\label{subsec:parametric_excitation_example}
\begin{figure}
 \centering
 \includegraphics[width=\columnwidth]{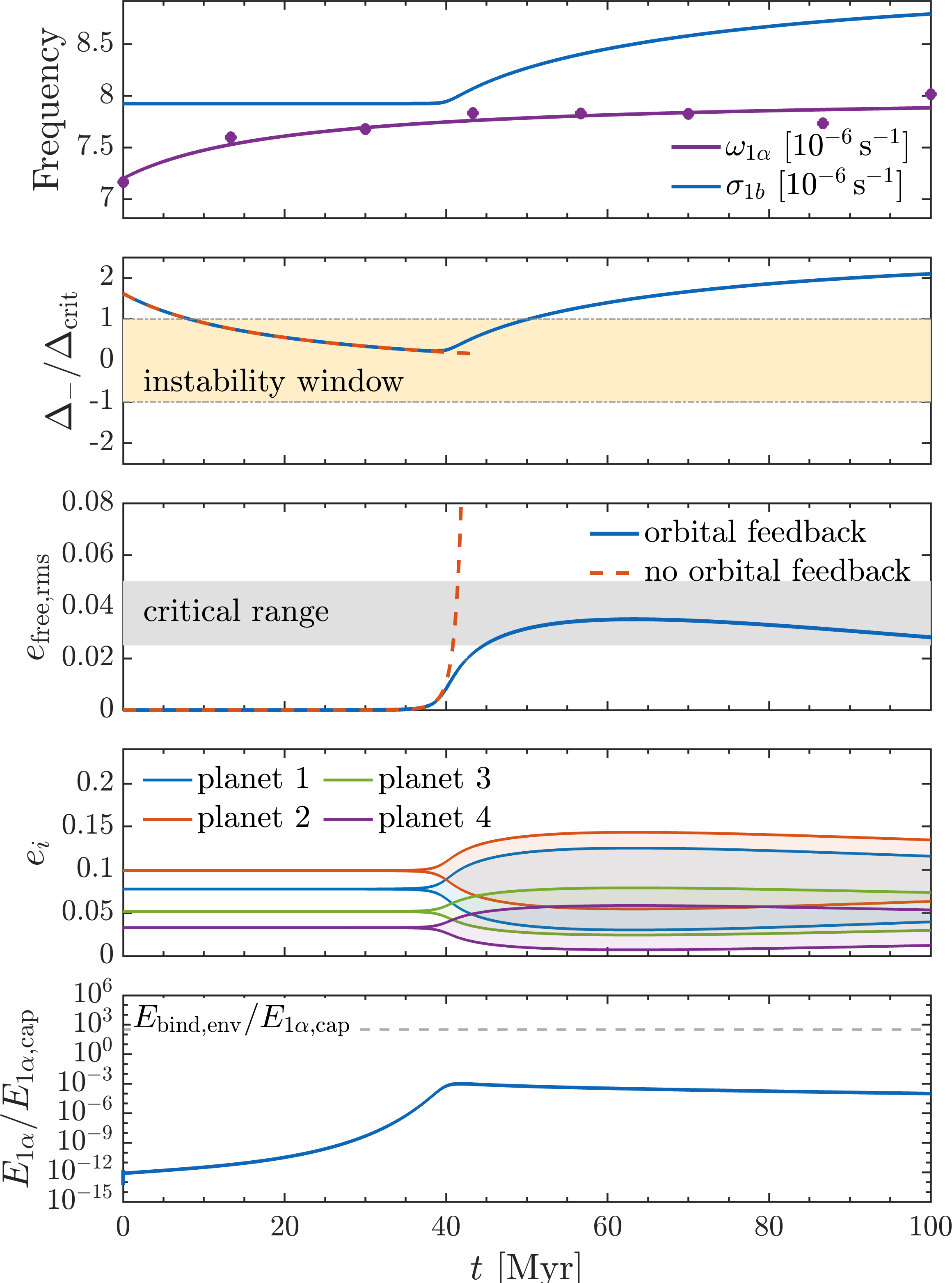}
 \caption{Parametric instability and chain excitation. Resonance encounter between a planetary inertial mode and chain mode $a=3$, as described in Section~\ref{subsec:parametric_excitation_example} and detailed in Appendix~\ref{app:parametric_excitation_example}.  From top to bottom: internal mode and tidal forcing frequencies, with $\omega_{1\alpha}$ computed using \texttt{MAQAM} at several snapshots of the thermal evolution in Figure~\ref{fig:orch_ps0_evolution} and fitted by the smooth curve; resonance detuning normalized by the instability window boundaries; rms scalar free eccentricity of the chain, with the adopted $0.025$--$0.05$ range shaded; the individual eccentricities of the four planets; and the internal mode energy normalized by its geometric capacity, with the envelope binding energy shown for comparison. Here $t=0$ corresponds to a planetary age of $10\,{\rm Myr}$. The eccentricities at $t=0$ are arbitrarily scaled; we choose initial eccentricities on the order of 0.1 so that the eccentricities after resonance passage are not too different from those before, in line with our linear analysis. Had we chosen initial eccentricities much smaller (as may have been the case in real-life chains; e.g.,~\citealt{choksi2026two}), the post-passage eccentricities and $e_{\rm free,rms}$ would be the same as shown here (data not shown). 
 }\label{fig:two_mode_simulation}
\end{figure}

We close with a case example. The gas envelope and fluid mantle support modes with frequencies comparable to those of the stellar tide and couplings sufficient to drive the instability (Fig.~\ref{fig:sminus_energy_capacity}). Cooling and contraction of the envelope (Fig.~\ref{fig:orch_ps0_evolution}) shifts mode frequencies, allowing the system to encounter a tidal sideband resonance. We illustrate such an encounter by integrating Eq.~\eqref{eq:sminus_linear_system} for one inertial mode in the inner planet's gas envelope, as it couples to our fiducial chain mode $a=3$. Appendix~\ref{app:parametric_excitation_example} contains the technical details underlying the model.

Figure~\ref{fig:two_mode_simulation}   follows the fiducial chain from $e_{\rm free,rms}=10^{-6}$ while the inner planet cools (note that starting with $e_{\rm free}=0$ at $t=0$ results in the same evolution, just delayed by $\sim$20 Myr). In the top panel, before resonance encounter, the tidal frequency $\sigma_{1b}$ remains nearly constant, while $\omega_{1\alpha}$ increases by ${\sim}7.5\%$. Such a change in $\omega_{1\alpha}$ occurs not because of changes in the planet's spin frequency $\Omega$ from contraction (which are smaller and offset by tidal synchronization with the orbit), but rather because 
of changes in the dimensions of the gas envelope (wave cavity) and its density profile \citep[e.g.,][]{fuller2016resonance}. As a baseline estimate, in a uniformly rotating unstratified fluid, inertial waves satisfy
\begin{equation}
\omega_{1\alpha}^2=4\Omega^2\frac{k_z^2}{k_z^2+k_\perp^2},
\end{equation}
where $k_z$ and $k_\perp$ are the wavevector components parallel and perpendicular to the rotation axis, respectively \citep[e.g.,][]{lin2018tidal}. 
For corresponding envelope dimensions $L_z$ and $L_\perp$,  $k_z\propto L_z^{-1}$ and $k_\perp\propto L_\perp^{-1}$. At fixed $\Omega$,
\begin{equation}
    d\ln|\omega_{1\alpha}|
    =-\left(1-\frac{\omega_{1\alpha}^2}{4\Omega^2}\right)
    d\ln\left(\frac{L_z}{L_\perp}\right).
\end{equation}
Thus the inertial wave frequency increases if $L_z \simeq R_{\rm p} - R_{\rm I}$ of the envelope contracts more than the transverse dimension $L_\perp \simeq R_{\rm p}$, which is plausible since the envelope contracts its outer radius $R_{\rm p}$ more than its inner radius $R_{\rm I}$ (mantle-atmosphere interface).

As cooling raises $\omega_{1\alpha}$, its separation from the nearly constant $\sigma_{1b}$ approaches $\omega_a$, reducing the detuning until the instability condition in Eq.~\eqref{eq:sminus_instability_threshold} is satisfied (see the second panel of Fig.~\ref{fig:two_mode_simulation}). The tidal pump then amplifies both mode actions, and planet eccentricities oscillate over a much wider range (fourth panel). The chain's rms free eccentricity reaches the adopted critical range $0.025-0.05$ after ${\sim}45$ Myr (third panel). With $\gamma_{1\alpha}/\gamma_{a}=2\times10^{5}$, the chain accumulates action while the more strongly damped planetary mode stores less.  The latter's energy, shown in the bottom panel, peaks at ${\sim}0.1\%$ of capacity (Eq.~\ref{eq:geometric_mode_energy_capacity}), which itself is only $\sim1/300$ of the entire envelope's binding energy. 

Without feedback on the orbital forcing frequency $\sigma_{1b}$, the increasing mode frequency $\omega_{1\alpha}$ from cooling carries the system closer to zero detuning, and the free eccentricity rapidly exceeds the critical range (third panel of Fig.~\ref{fig:two_mode_simulation}). When orbital feedback is accounted for, $\sigma_{1b}$ increases (top panel) as the inner planet's orbit decays from energy dissipation, 
detuning the system out of the instability window (second panel). Eccentricity growth is thus naturally arrested --- and within the desired range for the modes and damping coefficients chosen. Chain damping (always present) then outweighs the tidal driving, and the eccentricity envelopes slowly decay.

Our two-mode calculation suffers from many limitations:
\begin{itemize}

\item[i)] \textit{Linear chain dynamics:} 
We treat MMR libration as simple harmonic motion about a fixed equilibrium. In reality, when the MMR libration becomes large enough, a separatrix is crossed and libration gives way to circulation. Our model does not account for MMR circulation. Some if not most real-life resonant sub-Neptunes, including young chains, exhibit MMR angles that circulate (e.g.,~\citealt{choksi2026two}, and references therein).

\item[ii)] \textit{Single resonance passage:} We consider only one internal mode, one chain mode, and their parametric coupling. Other parametric resonances associated with other modes may be encountered as the system evolves  \citep[e.g.,][]{van2024non}. In connection with (i) above, if passage through one parametric resonance excites enough free eccentricity to cause MMR angles to circulate, how the now-circulating system responds to subsequent parametric resonance encounters (of either sideband) cannot be captured by our present theory.

\item[iii)] \textit{Mode saturation:} We define an internal oscillation mode's ``energy capacity'' $E_{1\alpha,{\rm cap}}$ using a somewhat arbitrary displacement criterion (Eq.~\ref{eq:geometric_mode_energy_capacity}). 
Nonlinear effects or energy transfer to daughter modes, not accounted for by our model, may saturate the mode well below this scale \citep[e.g.,][]{weinberg2012nonlinear}. 

\item[iv)] \textit{Tidal heating:} We neglect the effect of dissipated heat on planetary structure and mode properties. 

\end{itemize}

\section{Summary and Outlook}
\label{sec:Summary} 
Close-in sub-Neptunes occupy mean-motion resonant chains that break apart on a $\sim$100 Myr timescale 
\citep[][]{dai2024prevalence}. In the two-stage picture of \citet{choksi2026two}, excitation of free eccentricities to a few percent primes the chain (``rattles'' it; \citealt{hadden2026rattle}) for subsequent destruction on the observed timescale.  We have identified a tidal mechanism for their  first stage of eccentricity excitation. Chain librations modulate the stellar tide driving oscillations within the innermost planet, and the resulting deformation feeds back on the librations. This reciprocal interaction admits two sideband resonances, one which is stable and allows finite-amplitude exchange of action between the inner planet's internal oscillation and MMR oscillations, and a second resonance which is parametrically unstable   (Figures~\ref{fig:toy_model_integration}-\ref{fig:two_oscillator_phase_portrait}). We propose that parametric instability is the mechanism by which MMR librations can grow, even from zero, during stage one. The 
parametric resonance may be encountered as young planets cool; there is no need to invoke any reservoir of impactors or external perturbers.

We constructed a Hamiltonian formalism describing the coupling between the chain and internal planetary modes, and developed an eigenmode solver, \texttt{MAQAM}, to study pressure, gravity, and rotational modes in realistic planets modeled with \texttt{ORCHARD} \citep[][]{arevalo2026orchard}.
We applied this formalism to a four-planet chain, computing how the oscillations of the innermost sub-Neptune's gas envelope and molten mantle evolve as the planet cools. Our calculations taught us the following:
\begin{itemize}
    \item[i)] The sideband resonance allowing for coherent but bounded exchange of action (energy) between the internal planetary mode and the MMR modes cannot itself be responsible for chain disruption. For every internal mode tested,     
    storing the action needed to excite free eccentricities of a few percent required a mode energy one to four orders of magnitude above the mode's capacity (defined as the energy at which fluid displacements become comparable to the characteristic radius of the oscillating region; Figure~\ref{fig:energy_reqs}).

\item[ii)] By contrast, the parametrically unstable resonance, in connection with damping of the internal planet mode, overcomes this constraint. Action is drawn continuously from stellar tidal forcing into both internal planet oscillations and MMR librations; but because the internal oscillations damp more strongly than chain librations, the latter can grow without the 
former reaching capacity (Figure~\ref{fig:sminus_energy_capacity}). We found that inertial modes in the envelope and fluid mantle, unlike pressure and surface gravity modes, have low enough frequencies to be comparable to tidal frequencies (read: Keplerian orbital frequencies) and be driven into parametric resonance (Figure~\ref{fig:sminus_coupling}).

    \item[iii)] 
    As a sub-Neptune cools, its inertial mode frequencies change because of changes in the 
   volume 
    and density profile of its contracting envelope. When an inertial mode frequency matches the difference between the tidal harmonic frequency and an MMR libration frequency, parametric resonance is encountered, and chain librations = free eccentricities are excited. 
    The system eventually falls out of resonance because the tidal forcing frequency grows faster than the inertial mode frequency. We find, subject to assumptions about mode damping, that it is possible for resonance encounter with an inertial mode to occur on a $\sim$10 Myr timescale and to amplify free eccentricities to a few percent after resonance passage 
    (Figures~\ref{fig:sminus_growth_time} and \ref{fig:two_mode_simulation}). The inertial mode action stays well below capacity (Figure~\ref{fig:sminus_energy_capacity}).    
\end{itemize}

Our last finding (iii) is for passage through a single unstable resonance. What happens when multiple resonances (with different inertial modes, different tidal harmonics, and different MMR modes) are encountered in succession needs to be explored. In principle, the other, parametrically stable sideband resonance may also be encountered; this resonance could transfer action out of the chain and into the internal planet mode, where it would presumably be rapidly damped away. Note that this does not happen for the case example shown in Fig.~\ref{fig:two_mode_simulation} because orbital decay of the innermost planet detunes the system away from this sideband ($\Delta_-$ is prevented from crossing zero). Our theory is also incomplete because it does not account for how MMR libration gives way to circulation if the free eccentricities become large enough (larger than the forced eccentricities, as a rule of thumb). A chain with circulating MMR angles should couple more weakly to stellar tidal forcing than a chain in libration. This could provide another route by which parametric instability shuts off, by exciting enough free eccentricity to set the system into circulation. At least some young near-resonant systems 
have resonant angles that circulate \citep[e.g.,][]{hu2025unexpected}, and in the two-stage calculations of \citet{choksi2026two}, all resonant arguments are circulating after stage one eccentricity excitation.

Whether planet pairs are driven wide or narrow of perfect commensurability (i.e.~whether period ratios are greater or less than perfect integer ratios of 3/2 or 2/1) as a result of parametric resonance passage also needs to be examined. Observationally, some young pairs have period ratios that fall short of  commensurability by fractional amounts ranging up to $\Delta \simeq -2.7\%$ (e.g.~V1298~Tau, AU~Mic; \citealt{david2019four,wittrock2023validating}); such negative $\Delta$'s are not expected from capture into mean-motion resonance by convergent migration, and have been interpreted to imply collisional impacts \citep{choksi2026two}. How individual planet semimajor axes change from parametric resonance passage depends on which collective chain mode is parametrically resonant. A preliminary orbital integration of our fiducial 4-planet chain using chain mode $a=6$ produces period ratio offsets $\Delta \sim -1\%$, but only momentarily; over the MMR libration period of 1--2 years, the offsets  librate back to $\Delta \sim +1\%$. 
Again, extending our theory to account for circulation would be a natural next step. Conceivably both permanent $\Delta < 0$ and $\Delta > 0$ can be generated depending on the sign of $\Delta$ when circulation shuts off parametric growth.

\begin{acknowledgements}
M.F.~and E.C.~are supported by the Miller Institute for Basic Research in Science, University of California Berkeley, and  E.C.~by a Simons Investigator grant. We thank Roberto Tejada Arevalo for providing 
access to an early-release version of $\texttt{ORCHARD}$, and Nick Choksi for informative exchanges. We acknowledge the use of A.I.~tools (Claude.ai and ChatGPT) for assistance in coding, and pointers to and discussion of the literature. The authors take full responsibility for the content of this paper; no text was written by an A.I.
\end{acknowledgements}
 
\appendix

\section{Canonical Normalization of the Chain Modes}
\label{app:chain_mode_normalization}
At fixed $\boldsymbol\eta$, let $\mathbf A_{\rm ch}=\mathbf J_{\rm ch}\mathbf K_{\rm ch}$ be the linearized orbital evolution matrix. For each imaginary eigenvalue pair, choose a complex eigenvector $\mathbf e_a$ satisfying
\begin{equation}
\mathbf A_{\rm ch}\mathbf e_a=-i\omega_a\mathbf e_a,\qquad \mathcal N_a=\frac{i}{2}\mathbf e_a^\dagger\mathbf J_{\rm ch}\mathbf e_a,
    \label{eq:chain_eigenvector_norm}
\end{equation}
where $\omega_a$ is real and nonzero, $\dagger$ denotes conjugate transpose, and $\mathcal N_a$ is real. We select the member of each conjugate pair with $\mathcal N_a>0$, replacing
$(\mathbf e_a,\omega_a)$ by $(\mathbf e_a^\ast,-\omega_a)$ when necessary, and rescale $\mathbf e_a$ to $\mathbf{e}_a\sqrt{\mathcal{N}_a}$. Define the real mode vectors by
\begin{equation}
\mathbf u_a=\mathfrak{Re}\{\mathbf e_a\},\qquad \mathbf v_a=\mathfrak{Im}\,\mathbf \{e_a\}.
    \label{eq:chain_real_mode_vectors}
\end{equation}
Their symplectic products satisfy
\begin{equation}
\mathbf u_a^{\mathsf T}\mathbf J_{\rm ch}\mathbf u_b=\mathbf v_a^{\mathsf T}\mathbf J_{\rm ch}\mathbf v_b=0,\qquad \mathbf u_a^{\mathsf T}\mathbf J_{\rm ch}\mathbf v_b=-\delta_{ab},
    \label{eq:chain_symplectic_normalization}
\end{equation}
where $\delta_{ab}$ is the Kronecker delta. In these normalization equations, $a$ and $b$ label chain modes. The eigenvalue equation also gives
\begin{equation}
\mathbf A_{\rm ch}\mathbf u_a=\omega_a\mathbf v_a,\qquad \mathbf A_{\rm ch}\mathbf v_a=-\omega_a\mathbf u_a,
    \label{eq:chain_real_eigenrelations}
\end{equation}
and hence
\begin{equation}
\mathbf u_a^{\mathsf T}\mathbf K_{\rm ch}\mathbf u_b=\mathbf v_a^{\mathsf T}\mathbf K_{\rm ch}\mathbf v_b=\omega_a\delta_{ab},\qquad \mathbf u_a^{\mathsf T}\mathbf K_{\rm ch}\mathbf v_b=0.
    \label{eq:chain_energy_normalization}
\end{equation}
We introduce real oscillator coordinates $Q_a$ and momenta $P_a$ through
\begin{equation}
    \delta\mathbf Z=\sum_a(\mathbf u_aQ_a-\mathbf v_aP_a),
\end{equation}
where
\begin{align}\nonumber
Q_a&=\sqrt{2J_a}\cos\psi_a,\\
P_a&=-\sqrt{2J_a}\sin\psi_a.
    \label{eq:chain_cartesian_to_action}
\end{align}
Equation~\eqref{eq:chain_symplectic_normalization} makes $(Q_a,P_a)$ canonical. The polar transformation then gives the canonical pair $(\psi_a,J_a)$ with $J_a\geq0$ and reproduces Eq.~\eqref{eq:chain_mode_expansion}. Equation~\eqref{eq:chain_energy_normalization} yields
\begin{equation}
\mathcal H_{\rm ch}^{(2)}=\frac{1}{2}\sum_a\omega_a(Q_a^2+P_a^2)=\sum_a\omega_aJ_a.
    \label{eq:chain_signed_quadratic_form}
\end{equation}
Thus the sign of the quadratic mode energy is fixed by the Hamiltonian and the canonical orientation. 

\section{Tidal Sideband Coefficients}
\label{app:chain_modulation_coefficients}

Let $\boldsymbol\beta$ denote the fast and cyclic angles conjugate to $\boldsymbol\eta$. At fixed $\boldsymbol\eta$, their linear response to a chain oscillation satisfies
\begin{align}
\delta\dot{\boldsymbol\beta}&=\mathbf C_{\eta Z}\delta\mathbf Z,
    \label{eq:restored_angle_evolution}\\
\mathbf C_{\eta Z}&=\left.\frac{\partial^2\mathcal H_{\rm ch}}{\partial\boldsymbol\eta\,\partial\mathbf Z}\right|_\star.
    \label{eq:orbital_mixed_derivative}
\end{align}
Using Eq.~\eqref{eq:chain_mode_expansion} and $\dot\psi_a=\omega_a$ gives
\begin{equation}
\delta\boldsymbol\beta=\sum_a\sqrt{2J_a}\left(\mathbf u_a^\beta\cos\psi_a+\mathbf v_a^\beta\sin\psi_a\right),
    \label{eq:restored_angle_mode_expansion}
\end{equation}
where
\begin{equation}
\mathbf u_a^\beta=-\frac{\mathbf C_{\eta Z}\mathbf v_a}{\omega_a},\qquad \mathbf v_a^\beta=\frac{\mathbf C_{\eta Z}\mathbf u_a}{\omega_a}.
    \label{eq:restored_angle_mode_vectors}
\end{equation}
For a real tidal quantity $f$, combine the cosine and sine responses into the  derivative
\begin{equation}
\mathcal D_a^{(s)}f=(\mathbf u_a-is\mathbf v_a)\cdot\nabla_{\mathbf Z}f+(\mathbf u_a^\beta-is\mathbf v_a^\beta)\cdot\nabla_{\boldsymbol\beta}f,
\label{eq:sideband_directional_derivative}
\end{equation}
where $s=\pm1$. These derivatives are evaluated on the reference motion at fixed $\boldsymbol\eta$ and spin variables. For the amplitudes in Eq.~\eqref{eq:F_fourier}, $\nabla_{\boldsymbol\beta}\epsilon_{i\alpha b}=0$. Resolving Eq.~\eqref{eq:H_mod_raw} into the sidebands gives
\begin{equation}
g^{(s)}_{i\alpha ab}e^{i\chi^{(s)}_{i\alpha ab}}=\frac{1}{\sqrt{2}}\Bigl[\mathcal D_a^{(s)}\epsilon_{i\alpha b}
+i\epsilon_{i\alpha b}^{\star}\mathcal D_a^{(s)}\Theta_{ib}\Bigr].
    \label{eq:gchi_def}
\end{equation}
The two terms are the amplitude and phase contributions. Their sum determines $g^{(s)}_{i\alpha ab}$ and $\chi^{(s)}_{i\alpha ab}$ as its modulus and argument.
\section{Hamiltonian of the Fiducial Resonant Chain}
\label{app:fiducial_chain_hamiltonian}
To complete the transformation in Eq.~\eqref{eq:laplace_angles_mixed}, define the fast and cyclic angles
\begin{equation}
\phi_3=\lambda_3-\lambda_4,\qquad\phi_4=4\lambda_4-3\lambda_3.
\label{eq:fast_cyclic_angles_mixed}
\end{equation}
The inverse transformation is
\begin{align}
\lambda_1&=\phi_1+2\phi_2+8\phi_3+\phi_4,\nonumber\\
\lambda_2&=\phi_2+6\phi_3+\phi_4,\nonumber\\
\lambda_3&=4\phi_3+\phi_4,\nonumber\\
\lambda_4&=3\phi_3+\phi_4,
\label{eq:lambda_inverse_mixed}
\end{align}
with $\gamma_i=\sigma_i-\phi_4$. Hence
\begin{equation}
\begin{aligned}
&\sum_{i=1}^{4}(\Lambda_i\,d\lambda_i+\Gamma_i\,d\gamma_i)=L_1\,d\phi_1+L_2\,d\phi_2+\mathcal J_{\rm sc}\,d\phi_3\\
\end{aligned}
\label{eq:one_form_mixed}
\end{equation}
which verifies the canonical pairs and conserved actions used in the main text. Define
\begin{equation}
\hat\Lambda_i=\Lambda_i-\bar\Lambda_i,\qquad\hat L_k=L_k-\bar L_k,
\label{eq:Lambda_hat_fiducial_def}
\end{equation}
where $k=1,2$, and
\begin{equation}
\eta_{\rm res}=\sum_{i=1}^{4}\Gamma_i-\delta_{\rm res}=\sum_{i=1}^{4}\hat\Lambda_i.
\label{eq:eta_res_mixed_def}
\end{equation}
Then the inverse action relations are
\begin{align}
\hat\Lambda_1&=\hat L_1,\nonumber\\
\hat\Lambda_2&=\hat L_2-2\hat L_1,\nonumber\\
\hat\Lambda_3&=\hat L_1-3\hat L_2-3\eta_{\rm res},\nonumber\\
\hat\Lambda_4&=2\hat L_2+4\eta_{\rm res}.
\label{eq:Lambdahat_mixed}
\end{align}
Thus the instantaneous AMD is $\sum_i\Gamma_i=\delta_{\rm res}+\eta_{\rm res}$. The nominal mean motions are proportional to the weights in $\mathcal J_{\rm sc}$, so $\sum_i\bar n_i\hat\Lambda_i=0$ at fixed scale. The quadratic Keplerian expansion therefore gives Eq.~\eqref{eq:Hkep_fiducial_mixed}.

In the reduced angles, Eq.~\eqref{eq:Hres_fiducial_compact} becomes
\begin{align}
\mathcal H_{\rm res}&=C_{1,2}\sqrt{\Gamma_1}\cos(\sigma_1-2\phi_2-3\phi_1)\nonumber\\
&\quad+C_{2,1}\sqrt{\Gamma_2}\cos(\sigma_2-2\phi_2-3\phi_1)\nonumber\\
&\quad+C_{1,3}\sqrt{\Gamma_1}\cos(\sigma_1-2\phi_2-\phi_1)\nonumber\\
&\quad+C_{3,1}\sqrt{\Gamma_3}\cos(\sigma_3-2\phi_2-\phi_1)\nonumber\\
&\quad+C_{2,3}\sqrt{\Gamma_2}\cos(\sigma_2-2\phi_2)\nonumber\\
&\quad+C_{3,2}\sqrt{\Gamma_3}\cos(\sigma_3-2\phi_2)\nonumber\\
&\quad+C_{2,4}\sqrt{\Gamma_2}\cos(\sigma_2-\phi_2)\nonumber\\
&\quad+C_{4,2}\sqrt{\Gamma_4}\cos(\sigma_4-\phi_2)\nonumber\\
&\quad+C_{3,4}\sqrt{\Gamma_3}\cos\sigma_3\nonumber\\
&\quad+C_{4,3}\sqrt{\Gamma_4}\cos\sigma_4.
\label{eq:Hres_fiducial_mixed}
\end{align}

For an inner-outer pair $(i,j)$ in a $(p+1):p$ resonance, let
\begin{equation}
\alpha_{ij}=\frac{\bar a_i}{\bar a_j}=\left(\frac{p}{p+1}\right)^{2/3}\left(\frac{\mu_i}{\mu_j}\right)^{1/3}.
\label{eq:alpha_pair_fiducial}
\end{equation}
The direct coefficients follow from the Laplace expansion \citep{laskar1995stability,delisle2017analytical}:
\begin{align}
f_{\rm in}^{(p)}(\alpha)&=-\frac{1}{2}\left[2(p+1)+\alpha\frac{d}{d\alpha}\right]b_{1/2}^{(p+1)}(\alpha),\label{eq:f_in_general_first_order}\\
f_{\rm out}^{(p)}(\alpha)&=\frac{1}{2}\left[2p+1+\alpha\frac{d}{d\alpha}\right]b_{1/2}^{(p)}(\alpha),
\label{eq:f_out_general_first_order}
\end{align}
where the Laplace coefficients, evaluated by quadrature, are
\begin{equation}
b_{1/2}^{(m)}(\alpha)=\frac{1}{\pi}\int_0^{2\pi}\frac{\cos(m\theta)\,d\theta}{\sqrt{1-2\alpha\cos\theta+\alpha^2}}.
\label{eq:laplace_coeff_def}
\end{equation}
Here $m\geq0$ is an integer Fourier order. The indirect interaction contributes only to the outer eccentricity term of the $2:1$ resonance at this order. In our canonical heliocentric convention its dimensionless coefficient is
\begin{equation}
\mathcal I_{ij}=\frac{\mathcal G M_\star}{\sqrt{\mu_i\mu_j\alpha_{ij}}}.
\label{eq:fiducial_indirect_coefficient}
\end{equation}
The retained pair Hamiltonian is therefore
\begin{equation}
\begin{aligned}
\mathcal H_{i,j}=-\frac{\mathcal Gm_im_j}{\bar a_j}\Big\{&f_{\rm in}^{(p)}(\alpha_{ij})e_i\cos\vartheta_{ij,i}\\
&+\big[f_{\rm out}^{(p)}(\alpha_{ij})-\delta_{p1}\mathcal I_{ij}\big]\\
&\qquad\times e_j\cos\vartheta_{ij,j}\Big\},
\end{aligned}
\label{eq:pair_res_disturbing_function}
\end{equation}
where $\delta_{p1}$ is one for $p=1$ and zero otherwise. Using $e_i\simeq\sqrt{2\Gamma_i/\bar\Lambda_i}$ gives
\begin{equation}
C_{i,j}=-\frac{\mathcal Gm_im_j}{\bar a_j}\sqrt{\frac{2}{\bar\Lambda_i}}\,f_{\rm in}^{(p)}(\alpha_{ij}),
\label{eq:C_in_fiducial}
\end{equation}
and
\begin{equation}
\begin{aligned}
C_{j,i}={}&-\frac{\mathcal Gm_im_j}{\bar a_j}\sqrt{\frac{2}{\bar\Lambda_j}}\\
&\times\big[f_{\rm out}^{(p)}(\alpha_{ij})-\delta_{p1}\mathcal I_{ij}\big].
\end{aligned}
\label{eq:C_out_fiducial}
\end{equation}
At the nominal $2:1$ commensurability and for $m_i,m_j\ll M_\star$, $\mathcal I_{ij}=\alpha_{ij}^{-1/2}=2\alpha_{ij}$, recovering the usual indirect correction \citep[e.g.,][]{morbidelli2002modern}.

\section{Evolutionary Model of the Inner Sub-Neptune}
\label{app:orch_ps0_reproducibility}
We use \texttt{ORCHARD} \citep[][]{arevalo2026orchard} version \texttt{v0.1.0} with 500 Lagrangian mass zones and an outer pressure of $1\,{\rm bar}$. The iron core, silicate mantle, and envelope masses are $1.4985$, $3.0015$, and $0.5000\,M_\oplus$, respectively. The envelope has the homogeneous composition specified in Section~\ref{subsec:orch_ps0_evolution}. Its initial specific entropy is $S_{\rm env}=5\,k_{\rm B}$ per baryon, while the mantle has $S_{\rm man}=0.58\,k_{\rm B}$ per baryon, where $k_{\rm B}$ is Boltzmann's constant. The initial core state provides approximate temperature continuity across the core--mantle boundary. The relaxed initial model has structural radius $3.636\,R_\oplus$, transit radius $3.832\,R_\oplus$, and intrinsic effective temperature $T_{\rm int}=336.6\,{\rm K}$.

The envelope combines the Chabrier-Debras H/He equation of state \citep{chabrier2021new} with the AQUA water equation of state \citep{haldemann2020aqua}, using additive volume mixing and the ideal entropy of mixing between the components. The mantle uses the forsterite Mg$_2$SiO$_4$ ANEOS model \citep{stewart2020shock}, with the melting prescription described by \citet{arevalo2026orchard}. The core uses the solid-liquid pure-iron equation of state of \citet{gonzalez2023ab}. Convection is allowed in all three regions, with radiative diffusion in the envelope and conduction in the mantle and core. The iron thermal conductivity is $100\,{\rm W\,m^{-1}\,K^{-1}}$. The model includes mantle radiogenic heating from $^{40}$K, $^{232}$Th, $^{235}$U, and $^{238}$U, together with latent heat from interior phase transitions.

The nominal seven  day orbit corresponds to $0.071615\,{\rm AU}$ around the adopted solar mass star. The cooling boundary uses the metallicity dependent irradiated atmosphere grid implemented as \texttt{f07}, described by \citet{arevalo2026orchard} and based on \citet{fortney2007planetary}. It is interpolated using the incident flux, surface gravity, composition, and temperature near $1\,{\rm kbar}$. 

Envelope-mantle overshoot, helium and heavy-element rain, and stellar luminosity evolution are ignored, in addition to mass loss. The maximum time step and output interval are each $0.02\,{\rm Myr}$ for the first $2\,{\rm Myr}$ and $2\,{\rm Myr}$ thereafter. The calculation extends to $5\,{\rm Gyr}$.

The mantle-top temperature is evaluated in the outermost silicate cell. It rises from approximately $5800$ to $6300\,{\rm K}$ during the first $100\,{\rm Myr}$. Stable stratification first appears in the modeled envelope near $0.9\,{\rm Gyr}$ and occupies only the outermost $2\%$ of the planetary radius at $5\,{\rm Gyr}$.

\section{Rotating Fluid Eigenproblem}
\label{app:gms_continuum}
\texttt{MAQAM} computes the conservative oscillations of a spherical \texttt{ORCHARD} snapshot \citep{arevalo2026orchard} at an assigned uniform rotation rate. The envelope and fluid mantle are separate eigenproblems, with full Coriolis coupling and perturbed self-gravity. We give the envelope problem here; Appendix~\ref{app:gms_magma_ocean} specifies the mantle boundaries. Forcing, damping, and coupling to the resonant chain enter after the free modes have been recovered.

\subsection{Background and Perturbation Equations}
\label{app:orchard_background}
Let $M_{\rm p}$ and $R_{\rm p}$ denote the planetary mass and structural radius, and $m_r$, $P$, $\rho$, and $\Phi_0$ the enclosed mass, pressure, density, and gravitational potential at radius $r$. The snapshot satisfies
\begin{equation}
\begin{aligned}
\frac{dm_r}{dr}&=4\pi r^2\rho,\qquad \frac{dP}{dr}=-\rho g,\\
g&=\frac{\mathcal Gm_r}{r^2}=\frac{d\Phi_0}{dr}.
\end{aligned}
\label{eq:gms_background_structure}
\end{equation}
The dynamical frequency, sound speed, and the squared Brunt V\"ais\"al\"a frequency are
\begin{equation}
\omega_{\rm dyn}=\left(\frac{\mathcal GM_{\rm p}}{R_{\rm p}^3}\right)^{1/2},\qquad c_s^2=\frac{\Gamma_1P}{\rho},
\label{eq:gms_background_acoustic}
\end{equation}
\begin{equation}
N^2=g\left(\frac{1}{\Gamma_1}\frac{d\ln P}{dr}-\frac{d\ln\rho}{dr}\right),
\label{eq:gms_background_buoyancy}
\end{equation}
where $\Gamma_1$ is the logarithmic pressure derivative with respect to density at fixed entropy and composition. Rotation is uniform, $\boldsymbol\Omega=\Omega\hat{\boldsymbol z}$. We retain Coriolis acceleration but omit centrifugal deformation, whose characteristic size is
\begin{equation}
\epsilon_{\rm rot}=\frac{\Omega^2R_{\rm p}^3}{\mathcal GM_{\rm p}}=q_{\rm rot}^2,\qquad q_{\rm rot}=\frac{\Omega}{\omega_{\rm dyn}}.
\label{eq:gms_rotation_parameter}
\end{equation}
Thus the background approximation requires $q_{\rm rot}^2\ll1$, although Coriolis forces remain leading order for inertial oscillations. Centrifugally deformed models are needed beyond this regime \citep{dewberry2021constraining,dewberry2022dynamical}.

The gas occupies $R_{\rm I}<r<R_{\rm p}$, where $R_{\rm I}$ is the mantle-envelope interface. Material below $R_{\rm I}$ contributes to the equilibrium gravity but carries no displacement or density perturbation. We continue the perturbed potential through a gravity-only domain $R_{\rm Fe}<r<R_{\rm I}$, with $R_{\rm Fe}$ the iron core radius, and impose its regular interior solution at $R_{\rm Fe}$.

Perturbations have dependence $\exp(im\phi-i\omega t)$, where $\phi$ is the rotating azimuth, $m$ the azimuthal order, and $\omega\equiv\omega_{\rm rot}$ the signed rotating-frame frequency. We use
\begin{equation}
\omega_{\rm in}=\omega+m\Omega,\qquad \sigma=\frac{\omega}{\omega_{\rm dyn}},\qquad \kappa=\frac{\omega}{\Omega}.
\label{eq:gms_frequency_conventions}
\end{equation}
The displacement $\boldsymbol\xi$ and Eulerian perturbations $\delta p$, $\delta\rho$, and $\delta\Phi$ obey \citep{lynden1967stability,unno1979nonradial,Friedman1978langrangian,schenk2001nonlinear}
\begin{align}
 -\omega^2\rho{\boldsymbol\xi}
 -2i\omega\rho{\boldsymbol\Omega}\mathbin{\times}{\boldsymbol\xi}
 &=-\boldsymbol\nabla\delta p-\delta\rho\boldsymbol\nabla\Phi_0
   -\rho\boldsymbol\nabla\delta\Phi,
 \label{eq:gms_momentum}\\
\delta\rho&=-\boldsymbol\nabla\cdot(\rho\boldsymbol\xi),
\label{eq:gms_continuity}\\
\Delta p&=c_s^2\Delta\rho,
\label{eq:gms_adiabatic}\\
\nabla^2\delta\Phi&=4\pi\mathcal G\delta\rho.
\label{eq:gms_poisson}
\end{align}
For any background scalar $f$, its Lagrangian perturbation is
\begin{equation}
\Delta f=\delta f+\boldsymbol\xi\cdot\boldsymbol\nabla f.
\label{eq:gms_lagrangian_perturbation}
\end{equation}
Hydrostatic balance and the adiabatic relation give
\begin{equation}
\delta\rho=\frac{\delta p}{c_s^2}+\frac{\rho N^2}{g}\xi_r,
\label{eq:gms_density_relation}
\end{equation}
where $\xi_r=\boldsymbol\xi\cdot\hat{\boldsymbol r}$. These equations retain pressure, buoyancy, Coriolis, and gravitational restoring forces without the traditional or Cowling approximations.

\subsection{Vector Spherical Harmonic Projection}
Let $Y_\ell^m(\theta,\phi)$ be the scalar spherical harmonics, with degree $\ell$, colatitude $\theta$, and normalization
\begin{equation}
\begin{aligned}
\nabla_\Omega^2Y_\ell^m&=-L_\ell Y_\ell^m,\qquad L_\ell=\ell(\ell+1),\\
\int Y_\ell^{m*}Y_{\ell'}^{m'}\,d\Omega&=\delta_{\ell\ell'}\delta_{mm'}.
\end{aligned}
\end{equation}
Here $d\Omega$ is solid angle, $\nabla_\Omega$ the gradient on the unit sphere, and the deltas are Kronecker symbols \citep{abramowitz1972handbook}. The tangential gradient is
\begin{equation}
\boldsymbol\nabla_\perp=(\boldsymbol I-\hat{\boldsymbol r}\hat{\boldsymbol r})\cdot\boldsymbol\nabla=\frac{1}{r}\boldsymbol\nabla_\Omega,
\label{eq:gms_tangential_gradient}
\end{equation}
where $\boldsymbol I$ is the identity tensor. Define radial, spheroidal, and toroidal vector harmonics by
\begin{equation}
\begin{aligned}
\boldsymbol Y^R_{\ell m}&=Y_\ell^m\hat{\boldsymbol r},\\
\boldsymbol Y^S_{\ell m}&=r\boldsymbol\nabla_\perp Y_\ell^m,\\
\boldsymbol Y^T_{\ell m}&=\hat{\boldsymbol r}\times r\boldsymbol\nabla_\perp Y_\ell^m.
\end{aligned}
\end{equation}
Their squared angular norms are $n_{\ell R}=1$ and $n_{\ell S}=n_{\ell T}=L_\ell$. We expand
\begin{equation}
\boldsymbol\xi=\sum_{\ell\geq|m|}\left(u_\ell\boldsymbol Y^R_{\ell m}+v_\ell\boldsymbol Y^S_{\ell m}+w_\ell\boldsymbol Y^T_{\ell m}\right),
\label{eq:gms_vsh_displacement}
\end{equation}
\begin{equation}
\begin{aligned}
\delta p&=\sum_{\ell\geq|m|}\delta p_\ell(r)Y_\ell^m,\\
\delta\Phi&=\sum_{\ell\geq|m|}\delta\Phi_\ell(r)Y_\ell^m,\qquad d_\ell=\frac{d\delta\Phi_\ell}{dr}.
\end{aligned}
\label{eq:gms_scalar_expansion}
\end{equation}
The coefficients $u_\ell$, $v_\ell$, and $w_\ell$ are displacement amplitudes depending on $r$; $\delta p_\ell$ and $\delta\Phi_\ell$ are scalar radial coefficients, and $d_\ell$ makes Poisson's equation first order in radius.

Full Coriolis coupling follows from the angular projection \citep{papaloizou1978non,lee1987low,savonije1995non,ogilvie2009tidal}
\begin{equation}
C_{ab}^{\ell\ell'}=\frac{1}{n_{\ell a}}\int\boldsymbol Y_{\ell m}^{a*}\cdot\left(\hat{\boldsymbol z}\times\boldsymbol Y_{\ell'm}^{b}\right)\,d\Omega,
\label{eq:gms_coriolis_projection}
\end{equation}
where $a,b\in\{R,S,T\}$ and an asterisk denotes complex conjugation. Its three projected components are
\begin{equation}
c_{\ell a}=\sum_{\ell'}\left(C_{aR}^{\ell\ell'}u_{\ell'}+C_{aS}^{\ell\ell'}v_{\ell'}+C_{aT}^{\ell\ell'}w_{\ell'}\right).
\end{equation}
At fixed $m$, equatorial symmetry separates the equations into two parity blocks. In the even block, $u_\ell$, $v_\ell$, $\delta p_\ell$, and $\delta\Phi_\ell$ have $\ell-|m|$ even, while $w_\ell$ has $\ell-|m|$ odd. The assignments reverse in the odd block. We solve both blocks at $m=2$, retaining degrees through $\ell_{\max}$.

Projecting the mass conservation equation \eqref{eq:gms_continuity}  onto $Y_\ell^{m*}$ and using the density relation in Eq.~\eqref{eq:gms_density_relation} gives the first equation below. Projecting the momentum equation \eqref{eq:gms_momentum}  onto the radial, spheroidal, and toroidal
vector harmonics gives the next three equations. Finally, projecting
Poisson's equation \eqref{eq:gms_poisson}, onto $Y_\ell^{m*}$ and using the definition $d_\ell=d(\delta\Phi_\ell)/dr$ gives the final two equations. The resulting system within the gas envelope is
\begin{equation}
\frac{du_\ell}{dr}+\left(\frac{2}{r}-\frac{g}{c_s^2}\right)u_\ell-\frac{L_\ell}{r}v_\ell+\frac{\delta p_\ell}{\rho c_s^2}=0,
\label{eq:gms_radial_continuity}
\end{equation}
\begin{equation}
\begin{aligned}
\frac{d\delta p_\ell}{dr}+\frac{g}{c_s^2}\delta p_\ell+\rho(N^2-\omega^2)u_\ell&\\
+\rho d_\ell-2i\omega\Omega\rho c_{\ell R}&=0,
\end{aligned}
\label{eq:gms_radial_momentum}
\end{equation}
\begin{align}
\frac{1}{r}\left(\frac{\delta p_\ell}{\rho}+\delta\Phi_\ell\right)-\omega^2v_\ell-2i\omega\Omega c_{\ell S}&=0,
\label{eq:gms_horizontal_momentum}\\
-\omega^2w_\ell-2i\omega\Omega c_{\ell T}&=0,
\label{eq:gms_toroidal_momentum}\\
\frac{d\delta\Phi_\ell}{dr}-d_\ell&=0,
\label{eq:gms_phi_definition}
\end{align}
\begin{equation}
\frac{dd_\ell}{dr}+\frac{2d_\ell}{r}-\frac{L_\ell}{r^2}\delta\Phi_\ell=4\pi\mathcal G\left(\frac{\delta p_\ell}{c_s^2}+\frac{\rho N^2}{g}u_\ell\right).
\label{eq:gms_radial_poisson}
\end{equation}
In the rigid interior only $\delta\Phi_\ell$ and $d_\ell$ remain, satisfying Eqs.~\eqref{eq:gms_phi_definition} and \eqref{eq:gms_radial_poisson} with zero density source.

\subsection{Boundary Conditions}
The rigid mantle makes the envelope base impenetrable:
\begin{equation}
u_\ell(R_{\rm I})=0.
\label{eq:gms_base_mechanical}
\end{equation}
Tangential displacement is unconstrained in an inviscid fluid. The regular gravitational multipole in the excluded core is proportional to $r^\ell$, giving
\begin{equation}
\left[d_\ell-\frac{\ell}{r}\delta\Phi_\ell\right]_{r=R_{\rm Fe}}=0.
\label{eq:gms_interior_gravity}
\end{equation}
Both $\delta\Phi_\ell$ and $d_\ell$ are continuous across $R_{\rm I}$, whose fixed boundary produces no displaced density sheet.

At the outer boundary, the exterior pressure is held fixed, so
\begin{equation}
\Delta p_\ell(R_{\rm p})=\left[\delta p_\ell-\rho g u_\ell\right]_{r=R_{\rm p}}=0.
\label{eq:gms_surface_deltap}
\end{equation}
Matching to the exterior potential proportional to $r^{-(\ell+1)}$ gives
\begin{equation}
\left[d_\ell+\frac{\ell+1}{r}\delta\Phi_\ell+4\pi\mathcal G\rho u_\ell\right]_{r=R_{\rm p}}=0.
\label{eq:gms_surface_gravity}
\end{equation}
The last term retains the gravitational effect of the displaced surface when its equilibrium density is nonzero.

\subsection{Quadratic Eigenproblem and Rotating Normalization}

After projecting the equations onto spherical harmonics and discretizing the radial coordinate, we collect
all gas envelope variables (along with the interior gravitational potential values) into
a single vector ${\boldsymbol x}$. Because the restoring terms are independent of $\omega$, the Coriolis terms are linear in $\omega$, and the inertial terms are quadratic in $\omega$, the discretized equations form a quadratic matrix polynomial, commonly called a ``quadratic pencil"
\citep[e.g.,][]{tisseur2001quadratic}:
\begin{equation}
 {\boldsymbol P}(\kappa){\boldsymbol x}
 =\left(
 {\boldsymbol A}_0
 +\kappa\Omega^2{\boldsymbol A}_1
 +\kappa^2\Omega^2{\boldsymbol A}_2
 \right){\boldsymbol x}=0.
 \label{eq:gms_quadratic_pencil}
\end{equation}
Here ${\boldsymbol P}(\kappa)$ is the discretized mode operator,
${\boldsymbol A}_0$ contains the background, radial derivative,
self gravity, and boundary terms, ${\boldsymbol A}_1$ contains the
Coriolis couplings, and ${\boldsymbol A}_2$ contains the inertial terms. Using $\omega=\kappa\Omega$, Eq.~\eqref{eq:gms_quadratic_pencil} may be
written equivalently in the standard rotating fluid form
\begin{equation}
\left(-\omega^2\boldsymbol{\mathcal M}-i\omega\boldsymbol{\mathcal B}+\boldsymbol{\mathcal K}\right)\boldsymbol x=0,
\label{eq:gms_physical_quadratic_problem}
\end{equation}
where
${\boldsymbol{\cal K}}={\boldsymbol A}_0$,
$-i{\boldsymbol{\cal B}}=\Omega{\boldsymbol A}_1$, and
$-{\boldsymbol{\cal M}}={\boldsymbol A}_2$.
The matrices ${\boldsymbol{\cal M}}$, ${\boldsymbol{\cal B}}$, and
${\boldsymbol{\cal K}}$ contain the inertial, Coriolis, and restoring
terms, respectively
\citep[e.g.,][]{schenk2001nonlinear,papaloizou2005oscillations}.

For displacement fields $\boldsymbol\eta$ and $\boldsymbol\zeta$, define the mass-weighted inner product over the oscillating fluid volume $V_{\rm f}$ by
\begin{equation}
\langle\boldsymbol\eta,\boldsymbol\zeta\rangle_\rho=\int_{V_{\rm f}}\rho\boldsymbol\eta^*\cdot\boldsymbol\zeta\,d^3r.
\label{eq:gms_mass_inner_product}
\end{equation}
A stable mode $\alpha$ of real frequency $\omega_\alpha$ has gyroscopic norm
\begin{equation}
b_\alpha=\left\langle\boldsymbol\xi_\alpha,2\omega_\alpha\boldsymbol\xi_\alpha+2i\boldsymbol\Omega\times\boldsymbol\xi_\alpha\right\rangle_\rho.
\label{eq:gms_gyroscopic_norm}
\end{equation}
For the canonical description we retain nonzero frequency modes with nonzero norm. Complex conjugation replaces $(\boldsymbol\xi_\alpha,\omega_\alpha,m)$ by $(\boldsymbol\xi_\alpha^*,-\omega_\alpha,-m)$ and reverses $b_\alpha$. From each pair we choose the member with $b_\alpha>0$. With physical displacement $\mathfrak{Re}[A_\alpha(t)\boldsymbol\xi_\alpha]$, the action and rotating  frame energy are then
\begin{equation}
J_\alpha=\frac{b_\alpha}{4}|A_\alpha|^2,\qquad E_\alpha^{\rm rot}=\omega_\alpha J_\alpha,
\label{eq:gms_canonical_action}
\end{equation}
consistent with Eq.~\eqref{eq:mode_gyroscopic_norm} \citep[see also, ][]{Friedman1978langrangian,schenk2001nonlinear}.

\section{Numerical Framework of \texttt{MAQAM}}
\label{app:gms_numerics}

\subsection{State Preparation and Coriolis Projection}

The numerical calculation begins by converting the \texttt{ORCHARD}
snapshot onto the radial grids used by \texttt{MAQAM}. The input profiles provide finite values of $\rho$, $g$, $c_s^2$, and $N^2$ throughout the gas envelope, together with $M_{\rm p}$, $R_{\rm p}$, and either $\Omega$
or $q_{\rm rot}$. The gravity only interior and gas envelope domains are interpolated separately, so no interpolation crosses the interface at $R_{\rm I}$. The strictly positive $\rho$ and $c_s^2$ profiles are interpolated logarithmically, while $g$ and the signed $N^2$ profile are interpolated using shape preserving cubic interpolation. 

After the radial background has been prepared, the Coriolis coefficients in Eq.~\eqref{eq:gms_coriolis_projection} are evaluated for each
azimuthal order, parity, and angular truncation $\ell_{\max}$. The angular integrals are computed directly from the spherical harmonics and their derivatives using Gauss-Legendre quadrature in
$\mu=\cos\theta$, while the azimuthal integral is evaluated analytically.  
\subsection{Radial Finite Difference Discretization}

To solve the projected system in
Eqs.~\eqref{eq:gms_radial_continuity}-\eqref{eq:gms_radial_poisson},
each radial domain is sampled on a finite grid and the radial derivatives
are replaced by finite difference matrices, as in coupled harmonic
calculations of rotating stellar and planetary oscillations
\citep[e.g.,][]{lee1987low,savonije1995non,lovekin2008radial}.

An ordinary centered derivative can fail to detect a grid scale pattern that alternates in sign between adjacent points. To avoid generating numerical artifacts because of this, we use the second order upwind summation by parts pair of
\citet{mattsson2017diagonal}. Summation by parts operators are then constructed
to reproduce the integration by parts identity at the discrete level.

Consider a uniform computational coordinate
$s_i=(i-1)h$, where $i=1,\ldots,N$ and $h=(N-1)^{-1}$. Associated with these finite difference operators is a diagonal
quadrature matrix ${\boldsymbol H}_s$, which approximates integrals over
the computational coordinate. For grid vectors ${\boldsymbol f}$ and
${\boldsymbol g}$,
\begin{equation}
 \int_0^1 f^*(s)g(s)\,ds
 \simeq
 {\boldsymbol f}^\dagger{\boldsymbol H}_s{\boldsymbol g},
\end{equation}
where
\begin{equation}
 {\boldsymbol H}_s
 =h\,\operatorname{diag}
 \left(
 \frac{1}{4},\frac{5}{4},1,\ldots,1,
 \frac{5}{4},\frac{1}{4}
 \right).
 \label{eq:gms_sbp_norm}
\end{equation}
and the complementary derivative matrices
${\boldsymbol D}_+$ and ${\boldsymbol D}_-$ satisfy
\begin{equation}
 {\boldsymbol H}_s{\boldsymbol D}_+
 +{\boldsymbol D}_-^\dagger{\boldsymbol H}_s
 =
 {\boldsymbol e}_N{\boldsymbol e}_N^T
 -{\boldsymbol e}_1{\boldsymbol e}_1^T.
 \label{eq:gms_complementary_sbp}
\end{equation}
Here $\dagger$ denotes the conjugate transpose, while
${\boldsymbol e}_1$ and ${\boldsymbol e}_N$ select the first and last
grid points. Equation~\eqref{eq:gms_complementary_sbp} is the discrete
counterpart of integration by parts. Away from the boundaries, the two derivative formulas are
\begin{align}
 (D_+f)_i
 &=
 \frac{-3f_i/2+2f_{i+1}-f_{i+2}/2}{h},
 \\
 (D_-f)_i
 &=
 \frac{f_{i-2}/2-2f_{i-1}+3f_i/2}{h},
 \label{eq:gms_sbp_interior}
\end{align}
for $3\leq i\leq N-2$. Near the ends of the radial grid, the finite difference formulas in
Eq.~\eqref{eq:gms_sbp_interior} cannot be applied because they would require grid points outside the computational domain. The first two and last two rows of each derivative matrix are therefore replaced by 
boundary formulas chosen to preserve the summation by parts
identity. To specify the discrete operators completely, the nonzero coefficients of ${\boldsymbol D}_+$ at rows $1$, $2$, $N-1$, and $N$ are
\begin{equation}
 \frac{1}{h}
 \left[
 (-3,5,-2),\
 \left(-\frac{1}{5},-1,\frac{8}{5},-\frac{2}{5}\right),\
 (-1,1),\
 (-1,1)
 \right],
 \label{eq:gms_sbp_plus_closure}
\end{equation}
respectively, with each coefficient vector acting on the adjacent grid
values. The corresponding boundary rows of ${\boldsymbol D}_-$ are
\begin{equation}
 \frac{1}{h}
 \left[
 (-1,1),\
 (-1,1),\
 \left(\frac{2}{5},-\frac{8}{5},1,\frac{1}{5}\right),\
 (2,-5,3)
 \right].
 \label{eq:gms_sbp_minus_closure}
\end{equation}
The four boundary difference formulas for $\boldsymbol D_+$ act, respectively, on columns $1$--$3$, $1$--$4$, $N-1$--$N$, and $N-1$--$N$. Those for $\boldsymbol D_-$ act on columns $1$--$2$, $1$--$2$, $N-3$--N, and $N-2$--$N$. Together with the finite difference formulas used away from the boundaries, these boundary rows completely specify the matrices ${\boldsymbol D}_+$ and ${\boldsymbol D}_-$.

The finite difference operators above act on the uniform computational
coordinate $s$. Each computational grid is mapped onto its corresponding
physical radial domain through a monotonic function $r=r(s)$. Let
${\boldsymbol r}$ denote the physical radii at the grid points, and define
\begin{equation}
 {\boldsymbol D}_c
 =\frac{{\boldsymbol D}_++{\boldsymbol D}_-}{2},
 \qquad
 {\boldsymbol J}
 =\operatorname{diag}({\boldsymbol D}_c{\boldsymbol r}).
\end{equation}
The diagonal entries of ${\boldsymbol J}$ approximate the Jacobian $dr/ds$ at each grid point. They must remain positive so that radius increases monotonically across the domain. Applying the chain rule, $d/dr=(dr/ds)^{-1}d/ds$, gives the radial derivative matrices
\begin{equation}
 {\boldsymbol D}_{r,\pm}
 ={\boldsymbol J}^{-1}{\boldsymbol D}_\pm.
\end{equation}
Likewise, with $dr=(dr/ds)\,ds$, the quadrature matrix in physical radius is
\begin{equation}
 {\boldsymbol H}_r
 ={\boldsymbol H}_s{\boldsymbol J}.
 \label{eq:gms_mapped_sbp}
\end{equation}

In \texttt{MAQAM}, we apply ${\boldsymbol D}_{r,+}$ to the radial derivatives in the continuity equation ~\eqref{eq:gms_radial_continuity}, and in the definition $d(\delta\Phi_\ell)/dr=d_\ell$, Eq.~\eqref{eq:gms_phi_definition}. We apply  ${\boldsymbol D}_{r,-}$ to the derivatives in the radial momentum equation ~\eqref{eq:gms_radial_momentum}, and the radial Poisson
equation ~\eqref{eq:gms_radial_poisson}. This complementary assignment preserves the discrete integration by parts relation and
prevents grid scale alternating patterns from appearing as nonphysical numerical solutions. The quadrature matrix ${\boldsymbol H}_r$ is also used to evaluate the radial integrals in the
mode normalization and tidal overlaps \citep[e.g.,][]{hicken2013summation}.
\subsection{Matrix Assembly and Eigenvalue Search}
The projected equations form sparse matrices $\boldsymbol A_0$, $\boldsymbol A_1$, and $\boldsymbol A_2$ in Eq.~\eqref{eq:gms_quadratic_pencil}. Boundary conditions replace the corresponding differential rows. The rigid interior carries only gravitational variables, matched to those in the envelope.

To reduce disparities in coefficient magnitude, we set $\boldsymbol B_0=\boldsymbol A_0$, $\boldsymbol B_1=\Omega^2\boldsymbol A_1$, and $\boldsymbol B_2=\Omega^2\boldsymbol A_2$, then equilibrate the equations and unknowns with positive, nonsingular diagonal matrices $\boldsymbol D_L$ and $\boldsymbol D_R$:
\begin{equation}
\boldsymbol E_j=\boldsymbol D_L\boldsymbol B_j\boldsymbol D_R,\qquad j=0,1,2.
\label{eq:gms_equilibration}
\end{equation}
With $\boldsymbol x=\boldsymbol D_R\boldsymbol z$, this leaves the eigenvalues unchanged and gives $(\boldsymbol E_0+\kappa\boldsymbol E_1+\kappa^2\boldsymbol E_2)\boldsymbol z=0$. We use its companion linearization \citep{tisseur2001quadratic,fuller2014non},
\begin{equation}
\begin{aligned}
\boldsymbol L&=\begin{pmatrix}\boldsymbol 0&\boldsymbol I\\-\boldsymbol E_0&-\boldsymbol E_1\end{pmatrix},\qquad \boldsymbol R=\begin{pmatrix}\boldsymbol I&\boldsymbol 0\\\boldsymbol 0&\boldsymbol E_2\end{pmatrix},\\
\boldsymbol L\boldsymbol y&=\kappa\boldsymbol R\boldsymbol y,\qquad \boldsymbol y=\begin{pmatrix}\boldsymbol z\\\kappa\boldsymbol z\end{pmatrix},
\end{aligned}
\label{eq:gms_companion}
\end{equation}
where $\boldsymbol I$ and $\boldsymbol 0$ are identity and zero matrices. Algebraic constraints make the mass block singular; only finite physical roots are retained.

Because the matrices are large, computing the complete spectrum would be unnecessarily expensive. Instead, we use a shift invert restarted Arnoldi search to find the eigenvalues nearest a chosen target frequency \citep{sorensen1992implicit}. Multiple overlapping targets are used to cover both positive and negative rotating frame frequencies. The rotational frequency range is specified using $\kappa=\omega/\Omega$, while the dynamical and acoustic ranges are specified using $\sigma=\omega/\omega_{\rm dyn}$. 

Overlapping searches may return the same mode more than once. Duplicate candidates are identified by comparing their frequencies, dominant harmonic content, radial node counts, and eigenfunction overlaps. Each candidate
must also satisfy the original quadratic equation. For a candidate mode $(\kappa,{\boldsymbol x})$, we define
the normalized polynomial residual
\begin{equation}
 \epsilon_{\rm poly}
 =
 \frac{
 \left\|
 {\boldsymbol A}_0{\boldsymbol x}
 +\kappa\Omega^2{\boldsymbol A}_1{\boldsymbol x}
 +\kappa^2\Omega^2{\boldsymbol A}_2{\boldsymbol x}
 \right\|
 }{
 \left\|{\boldsymbol A}_0{\boldsymbol x}\right\|
 +|\kappa|\Omega^2
  \left\|{\boldsymbol A}_1{\boldsymbol x}\right\|
 +|\kappa|^2\Omega^2
  \left\|{\boldsymbol A}_2{\boldsymbol x}\right\|
 }.
 \label{eq:gms_polynomial_residual}
\end{equation}
Here $\|{\boldsymbol y}\|$ denotes the norm of a vector ${\boldsymbol y}$.  Writing Eq.~\eqref{eq:gms_companion} in the compact form
\begin{equation}
 {\boldsymbol L}{\boldsymbol y}
 =\kappa{\boldsymbol R}{\boldsymbol y},
\end{equation}
we also define the normalized residual of the linearized problem as
\begin{equation}
 \epsilon_{\rm comp}
 =
 \frac{
 \left\|
 {\boldsymbol L}{\boldsymbol y}
 -\kappa{\boldsymbol R}{\boldsymbol y}
 \right\|
 }{
 \left\|{\boldsymbol L}{\boldsymbol y}\right\|
 +|\kappa|
 \left\|{\boldsymbol R}{\boldsymbol y}\right\|
 }.
 \label{eq:gms_companion_residual}
\end{equation}
The residual $\epsilon_{\rm comp}$ measures how accurately the candidate satisfies the linearized problem solved by the eigensolver, whereas $\epsilon_{\rm poly}$ tests the recovered, unscaled mode directly in the original quadratic problem.

\subsection{Mode Acceptance and Convergence}
\label{subsec:gms_acceptance}
For the gas envelope calculation, before beginning the eigenvalue search, we verify that the matrices defining both equatorial parity problems contain only finite entries, that their boundary rows reproduce the mechanical and gravitational conditions derived above, and that the fluid variables are uncoupled from the gravity-only interior. For each radial domain, we evaluate both sides of the summation by parts identity in Eq.~\eqref{eq:gms_complementary_sbp} and require their relative difference to be smaller than $10^{-12}$. For each Arnoldi target, we define an acceptance interval, expressed in $\kappa$ for the rotational frequency search and in $\sigma$ for the dynamical  and acoustic frequency searches. The union of these overlapping intervals defines the total frequency range examined. A candidate mode is retained only if its frequency and reconstructed eigenfunction are finite, its real frequency lies within the corresponding acceptance interval, $|\mathfrak{Im}\{\kappa\}|\leq10^{-7}$, and $\epsilon_{\rm poly}\leq10^{-6}$. The residual $\epsilon_{\rm comp}$ verifies the solution of the linearized problem, but it is not used in place of $\epsilon_{\rm poly}$: the latter is evaluated after transforming the eigenvector back to the original variables and therefore tests the quadratic system that represents the physical oscillation equations. Candidates whose eigenfunctions contain grid scale alternating patterns or radial structure unresolved by the adopted grid are rejected.

The examples below uses an \texttt{ORCHARD} snapshot at $75\,{\rm Myr}$ with $M_{\rm p}=5.00M_\oplus$, $R_{\rm p}=2.72R_\oplus$, envelope mass fraction $0.10$, $R_{\rm I}=0.569R_{\rm p}$, and $R_{\rm Fe}=0.299R_{\rm p}$. 

The initial resolution comparison uses 60 interior and 100 envelope points at $\ell_{\max}=8$, then 80 and 140 points at the same angular truncation, and calculations at $\ell_{\max}=10$. We compare isolated acoustic and surface-gravity modes individually.

We require a separate and more rigorous angular convergence for the quadrupolar response inside the inertial interval, $|\kappa|=|\omega_{\rm rot}/\Omega|\leq2$.  In a shell, reflections can concentrate inertial waves into attractors and beams near critical latitudes, producing fine spatial structure and sharp spectral features \citep{ogilvie2004tidal,ogilvie2005wave,goodman2009dynamical}. Individual peaks may remain sensitive to resolution and regularization after integrated measures have stabilized \citep{ogilvie2013tides}. We therefore test the quadrupolar response of the compressible, self-gravitating envelope at finite frequency broadening.

To compute the forced response, we apply a quadrupolar potential with $U_{22}(r)=U_{22}(R_{\rm p})(r/R_{\rm p})^2$. Writing $U_\ell=\delta_{\ell2}U_{22}$, the RHS of Eqs.~\eqref{eq:gms_radial_momentum} and \eqref{eq:gms_horizontal_momentum} become $-\rho\,dU_\ell/dr$ and $-U_\ell/r$, respectively. Poisson's equation and the gravitational boundary conditions continue to apply to the induced potential $\delta\Phi$, excluding the applied potential. Define the quadrupolar Love number by
\begin{equation}
k_{22}=\frac{\delta\Phi_{22}(R_{\rm p})}{U_{22}(R_{\rm p})},
\label{Eq:Love_Number}
\end{equation}
where $\delta\Phi_{22}$ is the induced quadrupolar potential coefficient. At $q_{\rm rot}=0.01$, we evaluate the response over $-2\leq\kappa\leq2$ for $\ell_{\max}=6$ through 52. The frequency is displaced to $\kappa_{\rm c}=\kappa+i\gamma/\Omega$, with $\gamma>0$, providing causal broadening for the $\exp(-i\omega t)$ convention.  We use $\gamma/\Omega=0.02$ and $0.01$.

To compare spectra whose narrow peaks shift with resolution, we divide the interval into bins $\Delta_j$ of width $\Delta\kappa=0.1$. For angular truncation $L$, define
\begin{equation}
w_j^{(L)}=\int_{\Delta_j}|\mathfrak{Im}k_{22}^{(L)}|\,d\kappa,\qquad p_j^{(L)}=\frac{w_j^{(L)}}{\sum_k w_k^{(L)}}.
\end{equation}
The total variation distance between normalized binned spectra is
\begin{equation}
D_{\rm bin}(L,L')=\frac12\sum_j|p_j^{(L)}-p_j^{(L')}|.
\label{Eq:D_bin}
\end{equation}
It vanishes for identical distributions and equals unity for disjoint support.

\begin{figure}
\centering
\includegraphics[width=\linewidth]{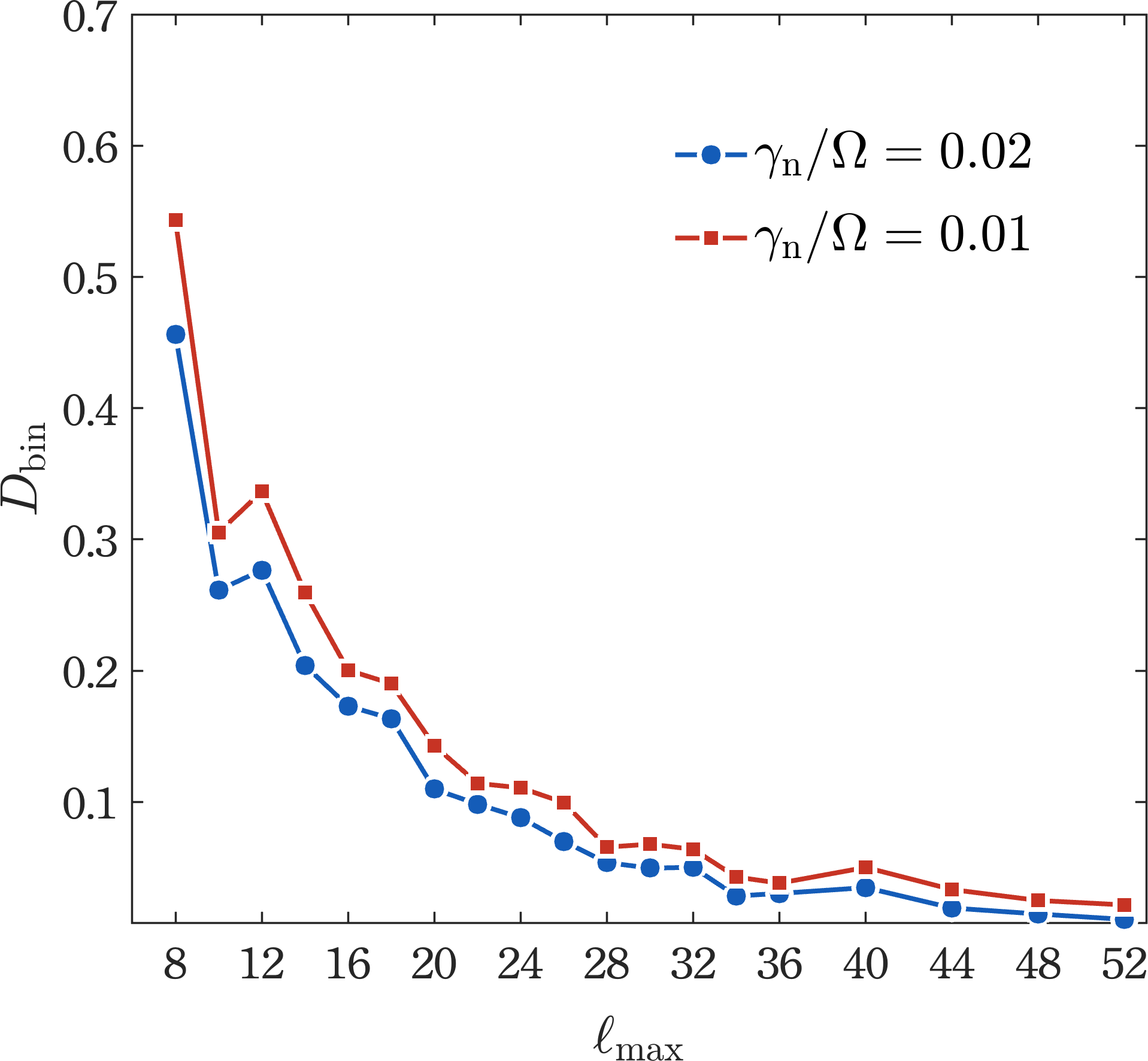}
\caption{Angular convergence of the quadrupolar inertial response. Total variation distance $D_{\rm bin}$ (Eq.~\ref{Eq:D_bin}) between the normalized distributions of $|\mathfrak{Im}\{k_{22}\}|$ in frequency bins of width $\Delta\kappa=0.1$, plotted against the upper angular truncation order, $\ell_{\max}$, for two numerical regularization parameters, $\gamma/\Omega=0.02$ and $0.01$.  The successive comparisons ending at $\ell_{\max}=48$ and 52 fall below $3\%$ for both regularizations.}
\label{fig:lmax_convergence}
\end{figure}

\begin{figure}
    \centering
\includegraphics[width=\linewidth]{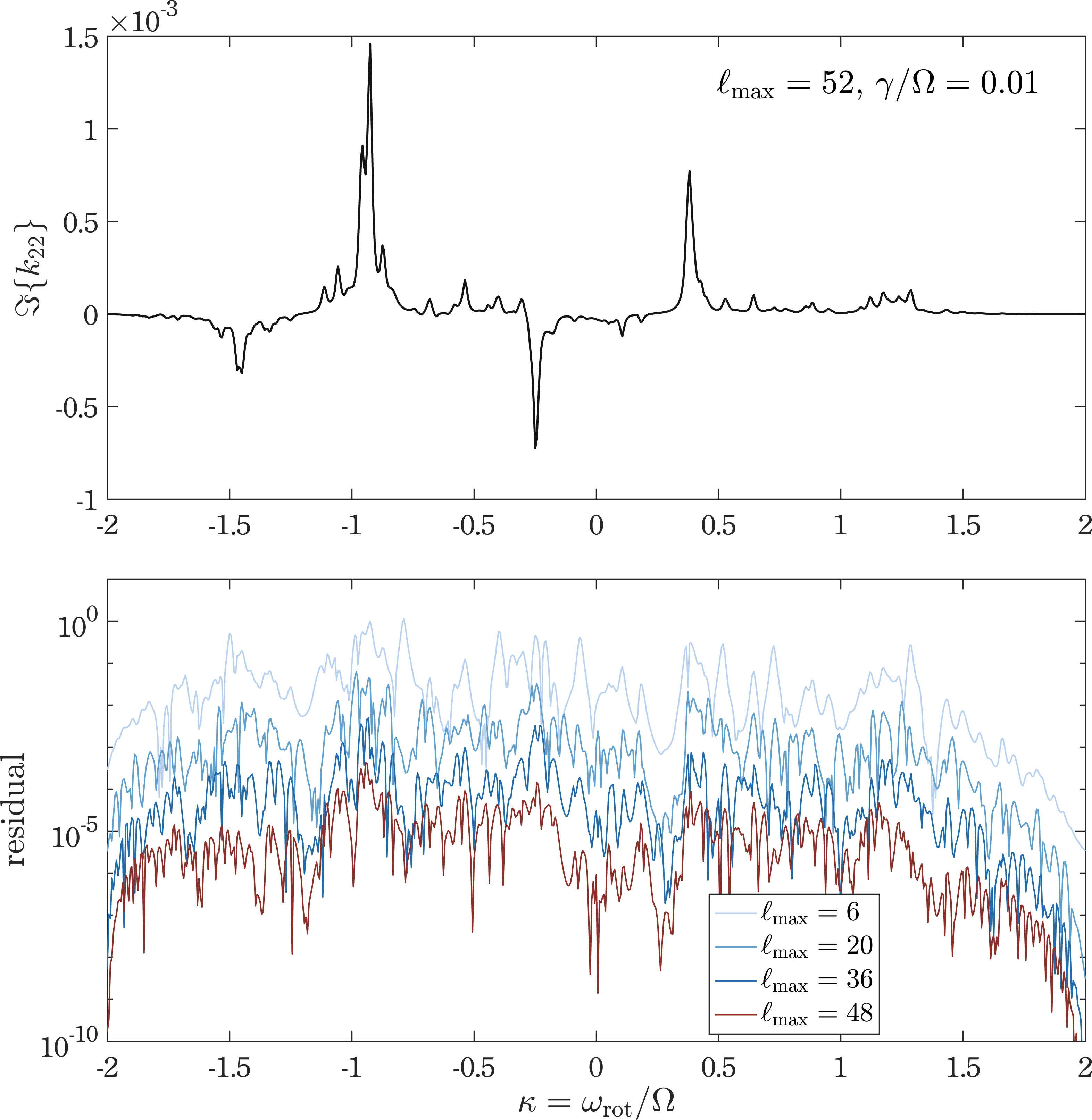}
    \caption{Resolved quadrupolar response and angular truncation order residuals in the inertial window.  Top: the spectrum of the imaginary part of the Love number (Eq.~\ref{Eq:Love_Number}) plotted for $\ell_{\max}=52$ and $\gamma/\Omega=0.01$.  Bottom: absolute differences from the $\ell_{\max}=52$ curve for selected lower truncations, normalized by $\max_\kappa|\mathfrak{Im}\{k_{22}^{(52)}\}|$. }
    \label{fig:Imk22_residuals}
\end{figure}

Figure~\ref{fig:lmax_convergence} shows that $D_{\rm bin}$ is not monotonic at low angular resolution, but falls below $3\%$ for the successive comparisons ending at $\ell_{\max}=48$ and 52. The upper panel of Figure~\ref{fig:Imk22_residuals} shows the Love number at $\ell_{\max}=52$, while the lower panel shows the residual between this reference curve and others with lower $\ell_{\max}$. A separate radial refinement test at $\ell_{\max}=52$ changes the full response curve by at most $1.2\%$ and changes $D_{\rm bin}$ by less than $0.4\%$. We therefore adopt $\ell_{\max}=52$ for the quadrupolar inertial-response calculation.  This test establishes convergence of the binned spectral distribution, not of the position and height of every narrow inviscid resonance.

\subsection{Benchmark Spectra}
\label{app:benchmark_GMSB0}
Having fixed the numerical resolution, we next illustrate the intrinsic normal-mode spectrum computed with \texttt{MAQAM}, independently of any orbit or tidal forcing.  We set $q_{\rm rot}=0.01$ and $m=2$.  Rotation couples several spherical harmonic degrees within one eigenfunction, so a rotating mode cannot in general be assigned a single $\ell$.  We therefore define its  cycle-averaged kinetic energy  at degree $\ell$, for unit-mode amplitude, by
\begin{equation}
 K_{\alpha\ell}
 =\frac{|\omega_\alpha|^2}{4}
 \int_{R_{\rm I}}^{R_{\rm p}}\!
 \rho r^2
 \left[
 |u_{\alpha\ell}|^2+
L_\ell\left(
 |v_{\alpha\ell}|^2+|w_{\alpha\ell}|^2
 \right)
 \right]\,dr,
\label{eq:gms_angular_kinetic_energy}
\end{equation}
for $ L_\ell=\ell(\ell+1),$
and the energy weighted effective degree
\begin{equation}
 \ell_{{\rm eff},\alpha}
 =\frac{\sum_\ell \ell K_{\alpha\ell}}
        {\sum_\ell K_{\alpha\ell}}.
 \label{eq:gms_effective_degree}
\end{equation}
Thus departures from an integer measure rotational mixing.

\begin{figure}
    \centering
    \includegraphics[width=\linewidth]{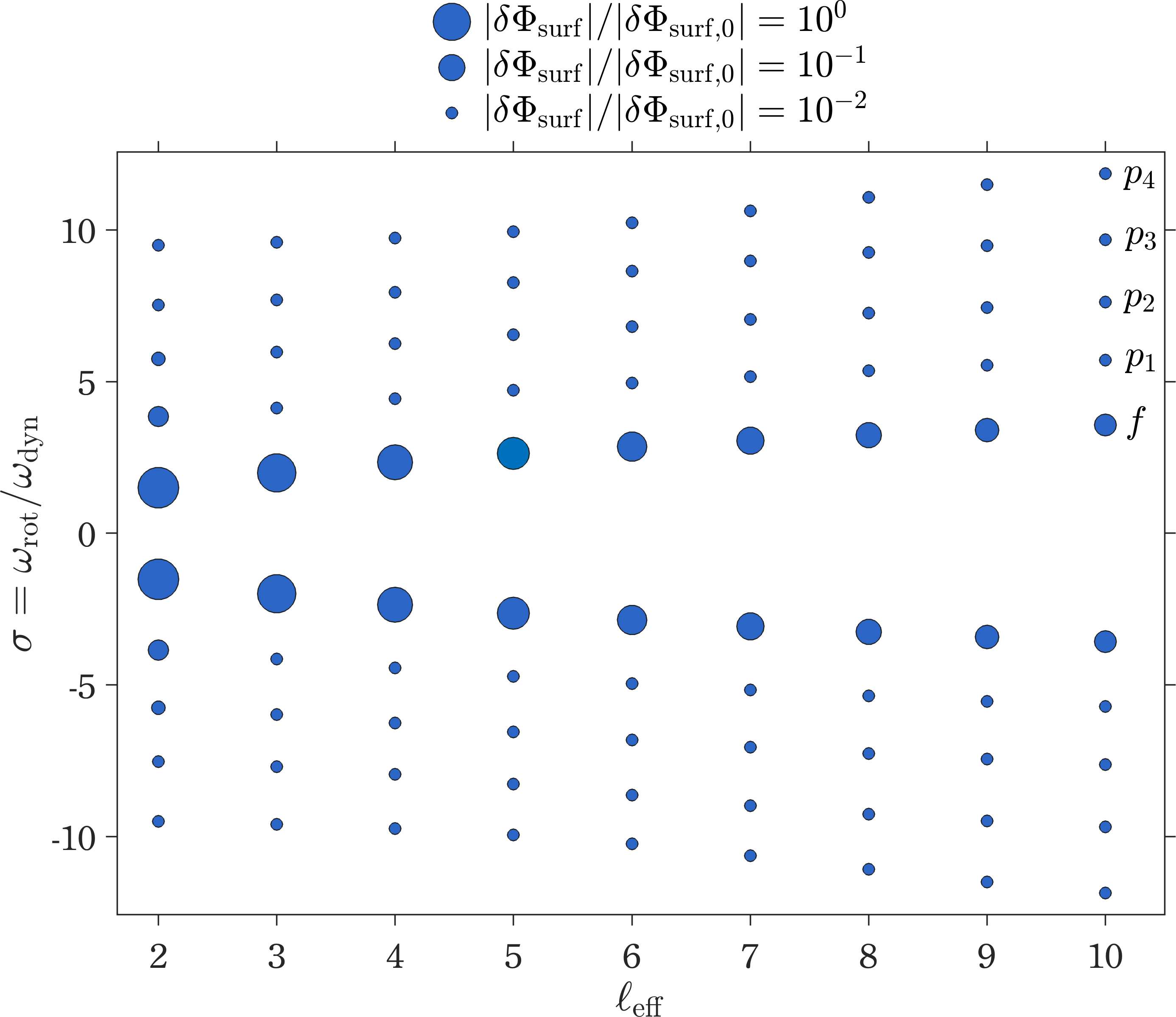}
    \caption{Dynamical and acoustic modes of the inner Sub-Neptune 
    $q_{\rm rot}=0.01$.  The signed rotating frame frequency
    $\sigma=\omega/\omega_{\rm dyn}$ is plotted against the
    effective degree defined in Eq.~\eqref{eq:gms_effective_degree}.  The labelled sequences are the surface gravity $f$ mode and the first four acoustic branches, $p_1-p_4$.  Symbol size gives the relative normalized surface gravitational perturbation from Eq.~\eqref{eq:gms_surface_gravity_proxy}.}
    \label{fig:gms_fp_spectrum}
\end{figure}

Figure~\ref{fig:gms_fp_spectrum} shows the dynamical and acoustic catalogue at $\ell_{\max}=48$ for $\ell_{\rm eff}\leq10$. All 45 positive frequency $f$--$p_4$ modes and the corresponding negative frequency branches are already recovered at $\ell_{\max}=10$. The median frequency change from 10 to 48 is $0.3\%$. At this slow rotation, $\ell_{\rm eff}$ remains close to an integer.

For this figure we compare gravitational signatures at equal maximum angular norm of fractional displacement,
\begin{align}\nonumber
 \Xi_\alpha&=
\max_{R_{\rm I}\le r\le R_{\rm p}}
\frac{1}{r}
\left(\int |\boldsymbol{\xi}_\alpha|^2\,d\Omega\right)^{1/2}
 \\
 &=\max_{R_{\rm I}\leq r\leq R_{\rm p}}
 \frac{1}{r}
 \left[
 \sum_\ell
 \left\{
 |u_{\alpha\ell}|^2+
L_\ell\left(
 |v_{\alpha\ell}|^2+|w_{\alpha\ell}|^2
 \right)
 \right\}
 \right]^{1/2}.
\end{align}
Define the displacement-normalized surface potential and its largest value among the plotted modes by
\begin{equation}
\begin{aligned}
|\delta\Phi_{{\rm surf},\alpha}|&=\Xi_\alpha^{-1}\left[\sum_\ell|\delta\Phi_{\alpha\ell}(R_{\rm p})|^2\right]^{1/2},\\
|\delta\Phi_{{\rm surf},0}|&=\max_\beta|\delta\Phi_{{\rm surf},\beta}|.
\end{aligned}
\label{eq:gms_surface_gravity_proxy}
\end{equation}
Symbol sizes in Figure~\ref{fig:gms_fp_spectrum} represent $|\delta\Phi_{{\rm surf},\alpha}|/|\delta\Phi_{{\rm surf},0}|$. The $f$ branch has the lowest frequencies within the displayed $f$--$p$ sequences and generally the largest surface gravitational perturbations; acoustic frequencies increase with degree and radial order \citep[e.g.,][]{fuller2014saturn,dewberry2021constraining}.

For the inertial spectrum, define the dimensionless, normalization-independent surface potential
\begin{equation}
\begin{aligned}
\widehat{\delta\Phi}_{{\rm surf},\alpha}&=\frac{\displaystyle\max_\theta\left|\sum_\ell\delta\Phi_{\alpha\ell}(R_{\rm p})Y_\ell^2(\theta,0)\right|}{\mathcal GM_{\rm p}/R_{\rm p}}\\
&\quad\times\left(\frac{2M_{\rm p}R_{\rm p}^2|\omega_\alpha|}{|b_\alpha|}\right)^{1/2}.
\end{aligned}
\label{eq:gms_normalized_surface_potential}
\end{equation}
This dimensionless quantity compares surface gravitational perturbations
at equal rotating mode normalization.  Here even and odd denote equatorial parity.  For $m=2$, the even block has
spheroidal components with even $\ell-m$ and toroidal components with odd $\ell-m$. These assignments are reversed in the odd block.

\begin{figure}
\centering
\includegraphics[width=\linewidth]{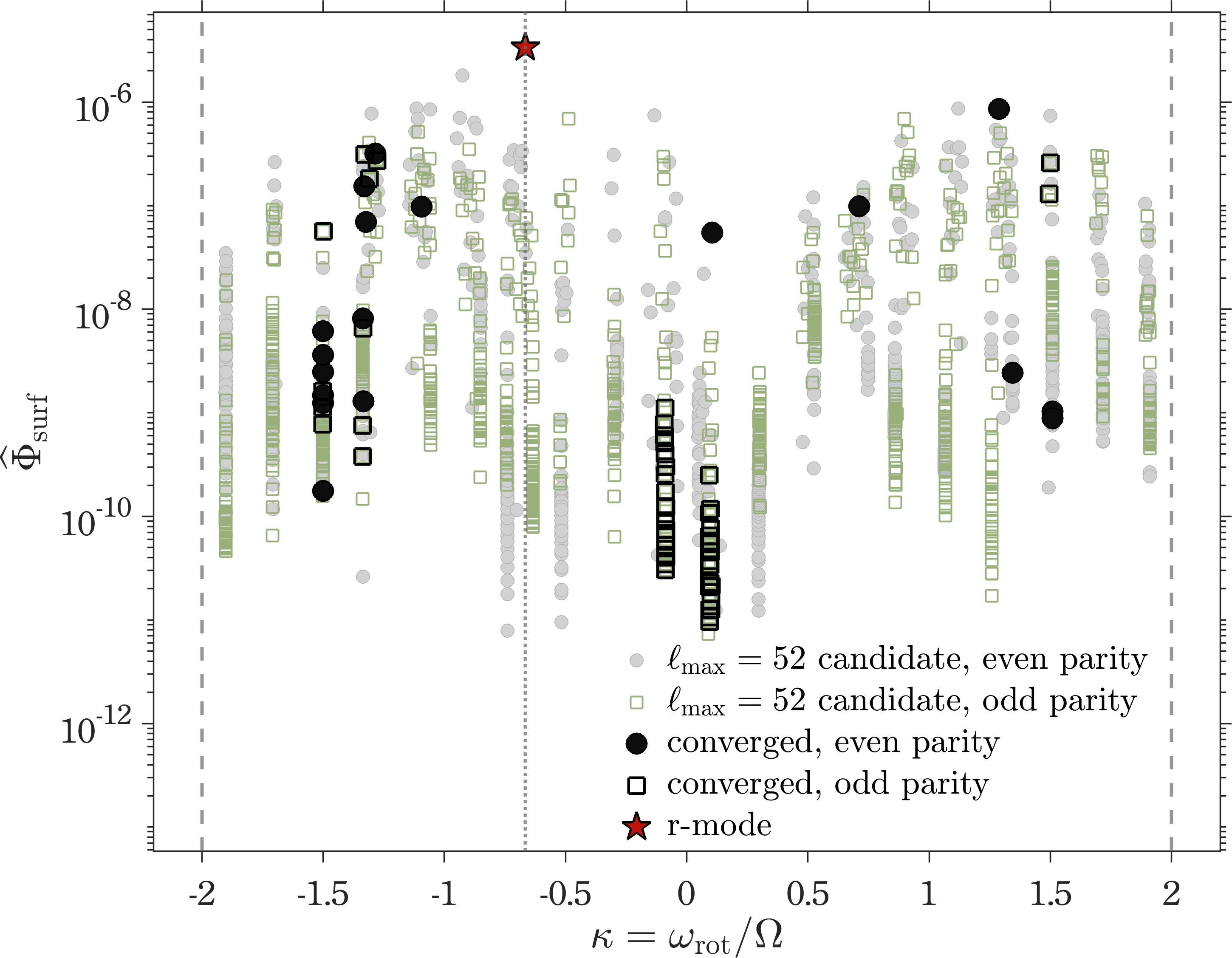}
\caption{Inertial modes computed with \texttt{MAQAM} at $q_{\rm rot}=0.01$ and $m=2$, with ordinate given by Eq.~\eqref{eq:gms_normalized_surface_potential}. Faint circles and squares show 1672 finite-grid candidates at $\ell_{\max}=52$ in the even and odd parity blocks. Black symbols identify 73 roots matched in frequency and eigenfunction structure between $\ell_{\max}=48$ and 52. Dashed lines mark $\kappa=\pm2$; the dotted line marks the sectoral Rossby frequency $\kappa=-2/3$.}
\label{fig:gms_inertial_surface_spectrum}
\end{figure}

Of the 1672 finite grid roots shown, 73 can be matched between $\ell_{\max}=48$ and 52 in frequency and radial, toroidal, and angular structure. We restrict individual interpretation to this subset. This reduction does not imply that the remaining
candidates are nonphysical, but is rather a conservative choice that limits individual interpretation to the roots that pass all convergence tests.  Their toroidal kinetic-energy fraction is
\begin{equation}
 f_{{\rm tor},\alpha}
 =
 \frac{
 \displaystyle
 \sum_\ell L_\ell
 \int_{R_{\rm I}}^{R_{\rm p}}
 \rho r^2|w_{\alpha\ell}|^2\,dr
 }{
 \displaystyle
 \sum_\ell
 \int_{R_{\rm I}}^{R_{\rm p}}
 \rho r^2
 \left[
 |u_{\alpha\ell}|^2+
 L_\ell\left(
 |v_{\alpha\ell}|^2+|w_{\alpha\ell}|^2
 \right)
 \right]\,dr
 }.
 \label{eq:gms_toroidal_fraction}
\end{equation}
The 72 ordinary inertial modes have $0.48\leq f_{{\rm tor},\alpha}\leq0.68$. The sectoral Rossby mode at $\kappa\simeq-2/3$ has $f_{{\rm tor},\alpha}>1-10^{-9}$ and is thus toroidal to numerical precision. Its odd equatorial parity excludes direct coupling to the aligned $Y_2^2$ tide, which excites only the even block \citep[e.g.,][]{papaloizou1978non,ogilvie2004tidal}.

\section{Fluid Mantle Modes}
\label{app:gms_magma_ocean}
The sampled \texttt{ORCHARD} models retain a fully molten mantle through the first $1\,{\rm Gyr}$. We treat it as an inviscid fluid shell over a rigid iron core, solving its oscillations separately from those of the gas envelope. The perturbation equations, harmonic expansion, and Coriolis projection of Appendix~\ref{app:gms_continuum} apply with the mantle profiles $\rho$, $g$, $c_s^2$, and $N^2$ on $R_{\rm Fe}<r<R_{\rm I}$.

The core contributes to the equilibrium gravity but has no displacement or density perturbation. Impenetrability and regular gravitational continuation require
\begin{align}
u_\ell(R_{\rm Fe})&=0,
\label{eq:gms_mo_core_mechanical}\\
\left[d_\ell-\frac{\ell}{r}\delta\Phi_\ell\right]_{r=R_{\rm Fe}}&=0.
\label{eq:gms_mo_core_gravity}
\end{align}
Tangential slip is unrestricted. At the mantle top we impose a free mechanical boundary and match the perturbed potential directly to vacuum:
\begin{equation}
\Delta p_\ell(R_{\rm I})=\left[\delta p_\ell-\rho g u_\ell\right]_{r=R_{\rm I}}=0,
\label{eq:gms_mo_top_pressure}
\end{equation}
\begin{equation}
\left[d_\ell+\frac{\ell+1}{r}\delta\Phi_\ell+4\pi\mathcal G\rho u_\ell\right]_{r=R_{\rm I}}=0.
\label{eq:gms_mo_top_gravity}
\end{equation}
Boundary coefficients are evaluated on the mantle side. These conditions omit the dynamical pressure and perturbed gravity of the overlying gas; the separate component spectra therefore exclude mantle--envelope interface and mixed modes.

The discretization, quadratic eigenproblem, companion linearization, and overlapping frequency searches follow Appendix~\ref{app:gms_numerics}. The mantle calculation requires a relative defect below $10^{-11}$ in the mapped SBP identity. Retained candidates have finite frequencies and eigenfunctions and $\epsilon_{\rm poly}\leq10^{-6}$; duplicates are identified by frequency and mass-weighted displacement overlap. Stable, nonzero-frequency modes enter the tidal calculation with the canonical normalization in Eq.~\eqref{eq:gms_canonical_action}.

\section{Reduced model of the parametric instability excitation}
\label{app:parametric_excitation_example}

We integrate Eq.~\eqref{eq:sminus_linear_system} for a gas envelope inertial mode and chain libration mode $a=3$, retaining their relative phase and computing $J_a=|c_a|^2$ and $J_{1\alpha}=|c_{1\alpha}|^2$. Cooling changes the internal-mode frequency $\omega_{1\alpha,{\rm rot}}$. We keep the reference chain frequency, eigenvectors, coupling coefficient, and damping rates constant as a local approximation. The chain begins with an rms free eccentricity amplitude of $10^{-6}$, and $t=0$ corresponds to a planetary age of $10\,{\rm Myr}$.

We assume rapid relaxation of the spin toward pseudo-synchronism. For a weak-friction background tide, we estimate the spin relaxation time as
\begin{equation}
t_{\rm syn}\simeq
\frac{7C_1}{2m_1a_{1}^2}\frac{f_a}{\gamma_a},
\end{equation}
where $f_a$ is the projection factor of Eq.~\eqref{eq:chain_damping_estimate}. For  $C_1/(m_1R_1^2)=0.25$, and $\gamma_a=10^{-9}\,{\rm yr}^{-1}$, $t_{\rm syn}\simeq2.8\times10^3\,{\rm yr}$, much shorter than the cooling and excitation timescales.

We use the prograde quadrupolar component with $m_{1\alpha}=2$ and $b=3$ corresponding to the dominant eccentricity tide, and whose phase is
$\Theta_{1b}=b\lambda_1-2\theta_1-(b-2)\varpi_1$.
Neglecting apsidal precession, its forcing frequency is
\begin{equation}
\sigma_{1b}=bn_1-m_{1\alpha}\Omega.
\label{eq:example_tidal_frequency}
\end{equation}
The pair interaction supplies equal action fluxes to both oscillators:
\begin{equation}
\mathcal F
=\dot J_a+2\gamma_aJ_a
=\dot J_{1\alpha}+2\gamma_{1\alpha}J_{1\alpha}.
\label{eq:example_orbital_action_flux}
\end{equation}
Internal mode damping transfers angular momentum
$2m_{1\alpha}\gamma_{1\alpha}J_{1\alpha}$ per unit time to the mean rotation. Maintaining pseudo-synchronism transfers the excess back to the orbit. We determine the orbital response from energy and angular momentum balance. The inner reference orbit has an independent scale $u(t)$, and we assume the outer three share scale $v(t)$:
\begin{equation}
a_{1}=u(t)a_{1,\star},
\qquad
a_{j}=v(t)a_{j,\star},
\quad j=2,3,4,
\label{eq:example_common_orbital_scale}
\end{equation}
with $u(0)=v(0)=1$. The orbital energy and angular momentum are
\begin{equation}
E_{\rm orb}=\left\langle\mathcal H_{\rm orb}\right\rangle,
\qquad
L_{\rm orb}=\left\langle\sum_i(\Lambda_i-\Gamma_i)\right\rangle,
\end{equation}
evaluated from the canonical reconstruction of both semimajor axis and eccentricity oscillations. The spin and inertial frame mode energies are
\begin{equation}
E_{\rm spin}=\frac12 C_1\Omega^2,
\qquad
E_{1\alpha,{\rm in}}
=(\omega_{1\alpha,{\rm rot}}+m_{1\alpha}\Omega)
J_{1\alpha}.
\end{equation}
Including the interaction energy $\mathcal{H}_-$, define the  mechanical energy
\begin{equation}
E
=E_{\rm orb}+E_{\rm spin}+E_{1\alpha,{\rm in}}+\mathcal H_-.
\end{equation}
The heating rate, to leading order in eccentricity, is
\begin{equation}
\mathcal D
=2\gamma_{1\alpha}\omega_{1\alpha,{\rm rot}}J_{1\alpha}
+\frac{GM_\star m_1}{a_1}
\frac{\gamma_a}{f_a}\left\langle e_1^2\right\rangle.
\end{equation}
The second term is the background circularization heat,
including the dissipation responsible for chain damping.
The two orbital scales are determined by
\begin{align}
\dot L_{\rm orb}
&=-\frac{d}{dt}
\left(C_1\Omega+m_{1\alpha}J_{1\alpha}\right),\\
\dot E_{\rm orb}
&=\mathcal P_{\rm cool}-\mathcal D
-\frac{d}{dt}
\left(E_{\rm spin}+E_{1\alpha,{\rm in}}+\mathcal H_-\right),
\label{eq:example_orbital_balances}
\end{align}
where $\mathcal P_{\rm cool}$ is the rate at which cooling and contraction change the energy stored in the planet’s spin and internal oscillation.

The inner mean motion evolves as $n_1=n_1(0)u^{-3/2}$. With the chain frequency fixed, the detuning is
\begin{align}
\Delta_-(t)={}&\Delta_-(0)+b[n_1(t)-n_1(0)]
\nonumber-m_{1\alpha}[\Omega(t)-\Omega(0)]
\nonumber\\
&-[\omega_{1\alpha,{\rm rot}}(t)
-\omega_{1\alpha,{\rm rot}}(0)].
\label{eq:example_time_dependent_detuning}
\end{align}
Orbital evolution therefore changes both the tidal and internal mode frequencies.

The third panel of Figure~\ref{fig:two_mode_simulation} shows the spatial rms scalar eccentricity displacement maximized over libration phase. The fourth shows the eccentricity oscillation amplitudes
$e_{i,\star}\pm\mathcal E_{ia}\sqrt{J_a(t)}$,
reconstructed using Eq.~\eqref{eq:Eia_mode_shape}. The bottom panel shows the rotating-frame energy
$E_{1\alpha}=\omega_{1\alpha,{\rm rot}}J_{1\alpha}$
relative to the reference capacity from
Eq.~\eqref{eq:geometric_mode_energy_capacity}.

\end{document}